\documentclass{aa} 
\usepackage{booktabs}
\usepackage{multirow}
\usepackage{adjustbox}
\usepackage{threeparttable}
\usepackage{placeins}
\usepackage{graphicx} 
\usepackage{xcolor}
\usepackage{txfonts}
\usepackage[breaklinks]{hyperref}
\usepackage{url}
\usepackage{flexisym}
\usepackage{float}
\usepackage{svg}
\usepackage[normalem]{ulem}
\newcommand{\orcid}[1]{\href{https://orcid.org/#1}{\includesvg[width=10pt]{plots/orcid.svg}}}
\let\svthefootnote\thefootnote
\newcommand\freefootnote[1]{%
  \let\thefootnote\relax%
  \footnotetext{#1}%
  \let\thefootnote\svthefootnote%
}

\begin{document}

   \title{16 new quasars at the end of the reionization unveiled by self-supervised learning}

   \author{L.N. Mart\'inez-Ram\'irez 
          \inst{1,2,3,4}\orcid{0009-0003-5506-5469} \and Julien Wolf  \inst{1, 10}\orcid{0000-0003-0643-7935} \and Silvia Belladitta  \inst{1,5, 10}\orcid{0000-0003-4747-4484}  \and Eduardo Bañados  \inst{1}\orcid{0000-0002-2931-7824}
          F. E. Bauer \inst{6}\orcid{0000-0002-8686-8737} \and Raphael E. Hviding  \inst{1}\orcid{0000-0002-4684-9005} \and Daniel Stern \inst{7}\orcid{0000-0003-2686-9241} 
          \and Chiara Mazzucchelli\inst{8}\orcid{0000-0002-5941-5214} \and Romain A. Meyer \inst{9}\orcid{0000-0001-5492-4522}
          \and Ezequiel Treister\inst{6}\orcid{0000-0001-7568-6412} \and Federica Loiacono\inst{5}\orcid{0000-0002-8858-6784}
          } 

   \institute{Max Planck Institut f\"ur Astronomie, K\"onigstuhl 17, D-69117, Heidelberg, Germany \\\email{lmartinez@mpia.de} 
\and
Instituto de Astrof\'isica, Facultad de F\'isica, Pontificia Universidad Cat\'olica de Chile Av. Vicu\~na Mackenna 4860, 7820436 Macul, Santiago, Chile
\and
Fakult\"at f\"ur Physik und Astronomie, Universit\"at Heidelberg Im Neuenheimer Feld 226, 69115 Heidelberg, Germany
\and 
Millennium Institute of Astrophysics (MAS), Nuncio Monseñor Sótero Sanz 100, Providencia, Santiago, Chile
\and
INAF – Osservatorio di Astrofisica e Scienza dello Spazio, Via Gobetti 93/3, I-40129, Bologna, Italy
\and
Instituto de Alta Investigaci{\'{o}}n, Universidad de Tarapac{\'{a}}, Casilla 7D, Arica, Chile
\and
Jet Propulsion Laboratory, California Institute of Technology, 4800 Oak Grove Drive, Pasadena, CA, 91109, USA
\and
Instituto de Estudios Astrof\'{\i}sicos, Facultad de Ingenier\'{\i}a y Ciencias, Universidad Diego Portales, Avenida Ejercito Libertador 441, Santiago, Chile.
\and
Department of Astronomy, University of Geneva, Chemin Pegasi 51, 1290 Versoix, Switzerland
}
   \date{Received ; accepted}

  \abstract{Luminous quasars at \textit{z}  $>$ 6 are key probes of early supermassive black hole (SMBH) growth, massive galaxy evolution, and intergalactic medium properties during cosmic reionization. However, their discovery is very challenging due to their scarcity and overwhelming contamination. Foreground ultracool dwarfs (UCDs) outnumber $z>6$ quasars by 2-4 orders of magnitude. In this work, we leveraged the extensive coverage of DESI Legacy Survey DR10 to conduct a self-supervised search for quasars at \textit{z}  $>$ 6, directly analyzing multiband optical images and minimizing the biases of the traditional catalog-driven color-color selection criteria.
  By applying a contrastive learning (CL) method followed by spectral energy distribution (SED) fitting prioritization, we identified 1139 high-priority quasar candidates, for which we expect a competitive 1:1 quasar-to-UCD ratio based on the literature samples. We spectroscopically confirm 16 new quasars at $z = 5.94-6.45$, achieving a $45\%$ success rate. 
  Remarkably, all 16 objects are relatively bright (M$_{1450} < $  $-25.5$) quasars, including several with unusual properties such as narrow \ion{Ly}{$\alpha$} emission (FWHM $\lesssim$ 2600 km\,s$^{-1}$), strong \ion{Ly}{$\alpha$}+\ion{N}{V} emission with an equivalent width $>100 \,\AA$, and a mild observed-frame red near-infrared (NIR) continua ($z - J > 0.4$). Notably, three of them would have been missed by traditional color–color selections.
  These results highlight the power of self-supervised machine learning, combined with SED fitting prioritization, to uncover rare, distant sources beyond the limitations of conventional techniques.
  Our approach offers a scalable and robust framework for data mining and can be readily extended to forthcoming wide-field surveys such as Rubin/LSST, 4MOST, \textit{Euclid}, and Roman.
  These applications will advance the census of high-redshift quasars, potentially extend the redshift frontier, and improve constraints on SMBH formation and evolution in the first billion years of the Universe.
  }

   \keywords{quasar --
                high-redshift--
                machine learning --
                imaging surveys
               }
   \maketitle
%
\titlerunning
\authorrunning
\section{Introduction}

\freefootnote{\hspace{-3mm}$^{10}$ The second and the third authors contributed equally to the manuscript.}

The accretion process in quasars makes them the most luminous non-transient sources in the Universe, outshining their host galaxies and dominating across the electromagnetic spectrum. The energy and momentum released by quasars strongly interact with the surrounding gas, influencing the evolution of their host galaxies and the environment \citep{fabian2012observational}. This process, referred to as active galactic nucleus (AGN) feedback, is crucial in cosmological simulations to reproduce realistic galaxy population properties \citep[e.g.,][]{kauffmann2000unified, granato2004physical, di2005energy, springel2005modelling, kaviraj2017horizon, bustamante2019spin}. Empirical correlations between supermassive black hole (SMBH) mass and galaxy properties, such as stellar bulge mass, velocity dispersion, and luminosity, support a coevolutionary scenario between quasars and their hosts \citep[e.g.,][]{tremaine2002slope, marconi2003relation, haring2004black, mcconnell2013revisiting, kormendy2013coevolution}. However, observations of $\gtrsim 100$ high-redshift quasars ($z\gtrsim6$; i.e., $\sim $ 1 Gyr after the Big Bang) suggest that this relation may evolve with redshift, although the nature and extent of this evolution remain uncertain. A subset of SMBHs at early cosmic times seems to be more massive than expected from the local SMBH mass and host galaxy total stellar mass (M$\bullet$–M${*}$) correlation \citep[e.g.,][]{wang2016probing, decarli2018alma, neeleman2021kinematics, harikane2023jwst}, potentially suggesting a faster growth of black holes relative to their host galaxies, or the existence of more massive black hole seeds. Expanding the quasar population at $z > 5$ and pushing the magnitude frontier to fainter sources, along with a comprehensive physical characterization are essential to test these coevolution scenarios at early cosmic times.

Neutral intergalactic medium (IGM) absorbs photons blueward of \ion{Ly}{$\alpha$}, producing the distinctive \ion{Ly}{$\alpha$} break in the quasars' spectra at rest frame $\sim 1216$\,\AA. Quasars at high redshift appear undetectable or faint in photometric filters covering wavelengths below \ion{Ly}{$\alpha$}, creating a red color in photometric indices involving the dropout band: this is the basis of the dropout selection technique \citep[e.g.,][]{warren1987first}. However, foreground ultracool M, L, and T dwarfs also exhibit red spectra that can mimic the dropout signature and a point-like morphology resembling that of quasars. Moreover, while the number density of quasars declines exponentially at \textit{z} $>5$ \citep{jiang2016final}, dropping to fewer than one with M$_{1450}$ $<-26$ per comoving Gpc$^3$ at \textit{z} $\sim 6$ \citep[][]{schindler2023pan}; ultracool dwarfs (UCDs) associated with the thin and thick disk of the Milky Way remain 2–4 orders of magnitude more numerous \citep[e.g.,][]{barnett2019euclid}, making them the primary astrophysical contaminants in high-redshift quasar searches. Therefore, it is necessary to analyze a large area of the sky in order to identify high-redshift quasar candidates.

The advent of wide-field optical surveys in the early 2000s led by the Sloan Digital Sky Survey \citep[SDSS;][]{york2000others}, opened the way to quasar discoveries at $z>5$ \citep[e.g.,][]{fan1999high, fan2001survey} via color drop-out selection. Then surveys, such as the Canada-France-Hawaii Telescope Legacy Survey (CFHTLS), Pan-STARRS1 \citep[PS1,][]{chambers2016pan}, and Dark Energy Survey \citep[DES,][]{abbott2018dark}, significantly expanded the sample \citep[e.g.,][]{banados2016pan, banados2023pan, reed2017eight, mazzucchelli2017no, wolf2024srg,belladitta2019extremely, belladitta2020first, belladitta2025discovery}. The vast majority of these quasars have been selected by conservative color criteria to minimize contamination. As a consequence, most known quasars at $z > 5$ exhibit similar properties: typically unobscured, with massive black holes \citep[$\sim4\times10^{7} \mathrm{M}_\odot$ to $\sim10^{10}\mathrm{M}_\odot$;][]{fan23} and moderate-to-high accretion rates \citep[median  L$_{\mathrm{Bol}}$/L$_{\mathrm{Edd}}$ = 0.79 and 0.72, respectively;][]{fan23, mazzucchelli2023xqr}, leaving potentially important populations underrepresented, such as red or obscured quasars, and objects with unusual emission line properties.

Machine learning, particularly traditional supervised methods such as Random Forests and XGBoost, has increasingly been applied to quasar selection, demonstrating strong performance up to intermediate redshifts \citep[$z \sim 5$; e.g.,][]{nakoneczny2019catalog, yeche2020preliminary, jin2019efficient, schindler2019extremely}. Even unsupervised clustering approaches \citep[e.g.,][]{logan2020unsupervised} and comparative studies of unsupervised, semi-supervised, and fully supervised methods \citep[e.g.,][]{clarke2020identifying} have been explored for the common star-galaxy-quasar classification in large surveys. More recently, machine-learning techniques have been extended to a higher redshift regime ($z \gtrsim 5$). For example, \citet{wenzl2021random} applied a Random Forest classifier to Pan-STARRS1 and WISE data to identify $z > 4.7$ quasars, achieving successful spectroscopic confirmations. However, these efforts heavily rely on catalog-level photometric measurements and flux ratios, which are sensitive to contamination from artifacts and systematics in the source extraction process. Furthermore, supervised methods are often constrained by the availability of balanced and representative training samples; rare populations such as red high-redshift quasars are typically underrepresented, limiting classification reliability in these regimes.

Recent efforts to apply multiband imaging to the discovery of strongly lensed quasars have identified dozens of candidates using deep learning and spectral energy distribution (SED) modeling that successfully passed visual inspection. These systems, however, remain at the candidate stage pending spectroscopic confirmation \citep{andika2023spectral}. The work by \citet{byrne2024quasar} marks the first application of contrastive learning (CL), a fully self-supervised representation-learning method, to optical imaging \citep[DES DR2,][]{abbott2021dark}, leading to the discovery of three quasars at $z = 5.94-6.07$. Unlike conventional supervised methods, which require labeled training sets and can suffer from bias, CL learns by contrasting data samples against each other without any prior labels, making it well-suited to uncover rare or atypical quasar populations. Empirical studies show that CL can perform as well as, or even better than, supervised training, producing representations that transfer more effectively and remain robust across different datasets \citep{chen2020simple, karthik2021tradeoffs}. Also, the work by \cite{sarmiento2021capturing} demonstrated that CL is more robust against nonphysical correlations led by observational effects in astronomical data than unsupervised dimensionality reduction algorithms such as principal component analysis \citep[PCA;][]{pearson1901principal}. Building on this promise, we extend CL-based quasar selection to test whether it enables the recovery of diverse high-redshift quasars with lower contamination and fewer selection biases than supervised alternatives.

In this paper, we apply a self-supervised CL approach to imaging data from the DESI Legacy Survey DR10 \citep[hereafter LS DR10;][]{dey2019overview} to identify $z \gtrsim 5.5$ quasar candidates. We demonstrate the effectiveness of our method with the spectroscopic confirmation of 15 new quasars, along with the reidentification of 3 previously published quasars, out of 40 observed candidates, corresponding to a 45$\%$ success rate. Several of these quasars exhibit narrow \ion{Ly}{$\alpha$} emission lines, red near-infrared (NIR) continua, and high \ion{Ly}{$\alpha$} EWs, and thus they are part of populations typically underrepresented. Section \ref{sec:preselection} describes the catalog-based preselection of $i$-dropout candidates used for training. Section \ref{sec:CL} outlines the CL framework and its implementation, and presents the resulting low-dimensional embedded space. In Section \ref{sec:followup}, we detail the candidate prioritization strategy and follow-up observations, and in Section \ref{sec:new_qsos}, we present the newly spectroscopically confirmed quasars and UCD contaminants.

Throughout this paper we adopt $\Lambda$CDM cosmology with H$_0 = 67.4$ km s$^{-1} \mathrm{Mpc}^{-1}$, $\Omega_{\mathrm{m}} = 0.315$ and $\Omega_{\Lambda} = 0.685$ \citep{aghanim2020planck}. We report magnitudes in the AB system.

\section{Preselection of \textit{i}-dropout candidates}
\label{sec:preselection}

To ensure effective training of our CL model, we first apply a preselection to the input dataset. Although larger batch sizes typically improve CL performance \citep{chen2020simple}, the underlying neural network can easily get sidetracked by the most dominant features, such as sizes or extended morphology, if the training set is not properly delimited. To mitigate this and encourage the learning of the most fundamentally relevant features, we restrict the input to red and nearly point-like sources. This preselection filters out contaminants such as extended galaxies, asteroids, satellites, and diffraction spikes in the photometry while minimizing the impact of irrelevant features.  

The procedure to select $z > 5.5$ quasar candidates is divided in three major steps: (1) The selection of a preliminary set of candidates from photometric optical catalogs by applying loose morphological and color constraints; (2) the training of the CL algorithm with \textit{g}, \textit{r}, \textit{i}, and \textit{z} band optical images to produce an embedded low-dimensional representation of \textit{i}-dropout sources, and the subsequent identification of candidates and contaminants guided by labeled data; (3) SED fitting procedure using quasar, galaxy and UCD templates to prioritize for follow up observations the sources with reliable photometry and SEDs more consistent with a quasar model.

\subsection{Photometric catalog} 
\label{sec_photometric_cat}

We selected our candidates from LS DR10, which offers deep optical imaging over $> 20\,000$ deg$^2$ of sky, including substantial coverage of the underexplored southern hemisphere, where high-redshift quasar discoveries remain sparse \citep[e.g.,][]{reed2017eight, belladitta2019extremely, belladitta2025discovery, wolf2024srg, onken2022discovery} and next generation facilities are being built \citep[e.g., The Extreme Large Telescope (ELT) and The square kilometer array (SKA) radio telescope, respectively;][]{padovani2023extremely, hall2008square}. The DR10 release combines imaging of: the southern sky from the DECam Legacy Survey \citep[DECaLS,][]{dey2019overview} using the Dark Energy Camera \citep[DECam,][]{flaugher2015dark} on the Blanco 4m telescope and other programs using DECam such as the DECam eROSITAS survey \citep[DeROSITAS,][]{zenteno2025erosita}, the DECam Local Volume Exploration Survey \citep[DELVE,][]{drlica2021decam}, the Blanco Imaging of the Southern Sky Survey \citep[BLISS; PI: Soares-Santos][]{mau2019faint}, and the Dark Energy Survey \citep[DES,][]{abbott2018dark}; the northern sky from the Mayall $z$-band Legacy Survey \citep[MzLS,][]{silva2016mayall} and the Beijing-Arizona Sky Survey \citep[BASS,][]{zou2017project}; and full-sky mid-infrared (MIR) photometry from the Wide-field Infrared Survey Explorer \citep[WISE,][]{wright2010wide}: the unblurred coadds of WISE \citep[unWISE,][]{mainzer2014initial} and Near-Earth Object Wide-field Infrared Survey Explorer Reactivation Mission \citep[NEOWISE,][]{mainzer2014initial}. LS DR10 photometry reaches median $5 \sigma$ point-source depths of \textit{g}$_{\text{DECam}} = 24.7$, \textit{r}$_{\text{DECam}} = 23.9$, \textit{i}$_{\text{DECam}} = 23.6$, \textit{z}$_{\text{DECam}} = 23.0$  AB magnitudes in the southern hemisphere. Similar but slightly shallower depths are achieved in the northern hemisphere.
 
Since machine-learning algorithms face problems when dealing with the lack of data, we requested valid observations in all optical bands (n$_{\text{obs}} > 0$), limiting the selection of targets to an area of 15,342 deg$^2$ and declinations $< 30^\circ$ (due to the limited $i$-band coverage). We focused on candidates for which the SED fitting refinement step is possible and informative, thus requiring detection in at least one NIR band from the VISTA Hemisphere Survey DR7 \citep[VHS,][]{mcmahon2012vista}, UKIRT Infrared Deep Sky Surveys DR11 \citep[UKIDSS,][]{lang2014unwise} or UKIRT Hemisphere Survey DR2 \citep[UHS,][]{dye2018ukirt} surveys and a signal-to-noise ratio (S/N) $> 3$ in WISE W1. While WISE bands are already included in the LS DR10 catalog, NIR bands were added by cross-matching our catalog with the previously mentioned surveys within a 1.5\arcsec\ radius. Due to the previous criteria, the catalog is limited to $5 \sigma$ depths of VHS DR7 \textit{J} = 21.7-22.3 and \textit{Ks} = 21.8-22.1 mag, UKIDSS DR11 depth of \textit{H}= 20.2 mag, UHS DR2 depth of \textit{J}=20.5 and \textit{Ks}=20.0 mag, and unWISE depth of \textit{W1} = 20.72 mag. Besides, the low spatial resolution (2.75\arcsec/pixel) of WISE W1 renders this band prone to source blending and, therefore, potential flux overestimation.

The constraints accounting for the \textit{i}-dropout sources are as follow:

   \begin{align}
     \hspace{35mm} \mathrm{(S/N)}_{\mathrm{z}} & > 7 \hspace{2mm}  \\
     ( \hspace{2mm}(\textit{i}_{\mathrm{PSF}} - \textit{z}_{\mathrm{PSF}} > 1.5 \hspace{3mm} & \text{and} \hspace{3mm}  \mathrm{(S/N)}_{\mathrm{i}} > 3) \hspace{3mm} \text{or} \\
     \nonumber  (\textit{i}_{\mathrm{lim PSF}} - \textit{z}_{\mathrm{PSF}} > 1.5 \hspace{2mm} & \text{and} \hspace{2mm} \mathrm{(S/N)}_{\mathrm{i}} < 3 ) \hspace{2mm} ) \\
     \mathrm{(S/N)}_{\mathrm{g, r}} & < 3 \hspace{2mm} \\
      \textit{i}_{\mathrm{flux\_ivar}} & \neq 0,
   \end{align}

\noindent where $\textit{i}_{\mathrm{PSF}}$ and $\textit{z}_{\mathrm{PSF}}$ are $i$ and $z$ band magnitudes, and $\textit{i}_{\mathrm{flux\_ivar}}$ the inverse variance of the flux in the $i$ band. All of them are products from \textit{The Tractor} algorithm \citep{lang2016tractor} used in LS DR10 for the source extraction, based on the convolution of the images to a PSF model. The magnitude limit for the $i$-band ($\textit{i}_{\mathrm{lim PSF}}$) is computed as

\begin{eqnarray}
\centering
    \hspace{20mm} \textit{i}_{\mathrm{lim PSF}} = 22.5 -2.5 \log \left(\frac{3}{{\sqrt{\textit{i}_{\mathrm{flux\_ivar}}}}}\right).
\end{eqnarray}

\noindent We set the color threshold at $1.5$, which is 0.5 mag lower than in previous selections \citep[e.g.,][]{reed2017eight, banados2016pan}. This choice enabled us to recover 127 out of the 166 known quasars at $z > 5.3$ from the literature \citep{fan23} that lie within the LS DR10 footprint, have complete optical coverage, and satisfy conditions (1)–(5). Among the 39 missed sources, only 15 lie at $z > 6$, the redshift range on which this work is focused. 

The recovery rate decreased when imposing infrared constraints: from 127 to 106 when requiring a significant detection in WISE W1, and further to 69 when requiring both NIR and WISE W1 detections. In other words, the NIR+W1 criterion removed $\sim 46\%$ of known quasars. However, without any IR constraints, the candidate list would swell to 3,254,837 sources, which is beyond the computational scope of this work. Moreover, such a large sample would have far less reliable prioritization for follow-up, as most sources would have too few photometric points for meaningful SED fitting. Even requiring only the W1 requirement (losing only $\sim 17\%$ of known quasars) still results in 1,850,912 candidates, a size that remains challenging to process and likely to have a high contamination rate.

We also adopted maskbits that select clean sources (including masked point-like and/or unsaturated sources).\footnote{i.e., maskbits < 2 or = 32, 64, 96, 128, 160, 256, 512, 768, 2048, 2176.} We explicitly did not select objects by source type nor by morphology features. We note that adopting the type "PSF" to focus on the most point-like objects would only reduce the catalog by 17$\%$, but exclude 11 known $z \sim 6$ quasars from the literature. The resulting sample consists of $165\,253$ objects. 

\subsection{Multiband imaging tensor}
\label{tensor_assembly}

We compiled $10\arcsec\times10\arcsec$ sized cutouts from the \textit{g}, \textit{r}, \textit{i}, and \textit{z} band images for each of the $165\,253$ targets from the LS DR10 archive. Given the native DECam pixel scale of $0\farcs262$ per pixel, this resulted in $38 \times 38$ pixel$^2$ images and a tensor dimension of $165253\times 4 \times 38 \times 38$.

\begin{figure}
\centering
\includegraphics[width = 0.98\linewidth]{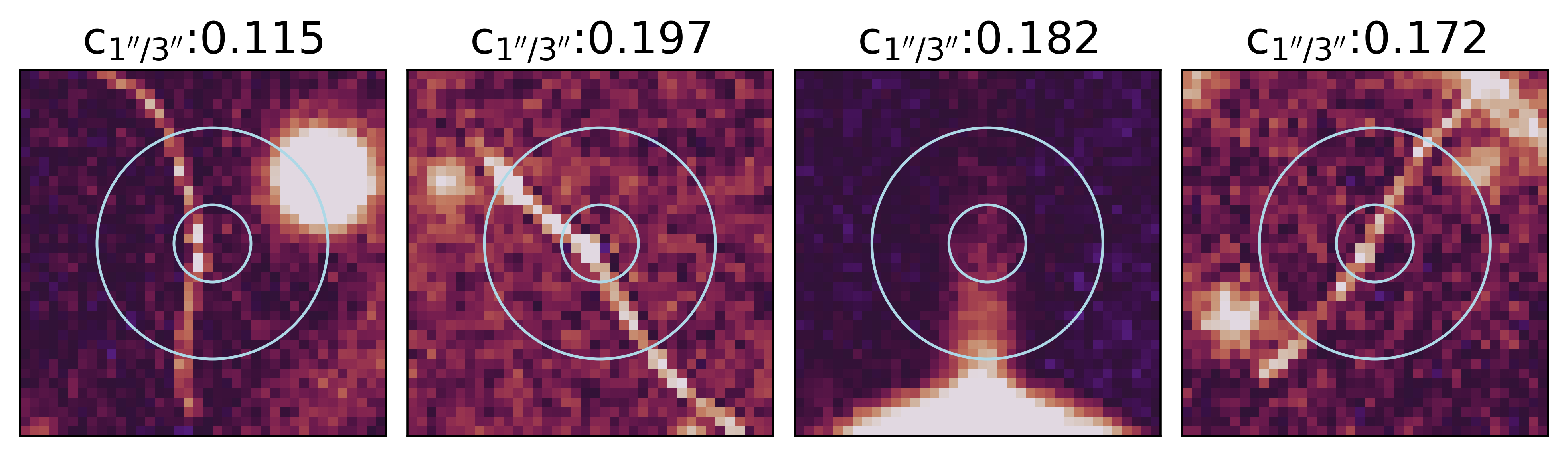}
\caption{\textit{z}-band $10\arcsec\times10\arcsec$ sized stamps illustrating examples of the artifacts identified in the photometry through visual inspection of the tensor. Light blue circles indicate the 1\arcsec\ and 3\arcsec\ radius apertures used in the compactness criteria calculation. The C$_{1\arcsec/3\arcsec}$ values are presented above the corresponding postage stamp.}
\label{artifacts}%
\end{figure}

During our inspection of tensor slices, we discovered clear artifacts, mainly cosmic rays and diffraction spikes (see Fig. \ref{artifacts} for examples), that the maskbits conditions at the catalog level did not filter out. To identify and remove these extended sources, we calculated the ratio of flux within a 1\arcsec\ radius aperture to the flux within a 3\arcsec\ radius aperture (C$_{1\arcsec/3\arcsec}$). If this ratio exceeded 0.3, the source was considered compact. Conversely, we rejected sources where 30\% or less of the total flux within the 3\arcsec\ aperture was confined to the inner 1\arcsec\ aperture, as this indicated an extended structure. This flexible 30\% threshold was calibrated on a sample of known quasars at \textit{z} $> 5.3$ \citep{fan23}, all with ratios above $45 \%$, except for the quasar pair J2037–4537 \citep{yue2021candidate} with a $32\%$ ratio, prompting us to lower our limit to avoid excluding similar systems. This step reduced the training sample used as input to the CL pipeline to 130472 sources by removing these extended unphysical sources in line with the literature.

\section{Contrastive learning for the selection of quasars candidates}
\label{sec:CL}

The selection method reported in this paper is based on CL, a self-supervised machine-learning algorithm mapping high-dimensional data sets into a low-dimensional feature space representation. This method learns the most informative correlations in the data by measuring the similarity between data points \citep{chopra2005learning, hadsell2006dimensionality} without any labels and taking advantage of data augmentation methods \citep{huertas2023brief}. As described in section \ref{tensor_assembly}, the data for this application of the CL consists of multiband image channels containing two-dimensional spatial information. This input is similar to that used in previous successful astronomical applications of CL \citep[e.g.,][]{sarmiento2021capturing, byrne2024quasar}.
 
With the imaging tensor input of the CL set, a preprocessing was carried out consisting of: 

\begin{itemize}
    \item[1)] masking nan-value pixels, if present, with zeros, given that the pixel flux distributions across all bands are dominated by background noise and peak at zero. 
    \item[2)] normalizing the tensor using a single global factor derived from the 99th percentile of the brightest band across all sources, which in our case corresponds to the \textit{z}-band. To compute this factor, we first built per-band distributions of pixel flux values (within a $1\arcsec$ radius from the center) across the full sample. We then determined the 99th percentile of each band’s distribution and identified the brightest among them. This value was adopted as the normalization factor and applied uniformly to all sources in the tensor. By using one global factor rather than source-by-source scaling, we preserved both the relative brightness distribution of the targets and the intrinsic SED shape across bands.
    \item[3)] random shuffling of the source indexes in the tensor to prevent any potential spatial or data quality bias, as the original tensor was generated in increasing right ascension order, and the training sample is divided into batches. 
\end{itemize}

\begin{figure*}[ht!]
   \centering
   \includegraphics[width = 0.95\linewidth]{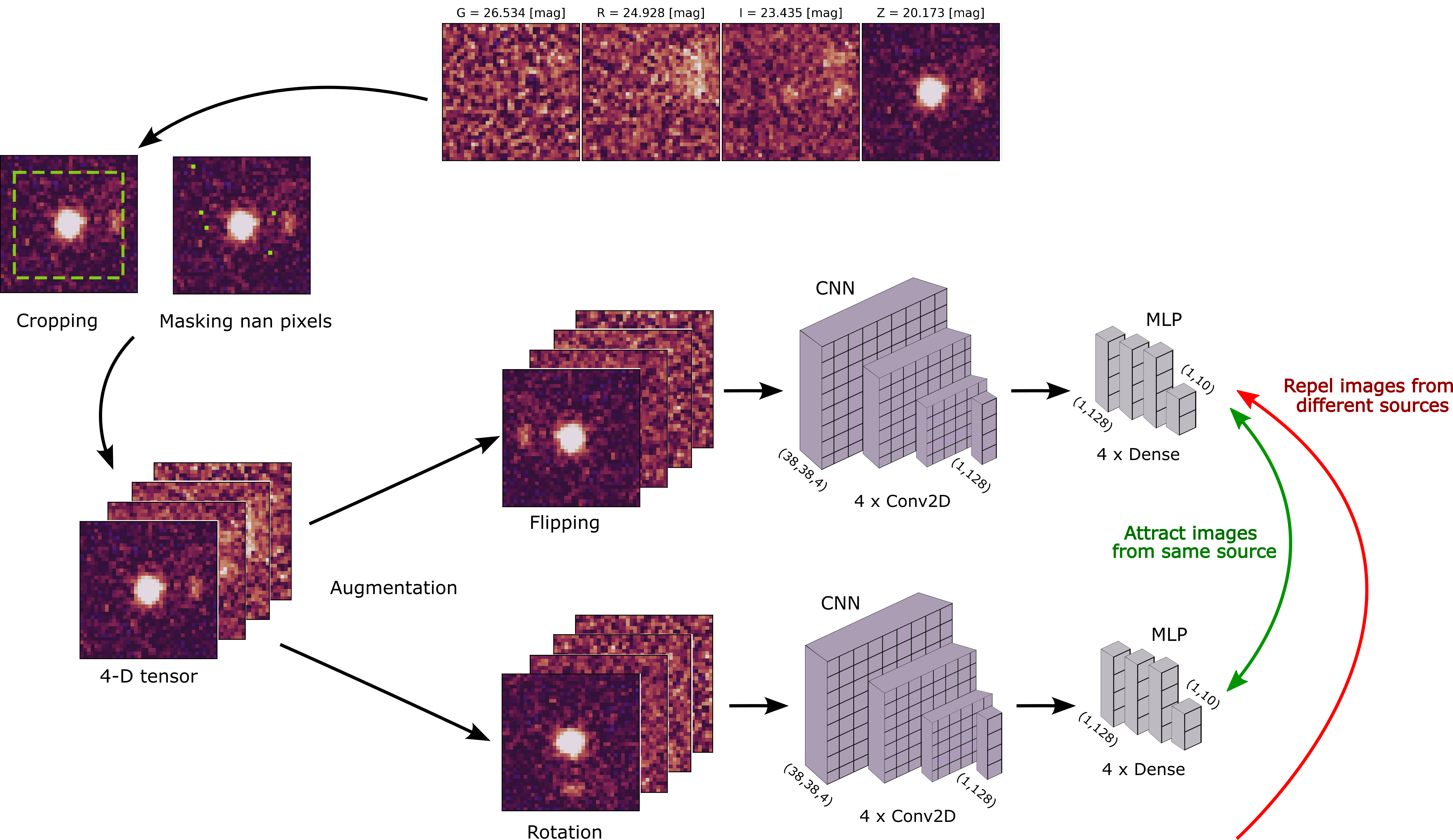}
   \caption{Architecture of the self-supervised CL framework used in this work. The top and left sections illustrate the image preprocessing and the tensor assembly steps. The structure of the network is shown sequentially from left to right, starting with data augmentation, then the encoder block, and lastly the projection head. Red and green arrows indicate the contrastive loss computation, which measures similarity between random pairs of augmented images.}
    \label{CL_architecture}%
\end{figure*}
  
We trained a modified version of the self-supervised model simCLR \citep[a simple framework for contrastive learning of visual representations,][]{chen2020simple}  built on \textit{Keras 2.15.0}. We implemented the following data augmentations: horizontal and vertical flips, and random rotation within $\pm 90$\textdegree, to produce transformed versions of each image. In principle, Gaussian noise perturbations can be included as an additional data augmentation, motivated by realistic calibration uncertainties and background fluctuations (e.g., zeropoint variations and sky noise). Such perturbations introduce mild pixel-level variations while preserving the physical interpretability of the encoded SED. In this work, we did not adopt this augmentation in order to keep the training scheme minimal and limit survey-dependent hyperparameters; its inclusion is left to future extensions explicitly modeling photometric uncertainties.

Two augmented views of the same input source are referred to as positive pairs, while negative pairs are augmented views of different input sources. These pairs drive the training, encouraging the network to cluster similar sources in the latent space while pushing away dissimilar ones. The encoder producing this low-dimensional representation of the images consisted of four convolutional neural network (CNN) layers that progressively reduced the spatial dimensions of the four channels to a 128-length feature vector. This was followed by a Flatten layer and a Dense layer, which applied a transformation to the vector, combining the local spatial information into a more compact and complex embedded representation. The stacked layers allow the network to capture increasingly complex visual features: from basic edges and shapes to higher-level structures such as smudges and PSF-like features. All convolutional and dense layers used a ReLU activation function to introduce nonlinearity and emphasize informative patterns.

The projection head mapped the 128-dimensional vectors resulting from the encoder into a 10-dimensional latent space, where the contrastive loss was computed by comparing the similarity between random pairs of sources. This small multilayer perceptron (MLP), composed of four dense layers, is designed to prevent overfitting of trivial features and enhance the separation between positive and negative pairs in the embedded space. Fig. \ref{CL_architecture} shows the architecture of our CL implementation. The combination of hyperparameters achieving the best results was: temperature of 0.1 for the normalized temperature-scaled cross entropy contrastive loss; width of 128 for the latent space; maximum batch size given by the limits of our GPU resources, 10000; and 3000 epochs. Other architecture choices, such as the number of CNN and Dense layers in the encoder and MLP, were selected to balance representation power and overfitting, following similar astronomy applications (e.g., 3 CNN + 3 Dense layers in \citealt{sarmiento2021capturing}; 4 CNN + 3 Dense layers in \citealt{byrne2024quasar}).

All the training experiments were carried out on one of the two GPU nodes of the \textit{Astronodes} cluster at the \textit{Max Planck Institute f\"ur Astronomie}. Each node is equipped with NVIDIA Tesla V100S GPUs (32 GB HBM2 memory), and runs CUDA 12.4 (driver version 550.90.07). The runtime was usually between $\sim 4.2$ hours for the optimized hyperparameter values, and $\sim 8$ hours for tests with more epochs, higher temperatures, batch sizes, or widths of the latent space.

\begin{figure*}
   \centering
   \includegraphics[width = 0.9\linewidth]{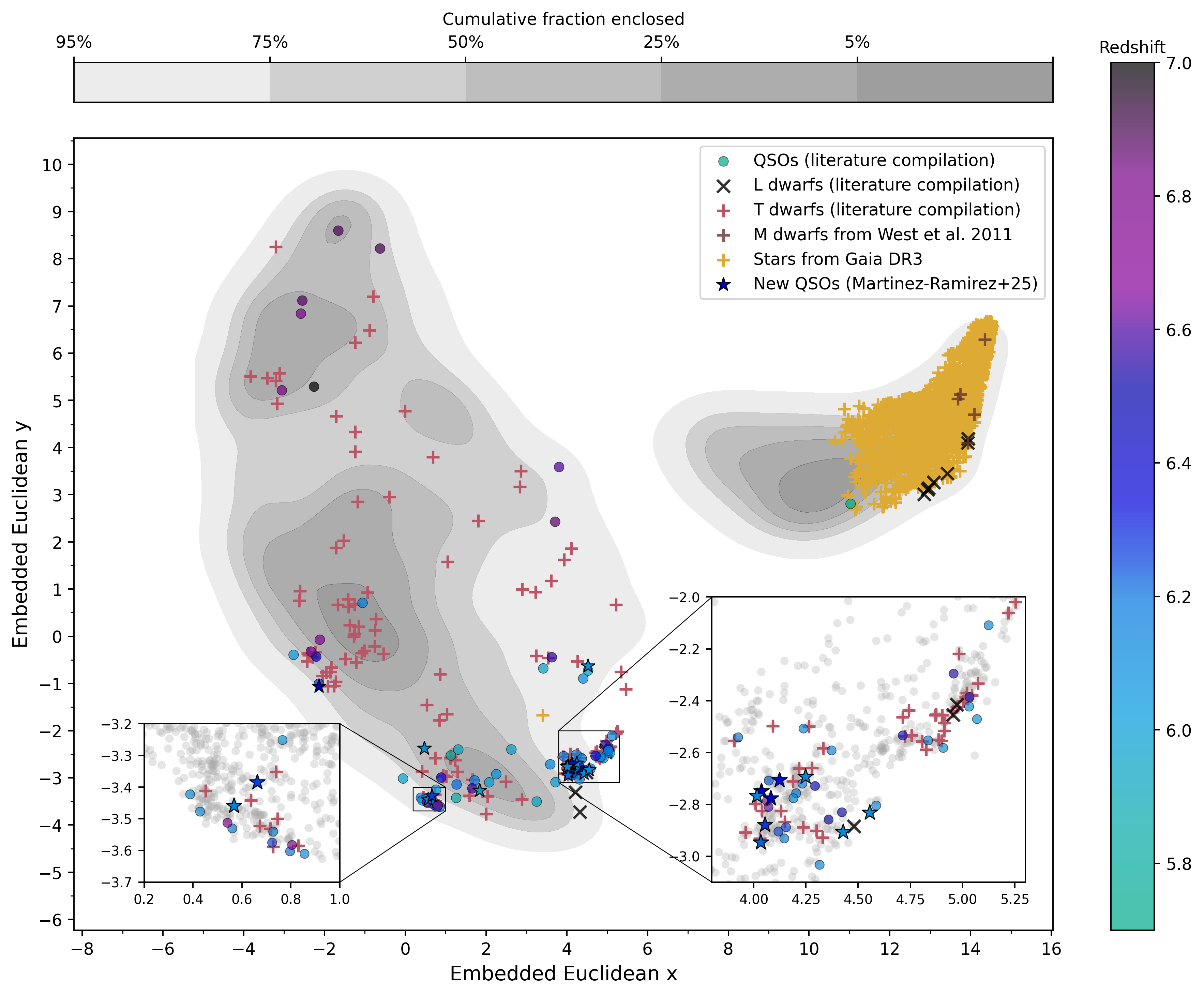}
   \caption{Embedded Euclidean space generated by UMAP for LS DR10 \textit{i}-dropout sources after training the CL algorithm. All sources are represented as gray countours given by the number density; known M, L and T dwarfs are: brown, black and red crosses, respectively; and Gaia DR3 stars are yellow crosses. Spectroscopically confirmed quasars from the literature and the new quasars from this work are represented by filled circles and stars, respectively, all of them color-coded by their redshift. Quasar candidates are selected from the lower part of the left island where the source density is low and the contamination ratio with UCDs is $\sim$ 1:1. The two regions with the highest concentration of quasars are presented as zoom-in panels and highlighted with black squares.}
    \label{Embedded_space}%
    \end{figure*}

\subsection{Embedded space}

We applied the unsupervised dimensionality reduction algorithm known as Uniform Manifold Approximation and Projection \citep[UMAP,][]{mcinnes2018umap} to reduce the representation dimensions from 128 to 2 in the embedded space generated by simCLR. 

As part of the downstream analysis, we crossmatched spectroscopically classified data within a $1.5$\arcsec\ radius to facilitate a comprehensive interpretation of the latent space, which helped us identify potential clusters of high-$z$ quasar candidates and contaminants. This included spectroscopically confirmed 
quasars at \textit{z} $ > 5.3$ from \cite{fan23}, M dwarfs from \cite{west11}, a compilation of LT dwarfs from the literature (\citealp[]{albert2011, burningham2008, burningham2010b, burningham2013, chiu2008, dayjones2013, deacon2011, kendall2007b, kirkpatrick2010, kirkpatrick2011, knapp2004, liu2002, lodieu2007, lodieu2009, lodieu2012, mace2013, matsuoka2011, pinfield2008, thompson2013, radigan2013, schmidt2010, scholz2010b, scholz2010, best2015, cardoso2015, marocco2015, tinney2018}) and stars from Gaia DR3 \citep{collaboration2023gaia} with a high probability of being a single star (PSS $ > 0.99$) and statistically significant proper motion (pm/e\_pm > 3). UMAP was tailored in order to optimize the 2D-separation between labeled quasars and contaminants, adopting the following parameters: n\_neighbors = 30, min\_dist = 0.02, metric = euclidean. While UMAP was used here for 2-dimensional visualization, the learned contrastive embeddings remain high-dimensional. Replacing UMAP with clustering algorithms operating directly in latent space is a natural path to further improve robustness in the candidate selection.

Figure \ref{Embedded_space} shows the resulting two-dimensional embedded space representation of the \textit{i}-dropout candidates, alongside known populations of stars, quasars, and MLT dwarfs. The multiple tests we performed to find the optimal UMAP parameters revealed that, regardless of their values, the embedded space naturally separated into two prominent regions: the right "island," dominated by stars and M and L dwarfs, and the left island, primarily populated by T dwarfs and high-redshift quasars. Notably, we identified a concentration of quasars at redshifts $z = 6$–$6.4$ near the bottom of the left island, while more sparsely distributed $z > 6.8$ quasars appear toward the top.

\subsection{Interpretation of the embedded space}
\label{inter_embedded_space}

In this section,  we search for the main drivers of the embedded space representation. We first started by color-coding the data points in the embedded space by catalog features such as magnitudes, color indices, depth, and PSF size. Here we summarize the most important results, but we are presenting additional details in the appendix section \ref{analysis_es_catalogmaps}. The catalog information (see Fig. \ref{photometric_properties}) reveals a strong $z$-magnitude gradient, a mild systematic effect due to the image quality, and that $z$–$J$ color is a strong separator of the embedded space into two distinct regions. This is puzzling, as the \textit{J}-band imaging was not included in the CL input. Therefore, some other feature present in the optical images must be driving this separation. To investigate further, we proceeded to inspect the image cutouts. 

\begin{figure*}[ht!]
   \centering
   \includegraphics[width = 0.95\linewidth]{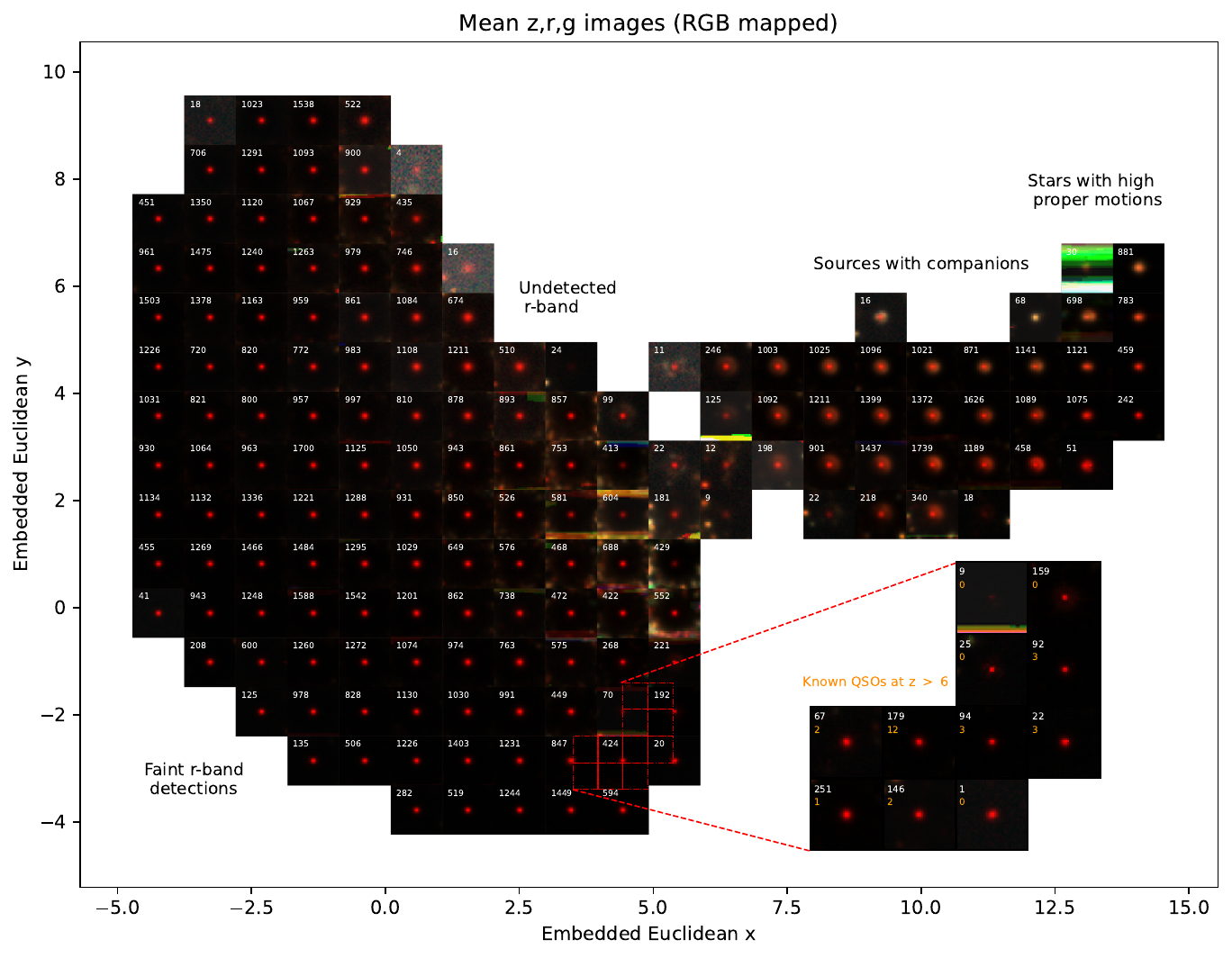}
   \caption{Mean pixel-by-pixel $z$, $r$, and $g$ band fluxes (RGB mapped) within $15 \times 15$ binned embedded space. The number of sources within each bin contributing to the mean is shown in white in the upper left side of each image. The red dashed lines highlight a region with the highest number of known quasars, with a zoom-in binned embedded space on the right side. The number of known quasars in each bin at the zoom-in plot is shown in orange, below the number of sources in white. }
    \label{average_images}%
    \end{figure*}

We divided the embedded space into $15\times 15$ bins, computed the mean pixel-by-pixel flux for each band for sources within each bin, and plotted the resulting mean $z$, $r$, and $g$ band channels (RGB mapped) at the location of each bin in the latent space (see Fig. \ref{average_images}). Since the LS DR10 images are weighted coadds of multiple epochs, temporal effects such as source motion or transient events may introduce spurious morphological features. These could influence the CL algorithm, potentially driving some of the trends observed in the embedded space. To assess this possibility, we inspected single-epoch cutouts of random sources in the subsequent analysis to validate our interpretations based on the coadded data.

The requirements of our catalog-based preselection are aimed at targeting sources with non-detections in the \textit{g} and \textit{r} bands but detectable emission in \textit{z}, without imposing any morphological constraints. When mapping the \textit{z}, \textit{r}, and \textit{g} bands into RGB colors, we noticed that the stacked-average images exhibit extended emission (red shadows) and contribution from the $r$-band (yellow shadows). The red structures populate the island on the right side of the embedded space and appear as faint rings or arcs around the central bright source, as well as elongated bright sources. 

Gaia DR3 revealed that the elongated sources are primarily stars with high proper motions (see Fig.~\ref{proper_motion}). Inspection of individual epochs showed that these stars are, in fact, point-like, but appear duplicated in the Legacy Survey DR10 catalog due to positional shifts across observing epochs. In these cases, the catalog erroneously lists two nearby sources: one faint object, centered in our stamps and detected only in \textit{z}, mimicking the color signature of a high-redshift quasar; and a second, brighter source detected in all bands but offset in position. Because LS DR10 performs forced photometry anchored to the \textit{z}-band detections, this positional mismatch leads to underestimation or loss of flux in the bluer bands, artificially producing a spectral break in otherwise stellar SEDs. This explains why some of these high proper motion stars were selected by our \textit{i}-dropout preselection.

The apparent elongation seen in Fig.~\ref{average_images} is not intrinsic to individual sources but a result of our visualization method. Each pixel in the figure represents a stack-averaged cutout of all sources within a given bin in the embedded space. In regions dominated by high proper motion stars, these small positional shifts accumulate in the averaging process, producing blurred or elongated structures. While some individual exposures show mild extension, likely due to motion, variability, or poor seeing, the dominant effect in the averaged images stems from the collective misalignment. Other moving objects, such as satellites and asteroids, can also produce real elongation in coadds, but their motion occurs on much shorter timescales (minutes).

Moving to the left side of the right island, we identified binary systems where the central target, detected in the \textit{z} band or both \textit{i} and \textit{z} bands, is the source selected in our analysis, while its companion is visible in all bands. In this region, the source density increases significantly, and the overlapping light from multiple companions creates a ring-like effect on average. Crossmatching with our labeled sources revealed the existence of one of the quasars from the pair at $z = 5.66$ \citep{yue2021candidate} with a spatial separation of 1\farcs24 ($7.3$ kpc at $z = 5.66$), along with stars exhibiting low proper motions, perhaps in binary systems or by chance alignment with a companion. We do not rule out the presence of additional quasar pairs or even gravitationally lensed sources in this region. However, given the high density of sources-over $2000$ within a small $1\times1$ latent-space unit region centered on the quasar pair and the significant contamination from stars (about $50 \%$, with $\sim 10 \%$ spectroscopically confirmed and the remaining having $> 90 \%$ probability of being stars), it is challenging to reliably identify quasars here. Consequently, we excluded candidates from this region for the time being. This demonstrates how environmental effects influence the latent space representation, even within a small 5\arcsec\ radius.

Since the large island on the left appears relatively homogeneous in the RGB map, we examined intensity maps of each band individually. We found that the left side of this island is dominated by faint central flux in the \textit{r}-band (comparable to the noise), and a clear detection in the \textit{i}-band. In contrast, the right side exhibits undetected \textit{r}-band, and fainter or nonexistent central emission in the \textit{i}-band. Given that the known quasars located there are at \textit{z} $\sim 6-6.4$, it is possible that a non-negligible Lyman-$\alpha$ forest can contribute to the \textit{i}-band below the Ly-$\alpha$ break. Based on the labels and the high source density in the center, we also anticipate significantly higher contamination from unlabeled T dwarfs in this region, yet to be confirmed.

\begin{figure*}
   \centering
   \includegraphics[width = 0.72\linewidth]{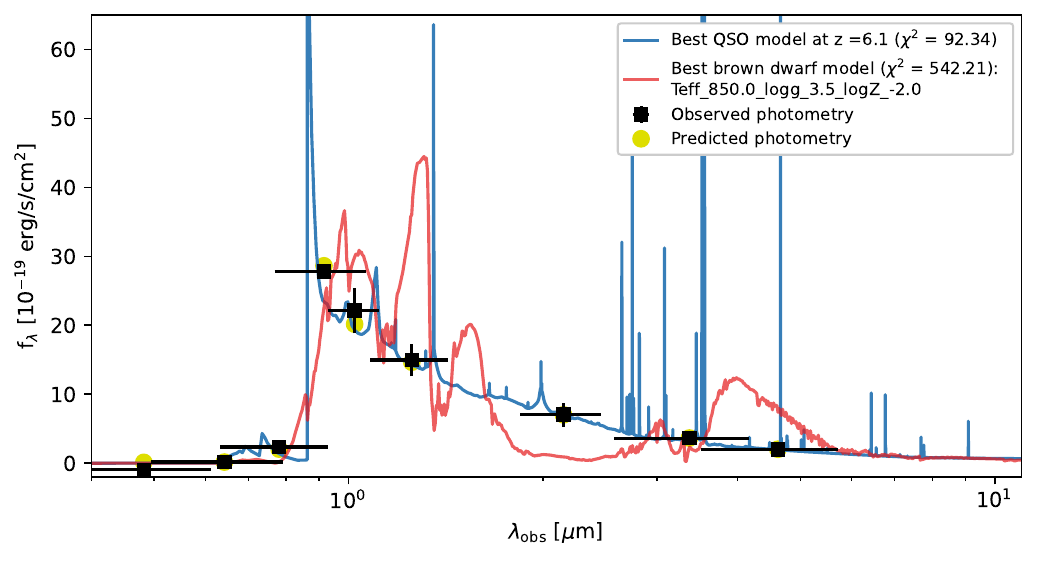}
   \includegraphics[width = 0.19\linewidth, trim = 0 {-8mm} 0 0]{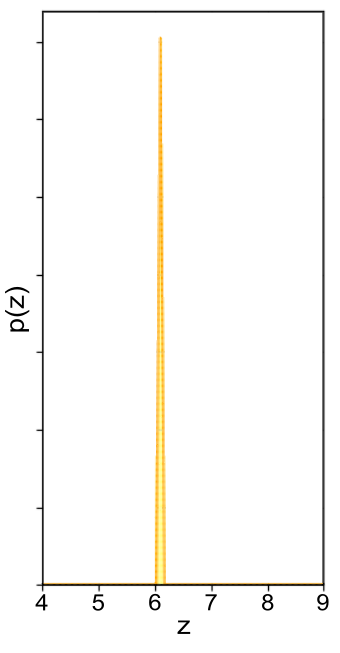} \\
   \includegraphics[width = 0.9\linewidth]{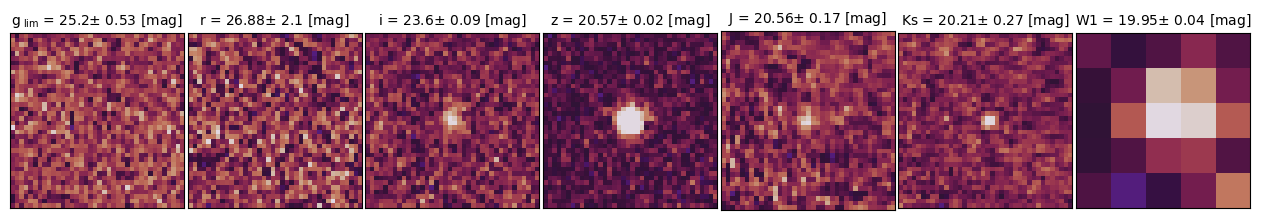}
   \caption{Example of an SED fitting result and the photometric redshift probability distribution function (\textit{top panels}). LS DR10, VHS DR7, and WISE postage stamps of the photometry used (\textit{bottom panel}) for the source LS J000-79. 
   \textit{Top left}: The blue curve represents the best-fit model given by the QSO1 template (see \citealt{salvato2022erosita} for details) at redshift 6.096, while the red curve shows the best-fit brown dwarf template from \cite{meisner2021new}. The observed photometry represented by black squares with error bars corresponds to the DECam, VHS, and WISE catalogs and their uncertainties, while the yellow circles are the expected flux densities assuming the best quasar model. \textit{Top right}: photometric redshift probability distribution function. Despite the distribution extending over $z = [0,12]$, we limited the plot to the range $z = [4,9]$ for better visualization.}
    \label{SEDfitting_PSO1.55}
\end{figure*}

We identified multiple isolated regions with high concentrations of known quasars and moderate contamination by UCDs. The explored latent space is large and rather poorly covered by labeled data, highlighting strong potential biases among currently confirmed objects. A comprehensive strategy would require spectroscopic confirmation of $>30000$ objects across the parameter space populated by quasars, given that the region with $\texttt{euclidean\_y} < 0$ alone already exceeds this number of candidates. As this is not currently feasible, we instead propose a more immediate, quantitative criterion that can be employed to find potentially similar objects to the existing labels by computing a 2D Gaussian Kernel Density Estimation (KDE) of quasars and UCDs based on the spectroscopically confirmed labels and normalized by the total number of labels. We iteratively adjusted the thresholds in quasar density ($\rho_{\mathrm{QSO}}$) and density ratio ($\rho_{\mathrm{QSO}}/\rho_{\mathrm{BD}}$) to define a quasar-candidate region that maximized the recovery of known quasars while limiting the number of candidates. The final selection, $\rho_{\mathrm{QSO}} > 0.0002$ $\text{per latent-space-unit$^{2}$}$ and $\rho_{\mathrm{QSO}}/\rho_{\mathrm{BD}} > 0.65$, included 54 of the 69 known quasars and resulted in 24,212 candidates. The quasar and brown dwarf KDE density maps, along with the density ratio map, are presented in Fig. \ref{KDEdensities}. 

While our initial candidate selection focused on the lower region of the embedded space, densely populated by known quasars, a more relaxed threshold (e.g., $\rho_{\mathrm{QSO}} > 0.0001$) would extend the search toward the upper region, where $z > 6.6$ quasars are more scattered, and the contamination from T dwarfs is lower. This part of the space, however, is itself densely populated (source density $> 0.01$) by faint $i$-dropout sources. Confirming them would require highly competitive observations with large-aperture ground-based telescopes and NIR spectrographs, making such targets more challenging and risky to pursue. Furthermore, the previously discussed approaches have largely been guided by existing spectroscopic samples and thus inevitably inherit their selection biases. A fully unbiased search will require probing less-explored regions of the latent space where spectroscopic labels are scarce or absent.  In this context, the analysis of Fig. \ref{photometric_properties} becomes essential: by revealing trends in catalog-based features such as red NIR slopes or $z$–W1 colors, it offers a pathway to identify promising regions even in the absence of spectroscopic labels. This paves the way for a more complete and less label-dependent quasar census in future work.

\section{Assessment of the quasar nature of promising candidates}
\label{sec:followup}
\subsection{Prioritization of quasar candidates through SED fitting }

To demonstrate the effectiveness of our CL approach and optimize the usage of valuable telescope time, we aimed to target candidates most likely to yield successful spectroscopic confirmations. This led us to implement an additional prioritization step to select the most promising sources from the high-density quasar regions in the embedded space. We performed SED fitting on the 24212 candidates with \textit{eazy-py} \citep{brammer2008eazy} version 0.6.9\footnote{See \url{https://github.com/gbrammer/eazy-py}} using quasar templates along with templates for common contaminants such as UCDs, stars, and galaxies. Detailed information about the contaminant population templates is presented in the appendix section \ref{contaminant_templates}.

The set of quasar and AGN emission models to fit was composed of a compilation of 35 individual and combined templates of the host galaxy and AGN emission with different contributions \citep[][see their Appendix B for more details and references]{salvato2022erosita}. To be more inclusive, we also considered less traditional objects such as the "little red dot" J0647-1045, whose NIRSpec spectrum is reproduced by templates of an obscured AGN at $z=4.50$ hosted by a star-forming galaxy \citep{killi2024deciphering}.

For all the targets, we included the following filters in the \textit{eazy-py} translate file: DECam $g$, $r$, $i$, and $z$ bands; WISE bands $W1$ and $W2$; and UKIDSS or VHS \textit{Y}, \textit{J}, \textit{H}, and \textit{Ks} bands, the latter subject to data availability. The setup parameters used are: the maximum photometric redshift \texttt{Z\_MAX} = 12, and the redshift grid interval \texttt{Z\_STEP} = 0.01. We ran the main fitting function in multiprocessing mode to fit galaxy and quasar templates, one template at a time, with a simultaneous photometric redshift estimation. To fit the stellar templates, redshift estimation is not required, and we simply ran the function \texttt{fit\_phoenix\_stars} with our customized templates mentioned above. The output contains $\chi^2$ and coefficients representing the contribution to the observed SED for all the templates in the set, the photometric redshift probability distribution (including the maximum-likelihood value $z_{\mathrm{ml}}$), rest-frame $U$, $B$, $V$ and $J$ band-magnitudes, luminosities, stellar masses, star formation rates, and other physical parameters associated to parametric FSPS and CORR\_SFHZ galaxy templates.

To calibrate our SED-fitting criterion, we tested the recovery of the 56 and 205 known quasars and UCDs, respectively, available in our catalog. While candidate selection was restricted to a smaller latent-space region, we used the full labeled set for calibration to ensure sufficient statistics. Recovery was quantified based on the ratio of the $\chi^2$ values obtained from quasar or AGN (quasar+host galaxy) templates, $\chi^2_{\mathrm{quasar}}$, compared to those from stellar contaminant templates (UCDs or stars), $\chi_{\mathrm{BD}}^2$. Applying a threshold of $\chi_{\mathrm{BD}}^2/\chi_{\mathrm{quasars}}^2 > 1$ yielded a precision of 0.72, a recall\footnote{Precision = TP/(TP+FP) and Recall = TP/(TP+FN), where TP are the true positives, FP the false positives, and FN the false negatives. They are metrics to evaluate the performance in a classification task.} of 0.86, and a contamination rate of $28 \%$, successfully recovering 48 quasars. In the opposite case, we selected UCDs with a precision of 0.96, a recall of 0.91 and a contamination rate of $4 \%$, recovering 185 objects.

To optimize our available resources for the follow-up observations, we focused our search on candidates with $\chi_{\mathrm{BD}}^2/\chi_{\mathrm{quasars}}^2 > 1$, maximum-likelihood photometric redshifts $z_{\mathrm{ml}}> 5.5$, and single narrow peaked redshift distribution given by z$_{84}$-z$_{16} < 1$, where z$_{84}$ and z$_{16}$ are the 84th and 16th percentiles. These constraints reduced the initial set of 24212 candidates from the high-density quasar loci in the embedded space to 1438 targets. 

To assess priorities, we evaluated the observability of our targets at the time and location of the spectroscopic follow-up observing runs available to us in past semesters. We also conducted visual inspection of the $g$, $r$, $i$, $z$, $J$, $Ks$, and $W1$ stamps and SED fitting results of the 1438 targets, sorted by decreasing $\chi_{\mathrm{quasars}}^2$. Fig. \ref{SEDfitting_PSO1.55} shows an example of the SED fitting, the photometric redshift probability distribution function, and the stamps of a high-priority candidate that ended up being confirmed as a quasar at $z = 6.21$ (see Section \ref{sec:new_qsos}). In this stage, we performed visual inspection of the stamps and removed 299 sources with spurious detections caused by cosmic rays, diffraction spikes, or luminous streaks (e.g., satellite trails). These artifacts are characterized by sharp and irregular shapes that usually show up in one band (the $z$-band in our case). Table \ref{selection_steps} summarizes the main steps for the selection and prioritization of candidates.

\begin{table}[ht]
      \caption[]{Downselection of candidates.}
      \begin{adjustbox}{max width=0.49\textwidth}
         \label{selection_steps}
         \centering
         \begin{tabular}{@{}ccc@{}}
         \toprule
         \multicolumn{1}{c}{\textbf{Condition}}  &  
         \multicolumn{1}{c}{\textbf{Size}}   &  
         \multicolumn{1}{c}{\textbf{Spec. C.}}
         \\
         \multicolumn{1}{c}{(1)}  &  
         \multicolumn{1}{c}{(2)}   &  
         \multicolumn{1}{c}{(3)}
         \\ \midrule
           \multicolumn{1}{c}{All LS DR10 sources}  &
         \multicolumn{1}{c}{2,827,055,986} &
         \multicolumn{1}{c}{0\%} 
         \\ \midrule
         \multicolumn{1}{c}{Color cut, maskbits conditions, and}  &
         \multicolumn{1}{c}{\multirow{2}{*}{165,253}} &
         \multicolumn{1}{c}{\multirow{2}{*}{0\%}} 
         \\
         \multicolumn{1}{c}{VHS, UKIDSS, or UHS, and W1 detection}  &
         \multicolumn{1}{c}{} &
         \multicolumn{1}{c}{} 
         \\ \midrule
         \multicolumn{1}{c}{Compactness criteria}  &
         \multicolumn{1}{c}{130,472} &
         \multicolumn{1}{c}{0\%}
         \\ \midrule
         \multicolumn{1}{c}{Density threshold in the embedded space}  &
         \multicolumn{1}{c}{24,212} &
         \multicolumn{1}{c}{1\%}
         \\ \midrule
         \multicolumn{1}{c}{SED fiting $\chi^2$ ratio, $z_{\text{ml}}$ and }  &
         \multicolumn{1}{c}{\multirow{2}{*}{1,438}} &
         \multicolumn{1}{c}{\multirow{2}{*}{4\%}}  
         \\ 
         \multicolumn{1}{c}{photometric redshift distribution}  &
         \multicolumn{1}{c}{} 
         \\ \midrule
         \multicolumn{1}{c}{No artifacts in optical stamps}  &
         \multicolumn{1}{c}{1,139} &
         \multicolumn{1}{c}{5\%}
           \\ \bottomrule
        \end{tabular}
        \end{adjustbox}
        \begin{tablenotes}
        \small
        \item \textbf{Notes:} Col (1): Condition defining the sample; Col (2): Number of sources in the sample; Col (3): Spectroscopic completeness: fraction of the sample with spectroscopic classification as quasar or UCD. The 0\% represents percentages $< 0.5\%$. 
        \end{tablenotes}
   \end{table}

We evaluated the performance of our full selection pipeline, CL coupled with SED fitting prioritization, in terms of precision and completeness, and compared it to representative literature color–color selections. The analysis was restricted to the 130472-size training sample, which defines the domain of applicability of the CL model, and used spectroscopically confirmed quasars and UCDs as reference samples. At each selection stage, only a very small fraction of the sample has spectroscopic classifications ($\lesssim$5\%; see Table \ref{selection_steps}), which limits the statistical robustness of these metrics. Precision quantifies the fraction of recovered quasars relative to all objects classified as quasar candidates, while completeness is the recovery fraction of known quasars that satisfy the selection criteria. Under these definitions, our pipeline achieves a purity of $84\%$ and a completeness of $75\%$.

Because most literature high-redshift quasar searches do not report precision or completeness in a form directly comparable to our work, we quantified these metrics by applying the published color-color selection criteria to our catalog. We emphasize that this comparison isolates only the color-based component of each pipeline while several studies apply additional selection steps such as X-ray detections (e.g., \citealt{wolf2024srg}), morphological filtering, proper-motion cuts, or stringent signal-to-noise requirements (e.g., \citealt{banados2016pan}), which effectively restrict their samples to specific quasar subpopulations and were not reproduced here. Due to heterogeneous photometric coverage, the color cuts can only be evaluated on subsets of our catalog with the required bands. Using PS1 data, we tested the selections of \cite{banados2016pan}, \cite{wang2019exploring}, \cite{yang2023desi}, and
the DELS+PS1 selection described in Section 2.3 of \cite{belladitta2025discovery}, which rely on $y_{P1}$ photometry and, in some cases, $J$-band data. Applied in this restricted form, these selections yield purities of $100\%$, $38\%$, $80\%$, and $5\%$, with corresponding completenesses of $18\%$, $50\%$, $48\%$, and $20\%$, respectively. We also tested the DES DR2 selection of \citet{wolf2024srg}, obtaining a purity of $43\%$ and a completeness of $54\%$. The low completeness arises from comparing narrow and highly targeted selections to our broader selection framework.

Then, to enable a consistent comparison across different photometric systems, we computed synthetic colors and adapted the literature color cuts to match the quasar color evolution in our photometric system. The adapted selections from \citet{banados2016pan}, \citet{wang2019exploring}, \citet{yang2023desi}, and \citet{wolf2024srg} achieve purities of $42\%$, $40\%$, $88\%$, and $25\%$, with corresponding completenesses of $90\%$, $11\%$, $74\%$, and $58\%$, respectively. In particular, the adapted selection of \citet{yang2023desi} reaches precision and completeness comparable to those of our pipeline, but at the cost of producing candidate lists more than an order of magnitude larger, implying substantial UCD contamination (e.g., compared to the expected number counts of $z \gtrsim 5.5$ quasars). This comparison highlights a key result of our approach: although the CL model was trained exclusively on optical imaging, the resulting representation naturally reproduces selection behavior similar to optical–NIR color criteria, while enabling a substantially more compact and efficient candidate prioritization for spectroscopic follow-up. Further details about the implementation and results are provided in Appendix~\ref{adapted color cuts}.

In addition to providing high purity and completeness, SED fitting prioritization exploits all available photometric bands to reduce the candidate list while retaining sources consistent with quasars at $5.7 < z < 7$ and providing photometric redshift estimates that directly inform follow-up strategies. Color-based prioritization is limited by incomplete photometry. For instance, $z$–$J$ cuts, using $J$ band with the most complete NIR coverage, cannot be applied to all candidates, and even among detected sources, a flexible threshold ($z$–$J < 1.5$) yields excessively large samples, while tighter cuts exclude known quasars at $z > 6.5$. Strengthening optical colors (e.g., $i$–$z > 2$) similarly reduces numbers but narrows the accessible redshift range. SED fitting is particularly effective at removing contaminants in the range $1 < z$–$J < 2$, where quasars overlap with UCDs and the most interesting high-redshift and red quasar populations are expected to reside.

\subsection{Follow-up observations}

   \begin{table*}
      \caption[]{Spectroscopic follow-up observations of the 16 new quasars.}
         \label{spectra obs}
         \centering
         \begin{tabular}{@{}cccccccc@{}}
         \toprule
         \multicolumn{1}{c}{\textbf{Name}}  &  \multicolumn{1}{c}{\textbf{RA}}    & \multicolumn{1}{c}{\textbf{Dec}}   & 
         \multicolumn{1}{c}{\textbf{Date}}   &
         \multicolumn{1}{c}{\textbf{Telescope/Instrument}} &
         \multicolumn{1}{c}{\textbf{Exp. time}} &
         \multicolumn{1}{c}{$\mathbf{z}$}
         \\
         \multicolumn{1}{c}{} &
         \multicolumn{1}{c}{deg} &
         \multicolumn{1}{c}{deg} &
         \multicolumn{1}{c}{} &
         \multicolumn{1}{c}{} &
         \multicolumn{1}{c}{seconds} &
         \\
         \multicolumn{1}{c}{(1)} &
         \multicolumn{1}{c}{(2)} &
         \multicolumn{1}{c}{(3)} &
         \multicolumn{1}{c}{(4)} &
         \multicolumn{1}{c}{(5)} &
         \multicolumn{1}{c}{(6)} &
         \multicolumn{1}{c}{(7)}
           \\ \midrule
          \multicolumn{1}{c}{LS J145109.79-044542.12}           &  \multicolumn{1}{c}{222.79078}             & \multicolumn{1}{c}{-4.76170}     & 
         \multicolumn{1}{c}{2022 May. 31$^{\mathrm{a}}$}   &
         \multicolumn{1}{c}{Mayall/DESI} &
         \multicolumn{1}{c}{3363} &
         \multicolumn{1}{c}{6.16}
         \\
         \multicolumn{1}{c}{LS J230129.72-153020.4}           &  \multicolumn{1}{c}{345.37384}             & \multicolumn{1}{c}{-15.50566}     & 
         \multicolumn{1}{c}{2024 Oct. 4}   &
         \multicolumn{1}{c}{Hale/DBSP} &
         \multicolumn{1}{c}{1200} &
         \multicolumn{1}{c}{6.32}
           \\
           \multicolumn{1}{c}{LS J020801.31-664713.7}           &  \multicolumn{1}{c}{32.00547}             & \multicolumn{1}{c}{-66.78714}     & 
         \multicolumn{1}{c}{2024 Oct. 25}   &
         \multicolumn{1}{c}{NTT/EFOSC2} &
         \multicolumn{1}{c}{5400} &
         \multicolumn{1}{c}{6.09}
           \\  
           \multicolumn{1}{c}{LS J222343.78-381526.8}           &  \multicolumn{1}{c}{335.93243}             & \multicolumn{1}{c}{-38.25753}     & 
         \multicolumn{1}{c}{2024 Oct. 25}   &
         \multicolumn{1}{c}{NTT/EFOSC2} &
         \multicolumn{1}{c}{4800} &
         \multicolumn{1}{c}{6.36}
         \\
           \multicolumn{1}{c}{LS J000622.24-793548.1}           &  \multicolumn{1}{c}{1.55765}             & \multicolumn{1}{c}{79.59670}     & 
         \multicolumn{1}{c}{2024 Oct. 28}   &
         \multicolumn{1}{c}{NTT/EFOSC2} &
         \multicolumn{1}{c}{3600} &
         \multicolumn{1}{c}{6.21}
           \\ 
           \multicolumn{1}{c}{LS J010449.12-685756.8}           &  \multicolumn{1}{c}{16.20466}             & \multicolumn{1}{c}{-68.96577}     &  
         \multicolumn{1}{c}{2024 Oct. 28}   &
         \multicolumn{1}{c}{NTT/EFOSC2} &
         \multicolumn{1}{c}{3000} &
         \multicolumn{1}{c}{6.38}
         \\ 
           \multicolumn{1}{c}{LS J103511.29-051537.9}           &  \multicolumn{1}{c}{158.79705}             & \multicolumn{1}{c}{-5.26053}     & 
         \multicolumn{1}{c}{2025 Feb. 16}   &
         \multicolumn{1}{c}{LBT/MODS2$^{\mathrm{b}}$} &
         \multicolumn{1}{c}{3000} &
         \multicolumn{1}{c}{6.09}
         \\ 
           \multicolumn{1}{c}{LS J113000.56+142043.97}           &  \multicolumn{1}{c}{172.50233}             & \multicolumn{1}{c}{14.34555}     & 
         \multicolumn{1}{c}{2025 Feb. 21}   &
         \multicolumn{1}{c}{Hale/NGPS} &
         \multicolumn{1}{c}{2700} &
         \multicolumn{1}{c}{6.28}
         \\ 
           \multicolumn{1}{c}{LS J133204.89+110208.94}           &  \multicolumn{1}{c}{203.02036}             & \multicolumn{1}{c}{11.03582}     & 
         \multicolumn{1}{c}{2025 Apr. 20}   &
         \multicolumn{1}{c}{LBT/MODS} &
         \multicolumn{1}{c}{3600} &
         \multicolumn{1}{c}{6.11}
         \\ 
           \multicolumn{1}{c}{LS J143510.65-105325.11}           &  \multicolumn{1}{c}{218.79436}             & \multicolumn{1}{c}{-10.89031}     & 
         \multicolumn{1}{c}{2025 Apr. 23}   &
         \multicolumn{1}{c}{LBT/MODS} &
         \multicolumn{1}{c}{2700} &
         \multicolumn{1}{c}{6.13}
         \\ 
           \multicolumn{1}{c}{LS J114156.14+100636.90}           &  \multicolumn{1}{c}{175.48392}             & \multicolumn{1}{c}{10.11025}     & 
         \multicolumn{1}{c}{2025 Apr. 23}   &
         \multicolumn{1}{c}{LBT/MODS} &
         \multicolumn{1}{c}{2700} &
         \multicolumn{1}{c}{5.94}
         \\  
           \multicolumn{1}{c}{LS J133014.01-402508.92}           &  \multicolumn{1}{c}{202.55836}             & \multicolumn{1}{c}{-40.41915}     & 
         \multicolumn{1}{c}{2025 May. 1}   &
         \multicolumn{1}{c}{Clay/LDSS3} &
         \multicolumn{1}{c}{1800} &
         \multicolumn{1}{c}{6.07}
         \\ 
           \multicolumn{1}{c}{LS J201119.04-443609.39}           &  \multicolumn{1}{c}{302.82931}             & \multicolumn{1}{c}{-44.60261}     & 
         \multicolumn{1}{c}{2025 May. 1}   &
         \multicolumn{1}{c}{Clay/LDSS3} &
         \multicolumn{1}{c}{1800} &
         \multicolumn{1}{c}{6.08}
         \\ 
           \multicolumn{1}{c}{LS J203704.37-515240.27}           &  \multicolumn{1}{c}{309.26819}             & \multicolumn{1}{c}{-51.87785}     & 
         \multicolumn{1}{c}{2025 May. 1}   &
         \multicolumn{1}{c}{Clay/LDSS3} &
         \multicolumn{1}{c}{2800} &
         \multicolumn{1}{c}{6.25}
         \\ 
           \multicolumn{1}{c}{LS J215501.13-511151.11}           &  \multicolumn{1}{c}{328.75473}             & \multicolumn{1}{c}{-51.19753}     & 
         \multicolumn{1}{c}{2025 May. 1}   &
         \multicolumn{1}{c}{Clay/LDSS3} &
         \multicolumn{1}{c}{1200} &
         \multicolumn{1}{c}{6.45}
         \\ 
            \multicolumn{1}{c}{LS J013938.24-520945.73}           &  \multicolumn{1}{c}{14.90935}             & \multicolumn{1}{c}{-52.16270}     & 
         \multicolumn{1}{c}{2025 Jul. 28}   &
         \multicolumn{1}{c}{NTT/EFOSC2} &
         \multicolumn{1}{c}{3600} &
         \multicolumn{1}{c}{6.05}
         \\
         \bottomrule
        \end{tabular}
        \begin{tablenotes}
        \small
        \item $^{\mathrm{a}}$ This source was confirmed as a quasar with DESI DR1 data release.
        \item $^{\mathrm{b}}$ Note that although MODS is composed of two identical two-channel spectrographs, at the time of this observing run only MODS2 was available.
        \newline \textbf{Notes:} Col (1): Quasar name with convention: "LS" for DESI Legacy Survey DR10, and then the coordinates in HMS and DMS format. Col (2) and Col (3): coordinates in degrees, Col (4): Date of the observations; Col (5): Telescope and instrument used for the optical spectroscopic follow-up; Col (6) Total exposure time; Col (7) spectroscopic redshift estimated with the spectrum fitting with an uncertainty of 0.03 (see Section \ref{sec:new_qsos}).
        \end{tablenotes}
   \end{table*}

We carried out spectroscopic follow-up observations between October 2024 and July 2025 with different telescopes and instruments: the Double Spectrograph (DBSP; \citealt{oke1982efficient} and the Next Generation Palomar Spectrograph (NGPS; \citealt{jiang2018preliminary}) mounted at the 200-inch Hale telescope at the Palomar Observatory, the ESO Faint Object Spectrograph and Camera (EFOSC2; \citealt{buzzoni1984eso}) at the New Technology Telescope at La Silla Observatory, the Multi-object Double Spectrograph (MODS; \citealt{pogge2010multi}) at the Large Binocular Telescope (LBT); and the Low Dispersion Survey Spectrograph (LDSS3) at the Magellan II Clay Telescope. All observations employed long-slit spectroscopy with slit widths of $1$\arcsec, 1\farcs2, or 1\farcs5. 

Standard calibrations such as bias subtraction, flat fielding, arc lamps for wavelength calibrations, and fluxing with spectrophotometric standard stars were included in the data reduction process with the Python package \textit{PypeIt} \citep{prochaska2020pypeit} and the Image Reduction and Analysis Facility (\textit{IRAF}; \citealt{tody1986iraf, tody1993iraf}). The resulting spectra were produced by combining individual science exposures after the background subtraction and calibration process, and before the spectra extraction, as an unweighted coadd. The final 1-D spectra were normalized to the LSDR10 \textit{z}-band magnitude.

The wavelength coverage of the optical spectra allowed us to identify the presence of the Lyman-$\alpha$ break, confirming the nature of the quasar candidates and assessing their redshift. We performed follow-up observations for 40 candidates, 22 of them turned out to be contaminants, likely UCDs (see Table \ref{ultracool_dwarfs}). We discovered 15 new quasars and 3 quasars that were already reported in the literature\footnote{DESI J011553.41+031829.3 discovered by \citealt{yang2023desi}, and J1111+0640 and J1257+1033 by \citealt{yang2024high}}. Additionally, one of our candidates was confirmed by DESI as a quasar at $z = 6.03$ in its first data release \citep{abdul2025data}. Since this source is not reported in the literature and is part of our selection, we included it and its spectrum in the following sections. The details of the follow-up campaigns where the new quasars were discovered are presented in Table \ref{spectra obs}.

   \begin{table*}
   \caption[]{Photometric properties of the new 16 quasars.}
   \begin{adjustbox}{max width=0.99\textwidth}
         \label{photometric_cat}
         \centering
         \begin{tabular}{@{}ccccccccccc@{}}
         \toprule
         \multicolumn{1}{c}{\textbf{Name}}  &  \multicolumn{1}{c}{\textbf{\textit{i}}}    & \multicolumn{1}{c}{\textbf{\textit{z}}}   & 
         \multicolumn{1}{c}{\textbf{\textit{Y}}}     & \multicolumn{1}{c}{\textbf{\textit{J}}}   &
         \multicolumn{1}{c}{\textbf{\textit{H}}} &
         \multicolumn{1}{c}{\textbf{\textit{Ks}}} &
         \multicolumn{1}{c}{\textbf{\textit{W1}}} &
         \multicolumn{1}{c}{\textbf{\textit{W2}}} &
         \multicolumn{1}{c}{$z_{\mathrm{ml}}$} &
         \multicolumn{1}{c}{$\chi_{\mathrm{BD}}^2/\chi_{\mathrm{quasars}}^2$}
           \\ \midrule
         \multicolumn{1}{c}{LS J1451-0445}           &  \multicolumn{1}{c}{23.69 $\pm$ 0.17}             & \multicolumn{1}{c}{20.83 $\pm$ 0.01}     & 
         \multicolumn{1}{c}{20.94 $\pm$ 0.23}       & \multicolumn{1}{c}{20.63 $\pm$ 0.17 }   &
         \multicolumn{1}{c}{20.21 $\pm$ 0.19} &
         \multicolumn{1}{c}{} &
         \multicolumn{1}{c}{20.02 $\pm$ 0.06} &
         \multicolumn{1}{c}{20.02 $\pm$ 0.14} &
         \multicolumn{1}{c}{6.1} &
         \multicolumn{1}{c}{4.37}
           \\
         \multicolumn{1}{c}{LS J2301-1530}           &  \multicolumn{1}{c}{24.75 $\pm$ 0.80}             & \multicolumn{1}{c}{20.65  $\pm$ 0.03}     & 
         \multicolumn{1}{c}{20.16 $\pm$ 0.08}       & \multicolumn{1}{c}{20.07 $\pm$ 0.10}   &
         \multicolumn{1}{c}{ } &
         \multicolumn{1}{c}{19.74 $\pm$ 0.18 } &
         \multicolumn{1}{c}{19.78 $\pm$ 0.05} &
         \multicolumn{1}{c}{19.91 $\pm$ 0.13} &
         \multicolumn{1}{c}{6.37} &
         \multicolumn{1}{c}{2.89}
           \\
           \multicolumn{1}{c}{LS J0208-6647}           &  \multicolumn{1}{c}{23.44 $\pm$ 0.14}             & \multicolumn{1}{c}{20.17 $\pm$ 0.01}     & 
         \multicolumn{1}{c}{}       & \multicolumn{1}{c}{20.72 $\pm$ 0.10}   &
         \multicolumn{1}{c}{ } &
         \multicolumn{1}{c}{19.85 $\pm$ 0.14} &
         \multicolumn{1}{c}{19.18 $\pm$ 0.02 } &
         \multicolumn{1}{c}{18.82 $\pm$ 0.03} &
         \multicolumn{1}{c}{6.11} &
         \multicolumn{1}{c}{5.53}
           \\  
           \multicolumn{1}{c}{LS J2223-3815}           &  \multicolumn{1}{c}{23.56 $\pm$ 0.20}             & \multicolumn{1}{c}{20.57 $\pm$ 0.03}     & 
         \multicolumn{1}{c}{ }       & \multicolumn{1}{c}{19.84 $\pm$ 0.04}   &
         \multicolumn{1}{c}{ } &
         \multicolumn{1}{c}{19.77 $\pm$ 0.12} &
         \multicolumn{1}{c}{19.92 $\pm$ 0.05} &
         \multicolumn{1}{c}{19.92 $\pm$ 0.11} &
         \multicolumn{1}{c}{6.30} &
         \multicolumn{1}{c}{2.25}
         \\
           \multicolumn{1}{c}{LS J0006-7935}           &  \multicolumn{1}{c}{23.60 $\pm$ 0.09}             & \multicolumn{1}{c}{20.57 $\pm$ 0.02}     & 
         \multicolumn{1}{c}{20.58 $\pm$ 0.16}       & \multicolumn{1}{c}{20.56 $\pm$ 0.17}   &
         \multicolumn{1}{c}{ } &
         \multicolumn{1}{c}{20.21 $\pm$ 0.27} &
         \multicolumn{1}{c}{19.95 $\pm$ 0.04} &
         \multicolumn{1}{c}{19.92 $\pm$ 0.08} &
         \multicolumn{1}{c}{6.10} &
         \multicolumn{1}{c}{5.87}
           \\ 
           \multicolumn{1}{c}{LS J0104-6857}           &  \multicolumn{1}{c}{23.75 $\pm$ 0.21}             & \multicolumn{1}{c}{20.50 $\pm$ 0.02}     & 
         \multicolumn{1}{c}{20.83 $\pm$ 0.19}       & \multicolumn{1}{c}{20.89 $\pm$ 0.22}   &
         \multicolumn{1}{c}{} &
         \multicolumn{1}{c}{19.81 $\pm$ 0.19} &
         \multicolumn{1}{c}{19.90 $\pm$ 0.04} &
         \multicolumn{1}{c}{19.33 $\pm$ 0.05} &
         \multicolumn{1}{c}{6.11} &
         \multicolumn{1}{c}{3.71}
         \\ 
           \multicolumn{1}{c}{LS J1035-0515}           &  \multicolumn{1}{c}{23.34 $\pm$ 0.22}             & \multicolumn{1}{c}{20.14 $\pm$ 0.01}     & 
         \multicolumn{1}{c}{20.51 $\pm$ 0.15}       & \multicolumn{1}{c}{19.68 $\pm$ 0.07 }   &
         \multicolumn{1}{c}{19.14 $\pm$ 0.07} &
         \multicolumn{1}{c}{18.55 $\pm$ 0.05} &
         \multicolumn{1}{c}{18.32 $\pm$ 0.01} &
         \multicolumn{1}{c}{18.19 $\pm$ 0.03} &
         \multicolumn{1}{c}{6.43} &
         \multicolumn{1}{c}{2.16}
         \\ 
           \multicolumn{1}{c}{LS J1130+1420}           &  \multicolumn{1}{c}{24.66 $\pm$ 0.39}             & \multicolumn{1}{c}{21.05 $\pm$ 0.03}     & 
         \multicolumn{1}{c}{21.13 $\pm$ 0.21}       & \multicolumn{1}{c}{}   &
         \multicolumn{1}{c}{20.61 $\pm$ 0.20} &
         \multicolumn{1}{c}{20.48 $\pm$ 0.24} &
         \multicolumn{1}{c}{20.31 $\pm$ 0.08} &
         \multicolumn{1}{c}{20.20 $\pm$ 0.16} &
         \multicolumn{1}{c}{6.20} &
         \multicolumn{1}{c}{21.09}
         \\ 
           \multicolumn{1}{c}{LS J1332+1102}           &  \multicolumn{1}{c}{24.17 $\pm$ 0.36}             & \multicolumn{1}{c}{20.80 $\pm$ 0.04}     & 
         \multicolumn{1}{c}{20.50 $\pm$ 0.13}       & \multicolumn{1}{c}{20.29 $\pm$ 0.11 }   &
         \multicolumn{1}{c}{20.03 $\pm$ 0.13} &
         \multicolumn{1}{c}{19.85 $\pm$ 0.12} &
         \multicolumn{1}{c}{19.95 $\pm$ 0.05} &
         \multicolumn{1}{c}{20.38 $\pm$ 0.18} &
         \multicolumn{1}{c}{6.22} &
         \multicolumn{1}{c}{2.92}
         \\ 
           \multicolumn{1}{c}{LS J1435-1053}           &  \multicolumn{1}{c}{24.64 $\pm$ 0.53}             & \multicolumn{1}{c}{21.27 $\pm$ 0.04}     & 
         \multicolumn{1}{c}{}       & \multicolumn{1}{c}{20.49 $\pm$ 0.21 }   &
         \multicolumn{1}{c}{} &
         \multicolumn{1}{c}{} &
         \multicolumn{1}{c}{20.51 $\pm$ 0.09} &
         \multicolumn{1}{c}{21.22 $\pm$ 0.42} &
         \multicolumn{1}{c}{6.21} &
         \multicolumn{1}{c}{1.24}
         \\ 
           \multicolumn{1}{c}{LS J1141+1006}           &  \multicolumn{1}{c}{24.08 $\pm$ 0.31}             & \multicolumn{1}{c}{21.14 $\pm$ 0.02}     & 
         \multicolumn{1}{c}{21.19 $\pm$ 0.22 }       & \multicolumn{1}{c}{}   &
         \multicolumn{1}{c}{20.95 $\pm$ 0.27} &
         \multicolumn{1}{c}{20.58 $\pm$ 0.22} &
         \multicolumn{1}{c}{20.93 $\pm$ 0.14} &
         \multicolumn{1}{c}{21.50 $\pm$ 0.56} &
         \multicolumn{1}{c}{6.07} &
         \multicolumn{1}{c}{2.70}
         \\ 
           \multicolumn{1}{c}{LS J1330-4025}           &  \multicolumn{1}{c}{22.80 $\pm$ 0.08}             & \multicolumn{1}{c}{20.13 $\pm$ 0.01}     & 
         \multicolumn{1}{c}{}       & \multicolumn{1}{c}{20.58 $\pm$ 0.25 }   &
         \multicolumn{1}{c}{} &
         \multicolumn{1}{c}{19.94 $\pm$ 0.25 } &
         \multicolumn{1}{c}{19.72 $\pm$ 0.04} &
         \multicolumn{1}{c}{19.46 $\pm$ 0.07} &
         \multicolumn{1}{c}{6.03} &
         \multicolumn{1}{c}{11.61}
         \\ 
           \multicolumn{1}{c}{LS J2011-4436}           &  \multicolumn{1}{c}{23.90 $\pm$ 0.26}             & \multicolumn{1}{c}{20.95 $\pm$ 0.03}     & 
         \multicolumn{1}{c}{}       & \multicolumn{1}{c}{20.44 $\pm$ 0.10 }   &
         \multicolumn{1}{c}{} &
         \multicolumn{1}{c}{19.59 $\pm$ 0.10 } &
         \multicolumn{1}{c}{19.32 $\pm$ 0.03} &
         \multicolumn{1}{c}{19.26 $\pm$ 0.06} &
         \multicolumn{1}{c}{6.31} &
         \multicolumn{1}{c}{2.90}
         \\ 
           \multicolumn{1}{c}{LS J2037-5152}           &  \multicolumn{1}{c}{25.55 $\pm$ 0.70}             & \multicolumn{1}{c}{21.08 $\pm$ 0.04}     & 
         \multicolumn{1}{c}{}       & \multicolumn{1}{c}{21.00 $\pm$ 0.17 }   &
         \multicolumn{1}{c}{} &
         \multicolumn{1}{c}{20.32 $\pm$ 0.21 } &
         \multicolumn{1}{c}{20.40 $\pm$ 0.08} &
         \multicolumn{1}{c}{19.75 $\pm$ 0.09} &
         \multicolumn{1}{c}{6.18} &
         \multicolumn{1}{c}{1.30}
         \\ 
           \multicolumn{1}{c}{LS J2155-5111}           &  \multicolumn{1}{c}{25.45 $\pm$ 0.34}             & \multicolumn{1}{c}{21.35 $\pm$ 0.03}     & 
         \multicolumn{1}{c}{}       & \multicolumn{1}{c}{21.05 $\pm$ 0.22 }   &
         \multicolumn{1}{c}{} &
         \multicolumn{1}{c}{} &
         \multicolumn{1}{c}{20.65 $\pm$ 0.09} &
         \multicolumn{1}{c}{20.41 $\pm$ 0.16} &
         \multicolumn{1}{c}{6.24} &
         \multicolumn{1}{c}{2.51}
         \\ 
           \multicolumn{1}{c}{LS J0139-5209}           &  \multicolumn{1}{c}{23.24 $\pm$ 0.08}             & \multicolumn{1}{c}{20.65 $\pm$ 0.01}     & 
         \multicolumn{1}{c}{}       & \multicolumn{1}{c}{20.71 $\pm$ 0.12 }   &
         \multicolumn{1}{c}{} &
         \multicolumn{1}{c}{20.27 $\pm$ 0.23} &
         \multicolumn{1}{c}{20.35 $\pm$ 0.07} &
         \multicolumn{1}{c}{20.64 $\pm$ 0.19} &
         \multicolumn{1}{c}{6.05} &
         \multicolumn{1}{c}{
1.83}
         \\  \bottomrule
        \end{tabular}
    \end{adjustbox}
    \begin{tablenotes}
    \small
    \item \textbf{Notes:} Col (1): Abbreviated quasar name; Col (2)-Col (3): optical magnitudes from LS DR10, Col (4)- Col (7): NIR magnitudes from UKIDSS DR11, VHS DR5 or UHS DR2; Col (8) -Col (9): MIR magnitudes from WISE; Col (10): maximum-likelihood photometric redshift estimated by \textit{eazy-py}; Col (11): stellar contaminant-to-quasar $\chi^2$ ratio used for the prioritization of the candidates.
    \end{tablenotes}
   \end{table*}

\section{New \textit{z} > 6 quasars}
\label{sec:new_qsos}

\begin{figure*}
\begin{minipage}{\linewidth}
   \centering
   \includegraphics[width = 0.90\linewidth]{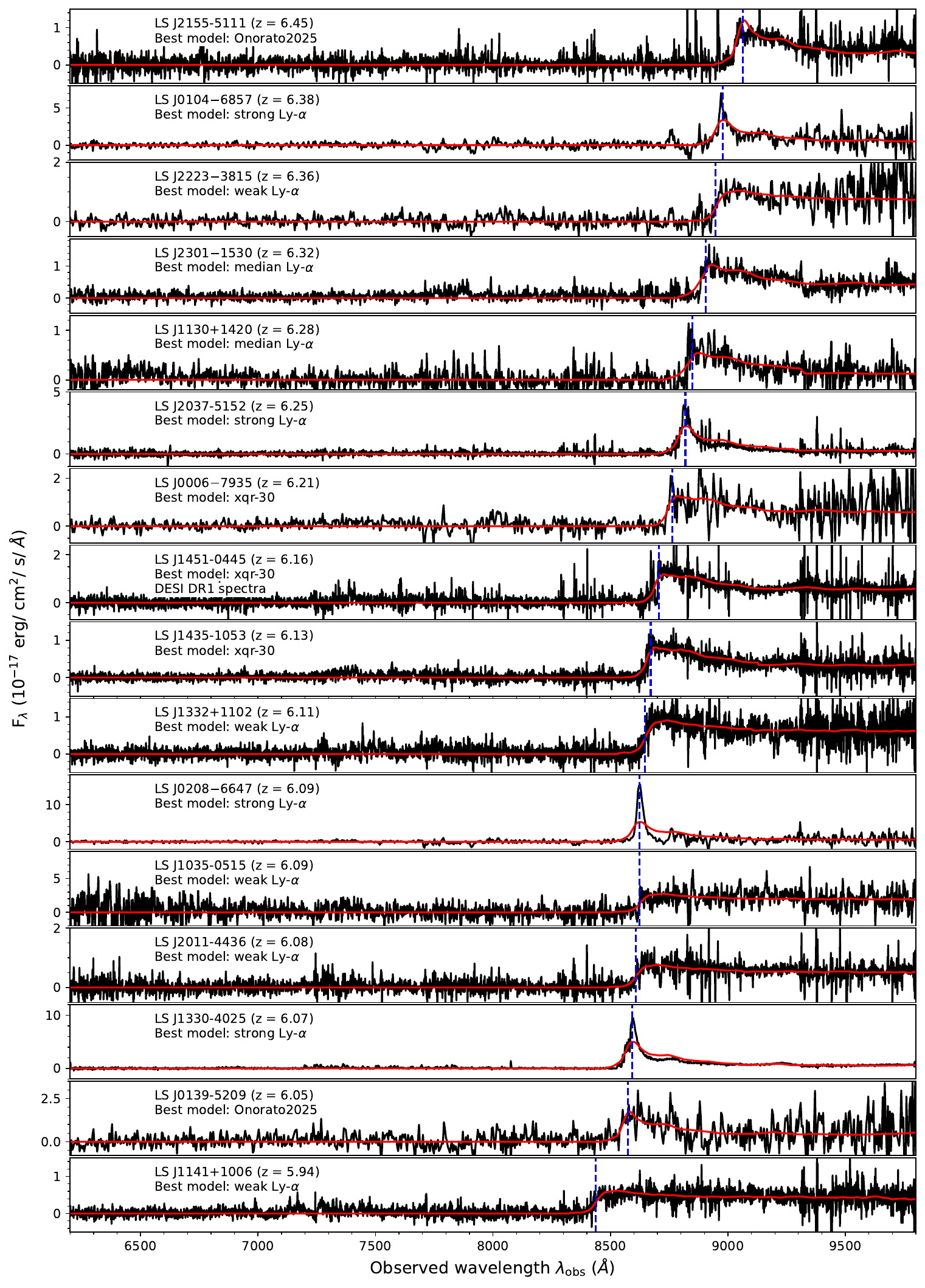}
   \caption{Optical discovery spectra of the 16 new quasars reported in this work, including the quasar LS J1451-0445 from the DESI data release \citep{abdul2025data}. The black curves show the observed spectra, the red curves the best-fit quasar templates used for redshift estimation, and blue dashed lines indicate the adopted best-fit position of Ly-$\alpha$. Each panel lists the source name, spectroscopic redshift, and the best-fit template in the upper left corner. The typical redshift uncertainty is $0.03$ (see Section~\ref{sec:new_qsos}). Quasars are sorted from top to bottom by decreasing redshift.}
    \label{all_Spectra}
\end{minipage}
\end{figure*}

During our spectroscopic follow-up campaign, we confirmed 15 new quasars at $z \geq 6$, and included one additional object from DESI DR1 public data that is part of our selection, for a total of 16. The photometric catalog properties, along with the main results from the \textit{eazy-py} run for the new quasars are presented in Table~\ref{photometric_cat}. In Fig. \ref{all_Spectra} we show the compilation of optical discovery spectra (observed frame), exhibiting a wide range of Ly-$\alpha$ line profiles and continuum slopes. For redshift estimation, we extended the approach presented in \cite{banados2023pan}. We fitted the spectra with a set of diverse quasar templates, but now including the SMC reddening law \citep{prevot1984typical} to account for obscuration. The free parameters of the $\chi^2$ minimization algorithm were the redshift, the normalization, and the color excess (E$_{\mathrm{B-V}}$). The template set was composed of the following:

\begin{itemize}
    
    \item \textit{median} Ly-$\alpha$, which is the median spectrum of 117 \textit{z} $\sim$ 6 quasars from the Panoramic Survey Telescope \& Rapid Response System 1 (PS1; \citealt{kaiser2002pan, kaiser2010pan}) presented in \cite{banados2016pan}.
    
    \item \textit{strong}  Ly-$\alpha$, which is the median PS1 spectrum of \textit{z} $\sim$ 6 quasars with the $10 \%$ largest rest-frame \ion{Ly}{$\alpha$} + NV equivalent width from \cite{banados2016pan}.
    
    \item \textit{weak}  Ly-$\alpha$, which is the median PS1 spectrum of the \textit{z} $\sim$ 6 quasars with the $10 \%$ smallest rest-frame \ion{Ly}{$\alpha$} + NV equivalent width from \cite{banados2016pan}.

    \item onorato2025, which is the weighted mean of 33 \textit{z} $> 6.5$ spectra presented in \cite{onorato2025optical}.

    \item xqr-30, which is the median of 42 \textit{z} $\sim 6$ high quality quasars spectra observed with X-Shooter and reported in \cite{d2023xqr}.

\end{itemize}

We selected the combination of parameters and template producing the lowest $\chi^2$ for each source in the wavelength range between $8000-8800$\,\AA, depending on the Ly-$\alpha$ location, and $\sim 9600$\,\AA, avoiding the highly noisy regions blueward of Ly-$\alpha$ and the red extreme of the spectra. The resulting redshift, $\chi^2$, and name of the best template are presented in Table \ref{physical_properties}. We do not include E$_{\mathrm{B-V}}$ as all the values are  $< 0.04$, indicating that none of the sources seems to be significantly reddened. We assumed a median uncertainty on the spectral redshift estimate with the fitting method of 0.03 as presented by \cite{banados2023pan}.

   \begin{table*}[ht]
      \caption[]{Redshift and physical properties of the new quasars.}
        \label{physical_properties}
         \centering
         \begin{tabular}{@{}cccccccccc@{}}
         \toprule
         \multicolumn{1}{c}{\textbf{Name}}  &  
         \multicolumn{1}{c}{\textbf{\textit{z}}}     & 
         \multicolumn{1}{c}{$\chi^2$/dof} &
         \multicolumn{1}{c}{\textbf{Template}} &  
         \multicolumn{1}{c}{\textbf{log $\lambda L_{\lambda} (1350 \,\AA)$}}  &
         \multicolumn{1}{c}{\textbf{log $L_{\mathrm{bol}}$}}  &
         \multicolumn{1}{c}{\textbf{m$_{1450}$}} &
         \multicolumn{1}{c}{\textbf{M$_{1450}$}} &
         \multicolumn{1}{c}{\textbf{FWHM (Ly-$\alpha$)}} &
         \multicolumn{1}{c}{\textbf{EW$_{\text{Ly}\alpha+\text{NV}}$}}
         \\ 
         \multicolumn{1}{c}{}  &
         \multicolumn{1}{c}{} &
         \multicolumn{1}{c}{} &
         \multicolumn{1}{c}{} &
         \multicolumn{1}{c}{erg s$^{-1}$} &
         \multicolumn{1}{c}{erg s$^{-1}$} &
         \multicolumn{1}{c}{} &
         \multicolumn{1}{c}{} &
         \multicolumn{1}{c}{km/s} &
         \multicolumn{1}{c}{\,\AA} 
         \\
         \multicolumn{1}{c}{(1)}  &
         \multicolumn{1}{c}{(2)} &
         \multicolumn{1}{c}{(3)} &
         \multicolumn{1}{c}{(4)} &
         \multicolumn{1}{c}{(5)} &
         \multicolumn{1}{c}{(6)} &
         \multicolumn{1}{c}{(7)} &
         \multicolumn{1}{c}{(8)} &
         \multicolumn{1}{c}{(9)} &
         \multicolumn{1}{c}{(10)} 
           \\ \midrule
        \multicolumn{1}{c}{LS J2155-5111}  &
         \multicolumn{1}{c}{6.45} &
         \multicolumn{1}{c}{0.11} &
         \multicolumn{1}{c}{Onorato25} &
         \multicolumn{1}{c}{45.31} &
         \multicolumn{1}{c}{45.89} &
         \multicolumn{1}{c}{21.27} &
         \multicolumn{1}{c}{$-$25.53} &
         \multicolumn{1}{c}{} &
         \multicolumn{1}{c}{82} 
           \\ 
        \multicolumn{1}{c}{LS J0104-6857}  &
         \multicolumn{1}{c}{6.38} &
         \multicolumn{1}{c}{0.43} &
         \multicolumn{1}{c}{\textit{strong} Ly-$\alpha$}&
         \multicolumn{1}{c}{45.65} &
         \multicolumn{1}{c}{46.22} &
         \multicolumn{1}{c}{20.43} &
         \multicolumn{1}{c}{$-$26.36} &
         \multicolumn{1}{c}{1613} &
         \multicolumn{1}{c}{90} 
           \\ 
        \multicolumn{1}{c}{LS J2223-3815}  &
         \multicolumn{1}{c}{6.36} &
         \multicolumn{1}{c}{0.12} &
         \multicolumn{1}{c}{\textit{weak} Ly-$\alpha$} &
         \multicolumn{1}{c}{45.62} &
         \multicolumn{1}{c}{46.20} &
         \multicolumn{1}{c}{20.49} &
         \multicolumn{1}{c}{$-$26.29} &
         \multicolumn{1}{c}{} &
         \multicolumn{1}{c}{15} 
           \\
         \multicolumn{1}{c}{LS J2301-1530}  &
         \multicolumn{1}{c}{6.32} &
         \multicolumn{1}{c}{0.04} &
         \multicolumn{1}{c}{\textit{median} Ly-$\alpha$}&
         \multicolumn{1}{c}{45.59} &
         \multicolumn{1}{c}{46.17} &
         \multicolumn{1}{c}{20.58} &
         \multicolumn{1}{c}{$-$26.20} &
         \multicolumn{1}{c}{} &
         \multicolumn{1}{c}{160} 
         \\
         \multicolumn{1}{c}{LS J1130+1420}  &
         \multicolumn{1}{c}{6.28} &
         \multicolumn{1}{c}{0.04} &
         \multicolumn{1}{c}{\textit{median} Ly-$\alpha$} &
         \multicolumn{1}{c}{45.42} &
         \multicolumn{1}{c}{46.00} &
         \multicolumn{1}{c}{20.98} &
         \multicolumn{1}{c}{$-$25.78} &
         \multicolumn{1}{c}{} &
         \multicolumn{1}{c}{87} 
         \\
         \multicolumn{1}{c}{LS J2037-5152}  &
         \multicolumn{1}{c}{6.25} &
         \multicolumn{1}{c}{0.14} &
         \multicolumn{1}{c}{\textit{strong} Ly-$\alpha$} &
         \multicolumn{1}{c}{45.41} &
         \multicolumn{1}{c}{45.99} &
         \multicolumn{1}{c}{21.01} &
         \multicolumn{1}{c}{$-$25.74} &
         \multicolumn{1}{c}{2656} &
         \multicolumn{1}{c}{212} 
           \\
           \multicolumn{1}{c}{LS J0006-7935}   &   
         \multicolumn{1}{c}{6.21} &
         \multicolumn{1}{c}{0.28} &
         \multicolumn{1}{c}{xqr-30} &
         \multicolumn{1}{c}{45.61} &
         \multicolumn{1}{c}{46.19} &
         \multicolumn{1}{c}{20.51} &
         \multicolumn{1}{c}{$-$26.24} &
         \multicolumn{1}{c}{} &
         \multicolumn{1}{c}{32} 
         \\ 
           \multicolumn{1}{c}{LS J1451-0445}  &
         \multicolumn{1}{c}{6.16} &
         \multicolumn{1}{c}{0.22} &
         \multicolumn{1}{c}{xqr-30} &
         \multicolumn{1}{c}{45.60} &
         \multicolumn{1}{c}{46.19} &
         \multicolumn{1}{c}{20.51} &
         \multicolumn{1}{c}{$-$26.22} &
         \multicolumn{1}{c}{} &
         \multicolumn{1}{c}{153} 
         \\ 
           \multicolumn{1}{c}{LS J1435-1053}  &
         \multicolumn{1}{c}{6.13} &
         \multicolumn{1}{c}{0.03} &
         \multicolumn{1}{c}{xqr-30} &
         \multicolumn{1}{c}{45.32} &
         \multicolumn{1}{c}{45.91} &
         \multicolumn{1}{c}{21.21} &
         \multicolumn{1}{c}{$-$25.52} &
         \multicolumn{1}{c}{} &
         \multicolumn{1}{c}{247} 
         \\ 
           \multicolumn{1}{c}{LS J1332+1102}  &
         \multicolumn{1}{c}{6.11} &
         \multicolumn{1}{c}{0.08} &
         \multicolumn{1}{c}{\textit{weak} Ly-$\alpha$} &
         \multicolumn{1}{c}{45.51} &
         \multicolumn{1}{c}{46.09} &
         \multicolumn{1}{c}{20.74} &
         \multicolumn{1}{c}{$-$25.98} &
         \multicolumn{1}{c}{} &
         \multicolumn{1}{c}{22} 
         \\
           \multicolumn{1}{c}{LS J0208-6647}  &
         \multicolumn{1}{c}{6.09} &
         \multicolumn{1}{c}{2.82} &
         \multicolumn{1}{c}{\textit{strong} Ly-$\alpha$} &
         \multicolumn{1}{c}{45.76} &
         \multicolumn{1}{c}{46.34} &
         \multicolumn{1}{c}{20.12} &
         \multicolumn{1}{c}{$-$26.60} &
         \multicolumn{1}{c}{1146} &
         \multicolumn{1}{c}{170} 
         \\ 
           \multicolumn{1}{c}{LS J1035-0515}  &
         \multicolumn{1}{c}{6.09} &
         \multicolumn{1}{c}{3.51} &
         \multicolumn{1}{c}{\textit{weak}  Ly-$\alpha$} &
         \multicolumn{1}{c}{45.77} &
         \multicolumn{1}{c}{46.35} &
         \multicolumn{1}{c}{20.08} &
         \multicolumn{1}{c}{$-$26.63} &
         \multicolumn{1}{c}{} &
         \multicolumn{1}{c}{8} 
         \\ 
         \multicolumn{1}{c}{LS J2011-4436}  &
         \multicolumn{1}{c}{6.08} &
         \multicolumn{1}{c}{0.09} &
         \multicolumn{1}{c}{\textit{weak}  Ly-$\alpha$} &
         \multicolumn{1}{c}{45.45} &
         \multicolumn{1}{c}{46.03} &
         \multicolumn{1}{c}{20.90} &
         \multicolumn{1}{c}{$-$25.82}  &
         \multicolumn{1}{c}{} &
         \multicolumn{1}{c}{45} 
         \\ 
           \multicolumn{1}{c}{LS J1330-4025}  &
         \multicolumn{1}{c}{6.07} &
         \multicolumn{1}{c}{0.38} &
         \multicolumn{1}{c}{\textit{strong}  Ly-$\alpha$} &
         \multicolumn{1}{c}{45.78} &
         \multicolumn{1}{c}{46.36} &
         \multicolumn{1}{c}{20.08} &
         \multicolumn{1}{c}{$-$26.63} &
         \multicolumn{1}{c}{1713} &
         \multicolumn{1}{c}{149} 
         \\ 
           \multicolumn{1}{c}{LS J0139-5209}  &
         \multicolumn{1}{c}{6.05} &
         \multicolumn{1}{c}{0.49} &
         \multicolumn{1}{c}{Onorato25} &
         \multicolumn{1}{c}{45.41} &
         \multicolumn{1}{c}{45.99} &
         \multicolumn{1}{c}{21.00} &
         \multicolumn{1}{c}{$-$25.71} &
         \multicolumn{1}{c}{} &
         \multicolumn{1}{c}{167} 
         \\ 
           \multicolumn{1}{c}{LS J1141+1006}  &
         \multicolumn{1}{c}{5.94} &
         \multicolumn{1}{c}{0.04} &
         \multicolumn{1}{c}{\textit{weak}  Ly-$\alpha$} &
         \multicolumn{1}{c}{45.36} &
         \multicolumn{1}{c}{45.94} &
         \multicolumn{1}{c}{21.10} &
         \multicolumn{1}{c}{$-$25.58} &
         \multicolumn{1}{c}{} &
         \multicolumn{1}{c}{25} 
         \\ \bottomrule
        \end{tabular}
        
    \begin{tablenotes}
    \small
    \item \textbf{Notes:} Col (1): Abbreviated quasar name; Col (2): spectroscopic redshift estimated with the spectrum fitting with an uncertainty of 0.03 (see Section \ref{sec:new_qsos}), Col (3): reduced $\chi^2$ of the best-fitted template; Col (4): Name of the best-fit template; Col (5): monochromatic luminosity at rest frame 1350 \,\AA; Col (6): bolometric luminosity, Col (7-8): apparent and absolute magnitudes at rest frame 1450 \,\AA.
    \end{tablenotes}
   \end{table*}

The sample of quasars spans a redshift range $z = [5.94, 6.45]$ and displays diverse types of Ly-$\alpha$ emission, from weak to strong lines. Four quasars (LS J010-68, LS J203-51, LS J020-66, and LS J133-40) exhibit prominent and relatively narrow, high EW Ly-$\alpha$ lines. The reduced $\chi^2$ values indicate that some models underfit the data, notably for LS J0208-6647, where the fit fails redward of the Ly-$\alpha$ line, and for LS J1035-0515, which has the noisiest spectrum in the sample. Conversely, nine sources exhibit very low reduced $\chi^2$ values (below 0.2), suggesting uncertainties or degeneracies in the fitted parameters, likely due to low S/N and poor spectral quality.

\subsection{Physical properties of the newly discovered quasars}

The availability of optical, NIR, and MIR broadband photometry allowed us to carry out a first diagnostic on the quasar parameter space that we are exploring. We started with widely used physical quantities, such as the bolometric luminosity and the magnitudes at rest frame wavelength 1450\,\AA, which are useful for bolometric corrections and quasar characterization. However, we did not derive these quantities directly from the best-fit spectral templates due to several limitations. First, the spectra are generally noisy, and our fits exclude data at observed wavelengths $> 9600$\,\AA\ (corresponding to rest-frame wavelengths $> 1385$\,\AA), which affects the reliability of continuum slope estimates. Consequently, flux predictions in these regions are not robust. Furthermore, our current templates did not reproduce the sharp Ly-$\alpha$ emission features along with the underlying continuum successfully, often leading to underestimated fluxes in the rest-frame UV regime for the peaky Ly-$\alpha$ systems. Besides, we wanted to use a method that would be consistent for all the quasars, regardless of the quality of the single spectrum, and consistent with the literature for fair comparisons.

We followed the approach presented in \citet{banados2016pan}, \citet{mazzucchelli2017no} and \citet{belladitta2025discovery}, modeling the quasar continuum as a power-law with slope $\alpha = 0.44$ \citep{berk2001composite}. Since all our sources have reliable $z$-band photometry, we used this magnitude at the pivot wavelength 9168.85\,\AA\ (for DECam) to extrapolate the apparent magnitude at 1450\,\AA\ ($m_{1450}$) and the monochromatic luminosity at 1350\,\AA\ ($\lambda L_\lambda$(1350\,\AA)). While $m_{1450}$ traces the accretion disk output, $\lambda L_\lambda$(1350\,\AA) provides a basis for estimating the bolometric luminosity. To compute the absolute magnitude M$_{1450}$, we adopted the redshifts derived from our spectral fitting. To estimate the total radiative output due to SMBH accretion, we adopted a bolometric correction based on the monochromatic luminosity at 1350 \,\AA\ following \cite{shen2008biases}:

\begin{equation}
    \hspace{20mm} L_{\mathrm{bol}} = (3.81 \pm 0.38) \times \lambda L_{\lambda} (1350 \,\AA).
\end{equation}

\noindent All these quantities are listed in Table~\ref{physical_properties}.

For the sources exhibiting prominent \ion{Ly}{$\alpha$} emission (LS J0104–6857, LS J2037–5152, LS J0208–6647, and LS J1330–4025), we estimated the full width at half maximum (FWHM) of the \ion{Ly}{$\alpha$} line. First, we shifted the spectra to the rest frame. Then, we modeled the continuum emission following the approach of \cite{diamond2009high}, fitting a power law of the form $f_\lambda = \mathrm{C} \lambda^{\beta}$, where $\beta = \alpha -2$, to spectral regions free of strong emission lines (at rest-frame wavelengths 1285–1295, 1315–1325, 1340–1375, 1425–1470, 1680–1710, 1975–2050, and 2150–2250 \,\AA).  However, because our spectra are noisy at rest frame $> 1385$\,\AA, we cannot reliably constrain the continuum slope. Therefore, we fixed the power-law index to $\beta = -1.56$, consistent with the quasar slope we adopted to estimate $\mathrm{m}_{1450}$ and $\lambda L_\lambda$(1350\,\AA). After subtracting the continuum component, we located the peak of the \ion{Ly}{$\alpha$} emission and fit a single Gaussian profile to the red side of the line. The blue side is strongly affected by IGM absorption and resonant scattering, which distorts the line profile and makes it unsuitable for fitting. Fitting only the red wing provides a more reliable estimate of the line width.

The resulting FWHM values are in the range $1147-2657$ km s$^{-1}$. According to the classification of \citet{matsuoka2018subaru}, which defines narrow \ion{Ly}{$\alpha$} quasars as those with FWHM $<$ 500 km s$^{-1}$, these sources do not meet the criterion. However, the prominent \ion{Ly}{$\alpha$} emitters represent 25\% of our sample and display significantly narrower line widths compared to the rest. As virial black hole mass estimates scale with the square of the broad emission line width, narrower lines may indicate lower black hole masses. Near-infrared spectroscopy will be essential to confirm this interpretation by providing measurements of other broad emission lines and estimates of the black hole mass.

In order to systematically investigate the emission-line properties of our quasar sample, we computed the equivalent width (EW) of the blended emission from \ion{Ly}{$\alpha$} $\lambda$1216\,\AA, \ion{N}{V} $\lambda$1240\,\AA\ and \ion{Si}{II} $\lambda$1263\,\AA, following the approach by \citet{diamond2009high} and \citet{banados2016pan} applied to SDSS DR5 and PS1 high-redshift quasars. We first determined the continuum using the same power-law model described earlier and then computed the EW by integrating the flux excess over the continuum in the rest-frame wavelength range 1160–1290\,\AA.

\begin{figure}
   \centering
   \includegraphics[width = 0.98\linewidth]{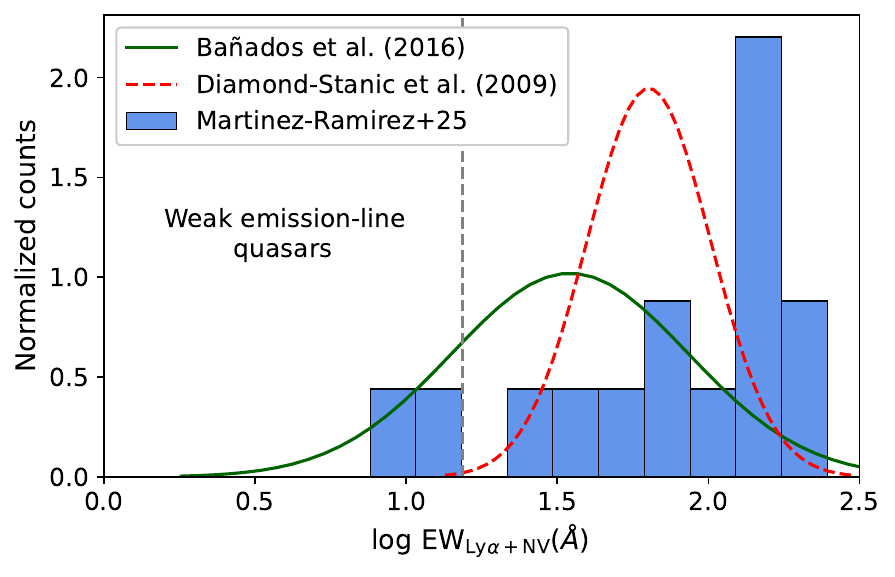}
   \caption{Normalized EW(\ion{Ly}{$\alpha$}+ \ion{N}{V}) distribution for the 16 quasars at $z= 5.93-6.45$ identified through our CL selection. The solid green and dashed red curves show the best-fit log-normal distributions for PS1 quasars \citep{banados2016pan} and SDSS quasars \citep{diamond2009high}, with mean values of $\log$(EW/\AA) = 1.542 and 1.803, and standard deviations of 0.391 and 0.205\,\AA, respectively. The vertical dashed gray line marks the limit for WLQs \citep[EW $<$ 15.4\,\AA;][]{diamond2009high}. While most of our sample overlaps with the PS1 distribution at $\log$(EW/\AA) $<$ 2, a substantial excess appears in the high-EW tail, potentially revealing a distinct population of strong-line quasars.}
    \label{EW-distributions_quasars}
\end{figure}

We found that only two of our quasars meet the definition of weak emission-line quasars (WLQs), with EW $<$ 15.4\,\AA\ as set by \citet{diamond2009high}. This represents $12.5\%$ of our sample, a fraction comparable to the $13.7 \%$ reported by \cite{banados2016pan} for 124 PS1 quasars at $5.6 \lesssim z \lesssim 6.7$. Although the size of our quasar sample prevents a robust statistical analysis, we compared our EW distribution of \ion{Ly}{$\alpha$}+ \ion{N}{V}+ \ion{Si}{II} to the log-normal fits derived for SDSS DR5 quasars at $3 < z < 5$ and PS1 quasars. Fig. \ref{EW-distributions_quasars} presents the normalized EW distribution of our 16 quasars, along with the SDSS and PS1 log-normal models from \citet{diamond2009high} and \citet{banados2016pan}, respectively. While $56.25$\% of our sample with $\log$(EW/\AA) $<$ 2 closely follows the PS1 distribution, a significant fraction ($43.75\%$) exhibits enhanced line strengths, occupying the high-EW tails of both reference distributions. This fraction is notably higher than the $17\%$ and $12\%$ predicted by the SDSS and PS1 fits, respectively.

\citet{diamond2009high} reported an increasing fraction of WLQs with redshift, which implies that our sample would be expected to show overall lower EW values than SDSS and be broadly consistent with PS1. We performed an Anderson–Darling test comparing our sample to the PS1 EW distribution, obtaining a statistic of $56.92$ and a p-value of $0.001$. This indicates a substantial difference between the two distributions and a very low probability of obtaining such a statistic if our sample were drawn from the PS1 distribution, leading us to reject the null hypothesis. The EW measurements therefore suggest that our selection method is recovering a population of quasars with systematically stronger \ion{Ly}{$\alpha$}+\ion{N}{V}+ \ion{Si}{II} emission than typically found in previous surveys. The EW values, along with the other physical quantities derived in this section, are summarized in Table~\ref{physical_properties}. A sample with at least several dozen quasars ($\gtrsim 50$) within this selection framework will be crucial to determine whether our selection is probing a different, less-explored region of quasar parameter space, or if the current results arise from small-number statistics.

\subsection{Comparison with the literature quasars}

\begin{figure}
   \centering
   \includegraphics[width = 0.99\linewidth]{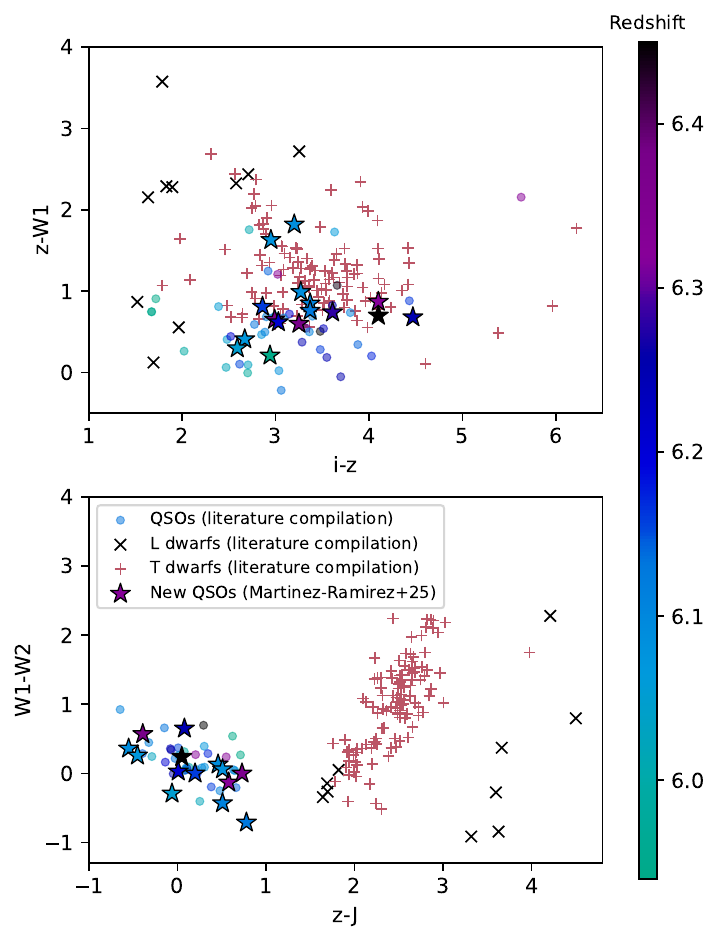}
   \caption{Color–color diagrams showing the positions of the newly discovered quasars (stars) and the known high-redshift quasars at the same redshift range of our discoveries $5.94 \leq z \leq 6.45$ (filled circles), both color-coded by redshift; and L and T dwarfs (black and red crosses, respectively) used as labels in the embedded space. Top: $i - z$ vs. $z - \mathrm{W1}$. Bottom: $z - J$ vs. $\mathrm{W1} - \mathrm{W2}$. While $i - z$ vs. $z - \mathrm{W1}$ shows a total overlap of the new quasars with the T dwarf population, the $z-J$ color shows a clear distinction.}
    \label{color-color}
\end{figure}

To place our newly discovered sources in context, we examined color–color diagrams to assess the parameter space occupied by them (see Fig. \ref{color-color}). Table \ref{photometric_cat} shows that all the sources from our sample have valid $i$, $z$, $W1$ and $W2$ band magnitudes,  while $J$ is available for $14$ of the sources. We therefore computed colors using these bands to enable a more homogeneous comparison.

13 of the 16 quasars exhibit strong spectral breaks with $i - z > 3$ and $z - \mathrm{W1} > 0.5$, overlapping entirely with the region populated by T dwarfs. While a significant fraction of known $z \sim 6$–6.7 quasars occupy the same region of color space, previous selections typically rely on additional data or strict constraints to separate them from contaminants. In contrast, our method successfully identifies quasars in regions heavily contaminated by UCDs by leveraging the $i$ and $z$ band imaging as input. This demonstrates the effectiveness of our CL approach in isolating quasars without requiring extensive multiwavelength information.

The $z - J$ colors indicate that $\sim 38\%$ of the newly discovered quasars exhibit mildly red NIR slopes ($z - J > 0.4$), though still consistent with the colors of known quasars at similar redshifts. We noticed a strong concentration of sources with $0.4 < z-J < 0.8$. For quasars within this color range, we computed an average redshift of $z = 6.27$ for the known sample from the literature, compared to $z = 6.18$ for our discoveries. This offset may suggest that our selection is recovering a population of quasars with slightly redder NIR colors. While this is not a direct outcome of the CL approach itself, it likely reflects the more flexible constraints adopted during our catalog-based preselection.

Quasars with redder NIR slopes are more likely to overlap with the locus of UCD in traditional color–color diagrams, and are therefore often excluded or deprioritized by traditional selections that require flat or blue continua redward of \ion{Ly}{$\alpha$} to minimize contamination \citep[e.g.,][]{banados2016pan, wang2017first, findlay2012selection}. Physically, redder slopes can arise from several effects: modest dust extinction in the quasar host galaxy, an intrinsically softer continuum from the accretion disk, or strong line contributions in the $J$ band (e.g., from \ion{C}{IV}$\lambda 1549$ or \ion{C}{III]}$\lambda1909$ depending on redshift). Previous studies have suggested that red quasars may represent either dust-obscured young transitional phases of SMBHs growth \citep[e.g.,][]{glikman2024accretion} or quasars viewed at larger inclinations \citep[e.g.,][]{martin2020peculiar}. Thus, the concentration of our new discoveries in this color regime may be pointing toward a less-explored population systematically missed by past color-cut selections.

To understand why our newly discovered quasars were overlooked by previous selections, we investigated the survey coverage and selection criteria applied to these sources. None of the 16 newly confirmed quasars have X-ray or radio detections\footnote{The X-ray and radio datasets checked are: eROSITA DR1, LOFAR DR2, EMU, RACS (low, mid, high), FIRST, NVSS, VLASS, and Meerkat-MALS.}, excluding them from selections relying on these bands. Also, of the 16, only three overlap with the DES footprint, three with PS1, and none with SDSS DR12, making incomplete spatial overlap a primary factor. All of our sources are classified as PSF-like, indicating that traditional selection methods based on point-source morphology would not discard them. Beyond coverage, color, and S/N requirements, along with prioritization strategies, also explain the missing sources. In the following, we provide a breakdown of the different selections from the literature:
\\

\paragraph{PS1-based selections:}
The DESI selection by \citet{yang2023desi} relied on PS1 $i$-band imaging, which immediately excludes 13 of our quasars due to lack of PS1 coverage. Of the remaining three, LS J1332+1102 failed their $W1 - W2 > -0.14$ color cut, and LS J2301-1530 and LS J1035–0515 were excluded by the flat continuum criterion ($z_{\mathrm{P1}} - y_{\mathrm{P1}} < 1$). Similarly, the PS1-based selection of \citet{banados2016pan} included only three of our quasars. LS J1332+1102 failed their $-0.2 < W1 - W2 < 0.85$ color prioritization criterion. Although LS J2301-1530 and LS J1035–0515 meet the selection thresholds, they may have been deprioritized due to low S/N or other quality cuts. The selection of $z \sim 7$ quasar candidates by \citet{wang2019exploring}, combining DESI Legacy imaging survey (DELS), PS1, UKIRT or UHS, and WISE data, also includes only three of our quasars. All of them (LS J2301-1530, LS J1332+1102, and LS J1035–0515) were excluded by the $z$-dropout requirement ($z_{\mathrm{P1}} - y_{\mathrm{P1}} > 1.5$).
\\

\paragraph{Comprehensive multi-method selection:} The recent work by \cite{belladitta2025discovery} explore a broader set of quasar selections strategies, including: 1) PS1 $i$-dropouts with NRAO VLA Sky Survey + ALLWISE detections \citep{belladitta2023powerful}, 2) Radio-selected quasars candidates from the Million Quasar catalog \citep{flesch2023million}, 3) PS1-based $i$-dropout selection from \citet{banados2023pan}, 4) $z \gtrsim 6.6$ selection using DELS + PS1 and 5) $z > 6.5$ selection from PS1 and ALLWISE by \cite{belladitta2025discovery}, 6) DES/VHS/CatWISE $i$-dropouts \citep{wolf2024srg}, and 7) DES-based photometric candidates \citep{yang2022southern}.

None of our quasars are detected in radio, excluding them from selections (1) and (2). For selection (3), LS J2301-1530, LS J1332+1102, and LS J1035–0515 failed the $z_{\mathrm{P1}} - y_{\mathrm{P1}} < 0.5$ cut. Selection (4) excluded LS J2301-1530 and LS J1332+1102 based on $z_{\mathrm{P1}} - z_{\mathrm{DE}} > 0.8$, and LS J1035–0515 based on $z_{\mathrm{DE}} - y_{\mathrm{P1}} > 0.5$. Selection (5) reproduces the criteria from \citet{wang2019exploring}, thus excluding the same sources.
Selection (6) missed LS J2037–5152 and LS J2155–5111 due to the absence of $i$-band detections, which are necessary for their color criteria. Despite LS J0139-5209 meeting the photometric constraints, it may have been rejected at the SED fitting stage or deprioritized due to quality flags. Finally, none of our sources were matched to the candidates catalog from selection (7).
\\

\paragraph{DECaLS and DES selections:} \citet{wang2017first} combined DECaLS with SDSS $i$-band photometry; however, their sample does not overlap with our sources. \citet{reed2017eight}, using DES data, included LS J2037–5152, LS J2155–5111 and LS J0139-5209 in their footprint, but the first two were excluded for being fainter than their $z_{\mathrm{PSF}} \leq 21$ magnitude threshold, while the last one met their color selection, they likely never followed up on it.

\subsection{Contaminant morphologies and classification notes}

Our catalog-based preselection did not impose strict morphological constraints on $i$-dropouts, allowing for the inclusion of both point-like and extended sources. All confirmed quasars in our sample are classified as point-like (“PSF” type) in the LS DR10 catalog. Among the contaminants, 19 are also point-like and are thus consistent with being UCDs. One contaminant is modeled with an exponential profile, and two are best fit by round exponential galaxy models, suggesting they could be compact, dusty red galaxies at low redshift or other types of extended contaminants. Spectroscopic analysis is required to confirm their nature.

Interestingly, about half of the contaminants have photometric redshifts $z_{\mathrm{phot}} \gtrsim 6.7$, a range where contamination from T dwarfs is particularly common. This further supports the interpretation that a significant fraction of our contaminants are UCDs misclassified due to their colors and morphology.

\section{Conclusions}

In this work, we report the discovery of 16 new quasars at $z = 5.94$–$6.45$ using a self-supervised CL approach applied to DESI Legacy Survey DR10 optical images. Our method combines a flexible catalog-based preselection of $i$-dropouts, free from strict morphological or color constraints, with representation learning, enabling us to recover sources that would otherwise be overlooked, including quasars with slightly red NIR slopes, softer \ion{Ly}{$\alpha$} breaks, and even potential quasar pairs. Such successful outcomes stem from the interplay of our modern self-supervised image-based method with traditional methods such as dropout color cuts, SED fitting, and aperture photometry constraints.

Starting from a sample of 165,253 $i$-dropout candidates, we trained a CL model using only four DESI LS DR10 bands and produced a low-dimensional latent space representation where quasars naturally cluster. Within this embedded space, we identified overdensities of spectroscopically confirmed quasars at $z \sim 6.0-6.3$, with contamination rates of $\sim50 \%$ from known M, L, and T dwarfs from the literature. We also uncovered trends linked to brightness and morphology, the primary drivers of the latent representation: point-like and elongated sources occupy distinct regions, and a smooth gradient in brightness is observed. Importantly, quasars appear across a broad range of $i-z \sim 1.5 - 2.5$ and $z-J$ colors, including regions heavily contaminated by UCDs, underscoring the strength of CL in disentangling complex parameter spaces.

To prioritize candidates for follow-up, we coupled our CL approach with SED fitting, achieving a $\sim45\%$ success rate in our spectroscopic campaigns. Among the confirmed quasars, we found a diversity in \ion{Ly}{$\alpha$} profiles, including four with relatively narrow lines (FWHM $\sim 1146$–$2656$ km s$^{-1}$), and six with red NIR slopes ($z-J > 0.4$) with slightly lower average redshifts compared to known quasars with similar colors. Our EW analysis of the \ion{Ly}{$\alpha$}+\ion{N}{V}+ \ion{Si}{II} complex reveals that $\sim 44\%$ of our sample exhibits enhanced line strengths. To robustly determine whether our selection systematically recovers quasars with high EWs, a larger spectroscopic sample, on the order of $50$–$100$ quasars, will be required. Such numbers will soon be within reach with DESI, Rubin/LSST, and 4MOST optical spectra, providing the statistical power to confirm whether the high-EW tail we observe reflects a genuine population rather than small-number statistics. With the currently available optical bands and a more flexible, data-driven search, we have shown that it is possible to probe parameter space heavily contaminated by T dwarfs and to recover quasars with systematically stronger emission features than those typically reported in the literature, which were overlooked by previous selections.

As discussed in Section \ref{sec_photometric_cat}, relaxing the NIR detection requirement in the input catalog for the CL pipeline could potentially double the number of quasar discoveries. Many of these currently missed candidates are likely to be revealed by ongoing surveys such as \textit{Euclid}, thanks to its deeper and more uniform NIR coverage. Notably, even within the more traditional regions of parameter space populated by known quasars, our selection moves beyond the standard color-cut paradigm by recovering sources with atypical properties. Alternative strategies, such as extending the search into less-explored regions of the latent space, where spectroscopic labels are scarce but trends in catalog properties offer guidance (see subsection \ref{inter_embedded_space}), hold the potential to uncover even more unusual populations, albeit at higher observational risk.

The set of artifacts and brown dwarf contaminants identified in our selection provides valuable training data to improve the algorithm’s performance, enhance the label-based interpretation of the embedded space, and refine the SED-fitting prioritization. Future iterations could make the algorithm more robust against artifacts by including them in the augmentation process or by incorporating them as a separate class in a semi-supervised framework. Similarly, by expanding the stellar training set with high-quality optical and NIR spectra of these newly discovered UCDs, and especially focusing on atmospheric compositions that mimic quasar SEDs, we can reduce contamination rates and explore alternative photometric or morphological indicators to help discriminate these sources more reliably.

Looking ahead, this approach can be naturally extended to ongoing and upcoming wide-field surveys such as Rubin/LSST, \textit{Euclid}, and Roman, where the combination of deep, multiband photometry and large-area coverage will dramatically increase the discovery potential for high-redshift quasars. With \textit{Euclid} already demonstrating the capability of the NISP spectroscopy to pinpoint rare and luminous quasars at $z > 5$ \citep[e.g.,][]{banados2025euclid}, our method is well-positioned to scale to these future datasets, leveraging both imaging and low-resolution spectral data to push the quasar frontier deeper into the reionization era and place stronger constraints on the formation and early growth of SMBHs.

\bibliographystyle{aa}
\bibliography{ref.bib}

\begin{appendix}

\section{Unraveling the embedded space with catalog features}
\label{analysis_es_catalogmaps}

\begin{figure*}
   \centering
   \includegraphics[width = 0.96\linewidth]{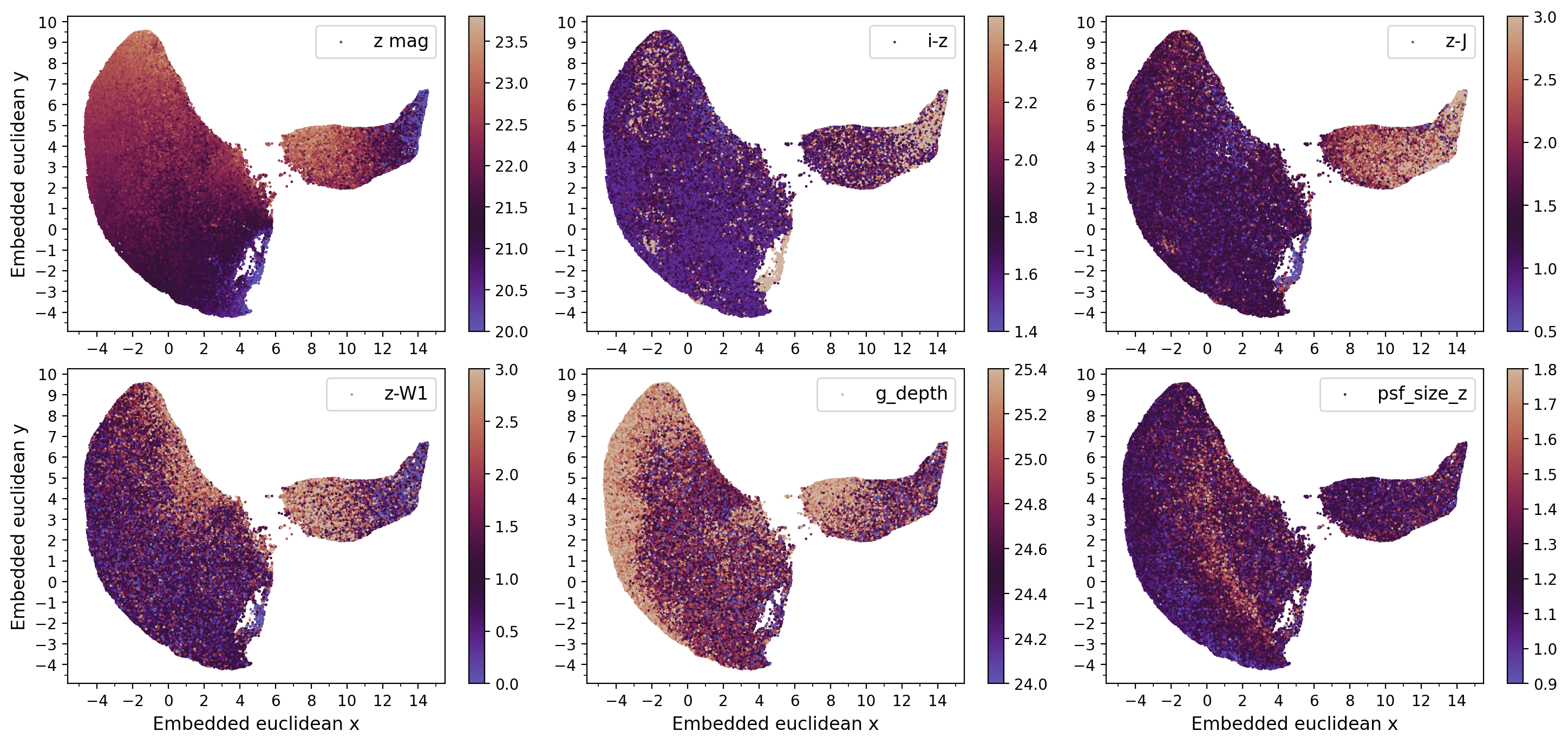}
   \caption{Color-coded embedded space maps showing catalog photometric features.  \textit{Top row} (left to right): $z$-band magnitude, $i-z$ color, and $z-J$ color. \textit{Bottom row}: $z$-W1 color, $g$-band depth, and $z$-band PSF size. \textit{Top left}: magnitude gradients across both regions underscore the crucial role of the brightness in shaping the low-dimensional representation. \textit{Top center}: Multiple regions of strong $i-z$ color indicate that while this feature may not fully dominate the embedding, it still plays a key role—evidenced by patterns and some clustering that overlap with the positions of spectroscopically confirmed quasars. \textit{Top right}: The $z-J$ color map reveals populations consistent with blue quasars ($z-J \leq 0.5$), quasars with intermediate colors ($z-J \sim 1.0$–1.5), and a population with red NIR slopes ($z-J >2 $), supporting the potential of the model to uncover diverse quasar types. \textit{Bottom left}: Red $z$-W1 colors seem to reflect the W1 blending due to companions and potential faint T dwarf populations. \textit{Bottom center}: Subtle observational systematics, such as variations in background noise texture, may influence the latent space structure as evidenced by the $g$-band depth pattern.  \textit{Bottom right}: Larger PSFs point to a mix of slightly extended or noisy $z$-band images.}
   \label{photometric_properties}
    \end{figure*}

To investigate the dominant properties influencing the arrangement of sources in the embedded space, we color-code the data points by catalog magnitudes, color indices, depth, and PSF size. Figure \ref{photometric_properties} highlights several relevant features: the $z$-band magnitude; $i-z$, $z-J$  and $z-W1$ colors; the $g$-band depth and $z$-band PSF size. These maps reveal trends that aid the interpretation of the algorithm. For instance, source brightness is a major driver of the representation, as evidenced by the gradient in $z$ magnitude: it increases from bottom to top within the left island and from right to left within the right island. The $i-z$ color map highlights regions with strong spectral breaks, where most confirmed quasars and stars are located in Fig. \ref{Embedded_space}. Notably, the area around (x,y) = (2, -3) containing quasars shows more moderate $i-z$ colors ($\sim 1.5$), suggesting that our more flexible selection criteria successfully include sources that might be excluded by stricter cuts. 

Although the NIR bands were not part of the CL input, we include $z-J$ color maps to investigate potential differences among quasar populations. The region near (x,y) = (5, -2.5) corresponds to typical blue quasars with flat NIR slopes ($z-J \leq 0.5$), whereas the region around (x,y) = (2, -3) contains a mix of intermediate colors ($z-J \sim 1.0$–1.5). Meanwhile, the region near (x,y) = (-2, -1) and the vertical structure toward the top of the map appear to host a distinct population of quasars with red NIR slopes. We also examine the $z$–W1 color, commonly employed for quasar selection, where colors $> - 0.2$ select quasars at $z \gtrsim 5.7$ \citep[e.g.,][]{yang2023desi}, with redder values corresponding to higher redshifts. Consistently, we observe $z$–W1 $\sim 0$ near (x,y) = (5, –2), a region containing several spectroscopically confirmed quasars and extreme $i$–$z$ breaks—suggesting a high-confidence selection region. Other areas hosting known quasars exhibit $z$–W1 values between $\sim 0 -0.6$, still consistent with $z > 6$ quasars. Anomalously red $z$–W1 colors ($\sim$ 3) appear on the left side of the small island, likely caused by W1 flux contamination from nearby sources, as the lower spatial resolution of WISE compared to LS DR10 can lead to blending. Meanwhile, the region with $z$–W1 $\sim 2.5$ on the right side of the large island coincides with fainter $z$-band fluxes, weaker $i$–$z$ breaks, and two known T dwarfs, suggesting that it may be hosting a larger population of T dwarfs.

The $g$-band depth map offers insight into the impact of data quality on the latent space representation. Sources with deeper $g$-band images, characterized by lower background noise, tend to cluster along the left edges of both latent-space islands. Upon visual inspection of the optical stamps of random sources from these regions, we did not find any trend in the background level; in fact, sources with both high and low background levels are often located adjacent to each other. However, we notice that the noise structure appears more peaked and irregular in the upper regions of the embedded space, becoming flatter and more homogeneous toward the bottom. This suggests that subtle observational systematics, particularly related to the noise texture, may influence the latent representation. Nevertheless, the effect is not dominant; there is no smooth gradient or sharply defined clustering of deep $g$-band sources, indicating that this feature likely plays only a minor role in shaping the embedded space. This effect of the imaging quality is also evidenced by the $z$-band PSF size map, as the smaller sizes indicating better resolution and less blurring overlap the deeper $g$-band areas and conversely.

\section{Proper motion of stars in the embedded space}

\begin{figure}[ht!]
   \centering
   \includegraphics[width = 0.95\linewidth]{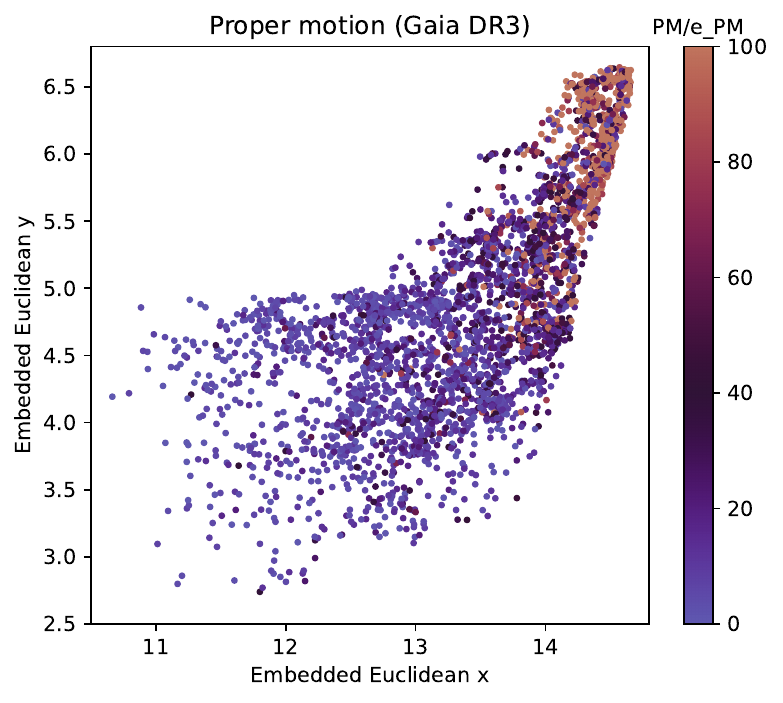}
   \caption{Embedded space distribution of spectroscopically confirmed stars, color-coded by S/N of Gaia DR3 proper motion measurements. Higher S/N values correspond to stronger detections of motion, highlighting a gradient in proper motion across the embedded space.}
    \label{proper_motion}
\end{figure}

As shown in Fig. \ref{average_images}, the region of the embedded space associated with elongated sources overlaps with the distribution of stars, suggesting the influence of high proper motions. To investigate this, we retrieved proper motion measurements from Gaia DR3 for all spectroscopically confirmed stars in the embedded space and color-coded them by their proper motion S/N, defined as the proper motion divided by its uncertainty (see Fig. \ref{proper_motion}). We find consistent trends in both RA and Dec components, with increasing proper motion S/N toward the upper right of the embedded space. This indicates that the CL algorithm is sensitive to the apparent motion of sources, placing those with the highest proper motions toward the periphery of the learned representation.

\section{2D Gaussian Kernel Density Estimation (KDE) maps}

We define a quantitative criterion to select candidates from regions in the latent space with a high density of quasars and a low contamination fraction from UCDs, using Gaussian kernel density estimation (KDE) maps. These maps in Fig. \ref{KDEdensities} are constructed from the currently available sample of spectroscopically confirmed quasars and MLT dwarfs, and are normalized by the total number of sources in each class. Although the limited number of labeled sources, particularly in the southern hemisphere, introduces potential biases, this KDE-based approach provides a reasonable first-order method for prioritizing quasar candidates.

\begin{figure}
   \centering
   \includegraphics[width = 0.95\linewidth]{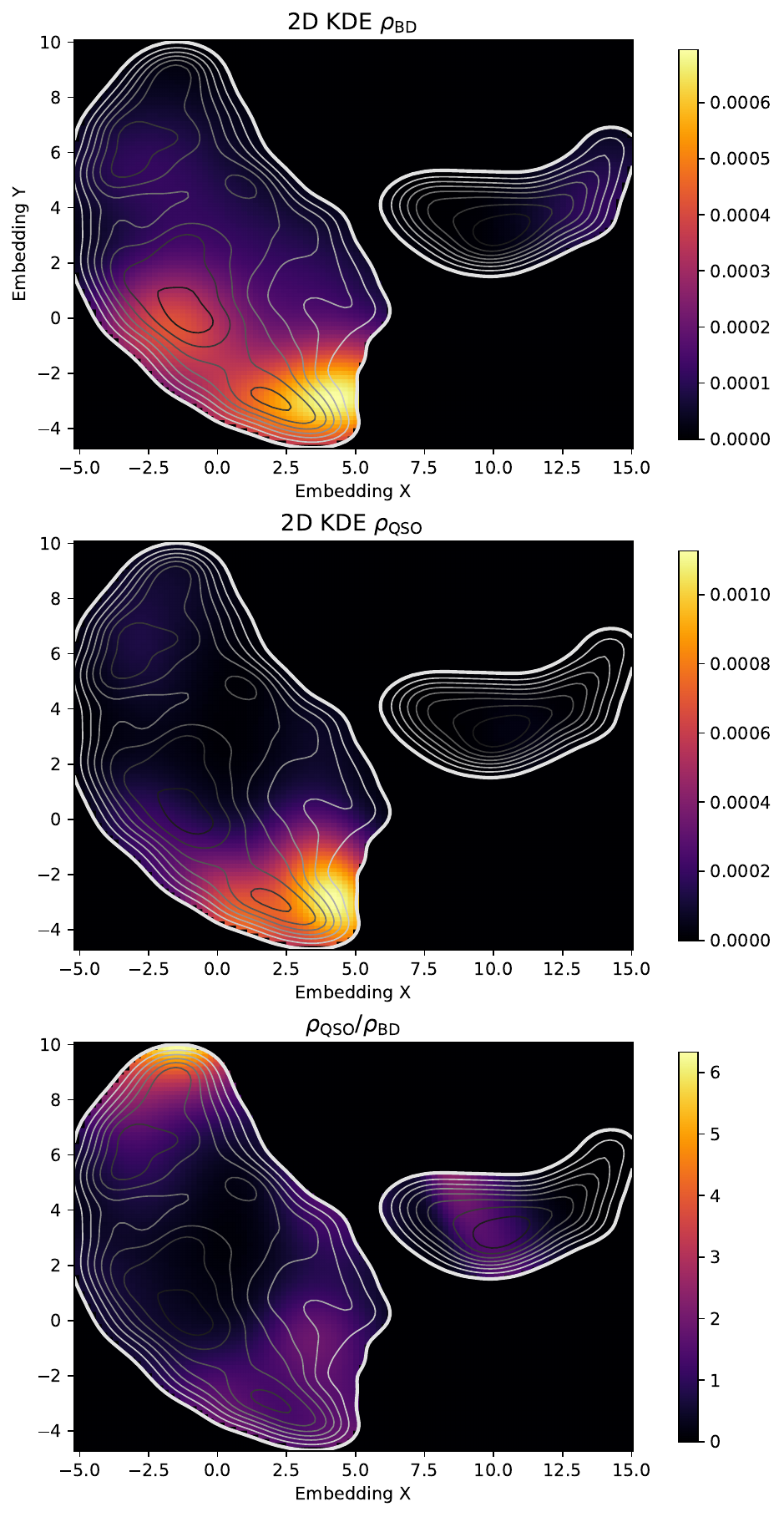}
   \caption{2D KDE density maps for MLT dwarfs (\textit{upper}), $z > 5.7$ quasar population (\textit{central}); and quasar-to-brown-dwarf density ratio map (\textit{lower}). White contours highlight the source density in the embedded space.}
    \label{KDEdensities}
\end{figure}

\section{Ultracool dwarf, star and galaxy templates included in \textit{eazy-py}}
\label{contaminant_templates}

The set of UCD templates includes: 1) cloud-free, solar-composition, and substellar atmosphere Sonora templates; 2) a compilation of low-resolution spectra of M3-M9, L0-L9, T0-T8 dwarfs from the IRTF SpeX spectrograph; 3) low temperature, low metallicity, and cloud-free atmosphere templates from the Lowz Library; 4) AMES-Dusty and 5) AMES-Cond atmosphere models with and without dust opacity, respectively; and 5) NextGen model atmosphere grid, which covers high-temperature low-mass stars and brown dwarfs. We also consider possible contamination by main-sequence stars by incorporating templates from the TRDS Pickles Atlas.  In Table \ref{stellar_templates} we provide technical details of the templates used for the SED fitting downselection step including: number of templates, type and wavelength coverage in $\mu$m for all the templates; and surface gravity ($\log g$), effective temperature (T$_{\mathrm{eff}}$) and metallicity ([M/H]) only for the theoretical ones. Also, we report on assumptions and the cleaning process we performed to reach the final number of templates.

\begin{table*}
      \caption[]{Templates included in \textit{eazy-py} to account for stellar contaminants.}
      \begin{adjustbox}{max width=0.99\textwidth}
         \label{stellar_templates}
         \centering
         \begin{tabular}{@{}ccccccccc@{}}
         \toprule
         \multicolumn{1}{c}{\textbf{Name}}  &  
         \multicolumn{1}{c}{\textbf{Reference}}  &  
         \multicolumn{1}{c}{\textbf{Type}}     & \multicolumn{1}{c}{\textbf{Number}}   &
         \multicolumn{1}{c}{$\mathbf{\log g}$ [cm s$^2$]} &
         \multicolumn{1}{c}{$\mathrm{\mathbf{T}}_{\mathrm{eff}}$ [K]} &
         \multicolumn{1}{c}{[M/H]} &\multicolumn{1}{c}{\textbf{Coverage} [$\mu$m]}
         \\
         \multicolumn{1}{c}{(1)} &
         \multicolumn{1}{c}{(2)} &
         \multicolumn{1}{c}{(3)} &
         \multicolumn{1}{c}{(4)} &
         \multicolumn{1}{c}{(5)} &
         \multicolumn{1}{c}{(6)} &
         \multicolumn{1}{c}{(7)} &
         \multicolumn{1}{c}{(8)} 
         \\ \midrule
        \multicolumn{1}{c}{Sonora 2018}  &
        \multicolumn{1}{c}{\cite{marley2018sonora}}  &
         \multicolumn{1}{c}{Theoretical} &
         \multicolumn{1}{c}{420} &
         \multicolumn{1}{c}{[3.25, 5.5]} &
         \multicolumn{1}{c}{[200, 2400]} &
         \multicolumn{1}{c}{0} &
         \multicolumn{1}{c}{[0.4, 40]}
         \\ \midrule
         \multicolumn{1}{c}{Lowz Library}  &
        \multicolumn{1}{c}{\cite{meisner2021new}}  &
         \multicolumn{1}{c}{Theoretical} &
         \multicolumn{1}{c}{374$^{\mathrm{a}}$} &
         \multicolumn{1}{c}{3.5, 5} &
         \multicolumn{1}{c}{[500, 1600]} &
         \multicolumn{1}{c}{[-2.5, 1]} &
         \multicolumn{1}{c}{[0.1, 10]}
         \\ \midrule
         \multicolumn{1}{c}{\multirow{4}{*}{AMES-Dusty}}  &
        \multicolumn{1}{c}{\cite{allard2001limiting}}  &
         \multicolumn{1}{c}{\multirow{4}{*}{Theoretical}} &
         \multicolumn{1}{c}{\multirow{4}{*}{16}} &
         \multicolumn{1}{c}{\multirow{4}{*}{4.5}} &
         \multicolumn{1}{c}{\multirow{4}{*}{[500, 2000]}} &
         \multicolumn{1}{c}{\multirow{4}{*}{0}} &
         \multicolumn{1}{c}{\multirow{4}{*}{[0.1, 10]}}
         \\
         \multicolumn{1}{c}{} &
         \multicolumn{1}{c}{\cite{baraffe2003evolutionary}} &
         \multicolumn{1}{c}{} &
         \multicolumn{1}{c}{} &
         \multicolumn{1}{c}{} &
         \multicolumn{1}{c}{} &
         \multicolumn{1}{c}{} &
         \multicolumn{1}{c}{} 
         \\
         \multicolumn{1}{c}{} &
         \multicolumn{1}{c}{\cite{grevesse1993revision}} &
         \multicolumn{1}{c}{} &
         \multicolumn{1}{c}{} &
         \multicolumn{1}{c}{} &
         \multicolumn{1}{c}{} &
         \multicolumn{1}{c}{} &
         \multicolumn{1}{c}{} 
         \\
         \multicolumn{1}{c}{} &
         \multicolumn{1}{c}{\cite{partridge1997determination}} &
         \multicolumn{1}{c}{} &
         \multicolumn{1}{c}{} &
         \multicolumn{1}{c}{} &
         \multicolumn{1}{c}{} &
         \multicolumn{1}{c}{} &
         \multicolumn{1}{c}{} 
         \\ \midrule
         \multicolumn{1}{c}{\multirow{4}{*}{AMES-Cond}}  &
        \multicolumn{1}{c}{\cite{allard2001limiting}}  &
         \multicolumn{1}{c}{\multirow{4}{*}{Theoretical}} &
         \multicolumn{1}{c}{\multirow{4}{*}{11}} &
         \multicolumn{1}{c}{\multirow{4}{*}{4.5 }} &
         \multicolumn{1}{c}{\multirow{4}{*}{[800,1900]}} &
         \multicolumn{1}{c}{\multirow{4}{*}{0}} &
         \multicolumn{1}{c}{\multirow{4}{*}{[0.1, 10]}}
         \\
         \multicolumn{1}{c}{} &
         \multicolumn{1}{c}{\cite{chabrier2000evolutionary}} &
         \multicolumn{1}{c}{} &
         \multicolumn{1}{c}{} &
         \multicolumn{1}{c}{} &
         \multicolumn{1}{c}{} &
         \multicolumn{1}{c}{} &
         \multicolumn{1}{c}{} 
         \\
         \multicolumn{1}{c}{} &
         \multicolumn{1}{c}{\cite{grevesse1993revision}} &
         \multicolumn{1}{c}{} &
         \multicolumn{1}{c}{} &
         \multicolumn{1}{c}{} &
         \multicolumn{1}{c}{} &
         \multicolumn{1}{c}{} &
         \multicolumn{1}{c}{} 
         \\
         \multicolumn{1}{c}{} &
         \multicolumn{1}{c}{\cite{partridge1997determination}} &
         \multicolumn{1}{c}{} &
         \multicolumn{1}{c}{} &
         \multicolumn{1}{c}{} &
         \multicolumn{1}{c}{} &
         \multicolumn{1}{c}{} &
         \multicolumn{1}{c}{} 
         \\ \midrule
         \multicolumn{1}{c}{\multirow{5}{*}{NextGen}}  &
        \multicolumn{1}{c}{\cite{allard1997model}}  &
         \multicolumn{1}{c}{\multirow{5}{*}{Theoretical}} &
         \multicolumn{1}{c}{\multirow{5}{*}{8}} &
         \multicolumn{1}{c}{\multirow{5}{*}{4.5}} &
         \multicolumn{1}{c}{\multirow{5}{*}{[2200, 9800]}} &
         \multicolumn{1}{c}{\multirow{5}{*}{0}} &
         \multicolumn{1}{c}{\multirow{5}{*}{[0.01, 921]}}
         \\
         \multicolumn{1}{c}{} &
         \multicolumn{1}{c}{\cite{baraffe1997evolutionary, baraffe1998evolutionary}} &
         \multicolumn{1}{c}{} &
         \multicolumn{1}{c}{} &
         \multicolumn{1}{c}{} &
         \multicolumn{1}{c}{} &
         \multicolumn{1}{c}{} &
         \multicolumn{1}{c}{} 
         \\
         \multicolumn{1}{c}{} &
         \multicolumn{1}{c}{\cite{grevesse1993revision}} &
         \multicolumn{1}{c}{} &
         \multicolumn{1}{c}{} &
         \multicolumn{1}{c}{} &
         \multicolumn{1}{c}{} &
         \multicolumn{1}{c}{} &
         \multicolumn{1}{c}{} 
         \\
         \multicolumn{1}{c}{} &
         \multicolumn{1}{c}{\cite{hauschildt1999nextgen}} &
         \multicolumn{1}{c}{} &
         \multicolumn{1}{c}{} &
         \multicolumn{1}{c}{} &
         \multicolumn{1}{c}{} &
         \multicolumn{1}{c}{} &
         \multicolumn{1}{c}{} 
         \\
         \multicolumn{1}{c}{} &
         \multicolumn{1}{c}{\cite{schryber1995computed}} &
         \multicolumn{1}{c}{} &
         \multicolumn{1}{c}{} &
         \multicolumn{1}{c}{} &
         \multicolumn{1}{c}{} &
         \multicolumn{1}{c}{} &
         \multicolumn{1}{c}{} 
         \\ \midrule
         \multicolumn{1}{c}{\multirow{2}{*}{SpeX Prism Library$^{\mathrm{b}}$}}  &
        \multicolumn{1}{c}{\cite{burgasser2006unified}}  &
         \multicolumn{1}{c}{\multirow{2}{*}{Empirical}} &
         \multicolumn{1}{c}{\multirow{2}{*}{149}} &
         \multicolumn{1}{c}{} &
         \multicolumn{1}{c}{} &
         \multicolumn{1}{c}{} &
         \multicolumn{1}{c}{\multirow{2}{*}{[0.6, 2.5]}}
         \\
         \multicolumn{1}{c}{} &
         \multicolumn{1}{c}{\cite{burgasser2014spex}} &
         \multicolumn{1}{c}{} &
         \multicolumn{1}{c}{} &
         \multicolumn{1}{c}{} &
         \multicolumn{1}{c}{} &
         \multicolumn{1}{c}{} &
         \multicolumn{1}{c}{} 
         \\ \midrule
         \multicolumn{1}{c}{\multirow{3}{*}{L and T dwarf data archive $^{\mathrm{c}}$}}  &
        \multicolumn{1}{c}{\cite{knapp2004}}  &
         \multicolumn{1}{c}{\multirow{3}{*}{Empirical}} &
         \multicolumn{1}{c}{\multirow{3}{*}{148}} &
         \multicolumn{1}{c}{ } &
         \multicolumn{1}{c}{} &
         \multicolumn{1}{c}{} &
         \multicolumn{1}{c}{\multirow{3}{*}{[0.1, 2.5]}}
         \\
         \multicolumn{1}{c}{} &
         \multicolumn{1}{c}{\cite{chiu2006seventy}} &
         \multicolumn{1}{c}{} &
         \multicolumn{1}{c}{} &
         \multicolumn{1}{c}{} &
         \multicolumn{1}{c}{} &
         \multicolumn{1}{c}{} &
         \multicolumn{1}{c}{} 
         \\
         \multicolumn{1}{c}{} &
         \multicolumn{1}{c}{\cite{golimowski2004and}} &
         \multicolumn{1}{c}{} &
         \multicolumn{1}{c}{} &
         \multicolumn{1}{c}{} &
         \multicolumn{1}{c}{} &
         \multicolumn{1}{c}{} &
         \multicolumn{1}{c}{} 
         \\ \midrule
         \multicolumn{1}{c}{TRDS Pickles Atlas$^{\mathrm{d}}$}  &
        \multicolumn{1}{c}{\cite{Pickles1998}}  &
         \multicolumn{1}{c}{Empirical} &
         \multicolumn{1}{c}{131} &
         \multicolumn{1}{c}{ } &
         \multicolumn{1}{c}{[2951.21, 39810.7]} &
         \multicolumn{1}{c}{} &
         \multicolumn{1}{c}{[0.115, 2.5]}
           \\ \bottomrule
        \end{tabular}
        \end{adjustbox}
        \begin{tablenotes}
        \small
        \item $^{\mathrm{a}}$ This subsample is defined by a carbon-to-oxygen ratio C/O $= 0.55$ and vertical eddy diffusion coefficient $\log_{10} \mathrm{K}_{\mathrm{zz}} = 2.0$.
        \item $^{\mathrm{b}}$ These templates were extended to 0.1 $\mu$m and 100 $\mu$m by assuming Wien and Rayleigh-Jeans emission tails, respectively.
        \item $^{\mathrm{c}}$ Find compilation at \url{http://svo2.cab.inta-csic.es/theory/newov2/index.php?}.
        \item $^{\mathrm{d}}$ TRDS Pickles Atlas account for all normal spectral types and luminosity classes at solar abundance, as well as metal-weak and metal-rich F-K dwarf and G-K giant stars. 
        \newline
        \textbf{Notes}: Col (1): Template set name; Col (2): references; Col (3): type templates: either from a theoretical grid or empirical observed spectra; Col (4): number of templates included; Col (5): logarithm of the surface gravity; Col (6): Effective temperature; Col (7): logarithm of the ratio of the mass fraction of metals to the mass fraction of hydrogen, compared to the solar value; Col (8): lower and upper limits of the wavelength coverage.
        \end{tablenotes}
\end{table*}

To account for red galaxies (early type spectra and high-z) as potential contaminants, we include: five templates from the Flexible Stellar Population Synthesis (FSPS) library, modeling the emission of a range of galaxy types such as star-forming, quiescent and dusty, with diverse star formation histories (SFHs; \citealt{conroy2009propagation, conroy2010propagation}); three galaxy templates with redshift-dependent SFHs and low dust attenuation from the CORR\_SFHZ\_13 library, based on JWST data of high-redshift galaxies \citep{larson2023spectral, carnall2023first}; and, to consider extreme emission line galaxies, the rest-frame version of the best-fit template to the NIRSpec spectrum of a $z=8.5$ galaxy (ID= 4590 in \citealt{carnall2023first}), which features high-equivalent-width [OIII] and H$\beta$ emission lines producing a distintive U-shaped SED in the F277W, F356W, and F444W bands.

\section{Comparison with adapted color cuts}
\label{adapted color cuts}

We compute synthetic photometry using quasar SED templates by \cite{temple2021modelling} convolved with the PS1, DECam, UKIDSS, VHS, and WISE filter transmission curves. We therefore tested the aforementioned color selections by substituting the PS1 or DES DR2 $y$-band with VHS or UKIDSS Y where available, and by adopting $i$ and $z$ photometry from Legacy Survey DR10. The original color cuts are adjusted to our photometric system to account for the modified color evolution induced by the different filter responses. We emphasize that these transformations introduce additional scatter, and the resulting purity and completeness estimates should therefore be interpreted as approximate rather than exact. 

For \cite{banados2016pan}, targeting quasars at $5.7 < z < 6.5$, the adapted criteria for our photometry are:

\begin{align*}
    \hspace{2.2cm} i-z > 0.8 \text{ and } z-y < 0.12
\end{align*}
for $5.7 < z < 6.2$, and 
\begin{align*}
    \hspace{2.2cm} i-z > 0.8 \text{ and } z-y > 0.12
\end{align*}

\noindent  for $6.2 < z < 6.5$, while the remaining signal-to-noise ratio and color constraints remain the same. This corresponds to the same selection as \cite{wolf2024srg}, with the additional constraint $y-J < 0.8$. The adaptation of the \cite{wang2019exploring} for $z \sim 7$ selection yields:

\begin{align*}
    \hspace{2.2cm} z-y > 0.5 \text{ and } y-J < 0.4,
\end{align*}

\noindent  with the original $J-W1 > 1.5$ cut. Finally, for \cite{yang2023desi}, the changed color cuts are:

\begin{align*}
    \hspace{1.5cm} i-z > 0.8, \hspace{1mm} z-y < 0.3 \text{, and } y-W1 > -0.1
\end{align*}

\noindent for $5.7 < z < 6.4$, and

\begin{align*}
    \hspace{1.5cm}  z-y > 0.3, \hspace{1mm} y-W1 > -0.1 \text{, and } z-W1 > 0.7
\end{align*}

\noindent at higher redshift, while rejecting sources with $y-J > 0.4$. The performance of these adapted selections is summarized in the following Table \ref{performance_colorcuts}.

\begin{table}[ht]
      \caption[]{Performance of color-based selections.}
      \begin{adjustbox}{max width=0.495\textwidth}
         \label{performance_colorcuts}
         \centering
         \begin{tabular}{@{}ccccccc@{}}
         \toprule
         \multicolumn{1}{c}{\textbf{Selection}}  &  
         \multicolumn{1}{c}{\textbf{TP}}   &  
         \multicolumn{1}{c}{\textbf{FN}}   & 
         \multicolumn{1}{c}{\textbf{FP}}   &
         \multicolumn{1}{c}{\textbf{P}}   &  
         \multicolumn{1}{c}{\textbf{C}}   &  
         \multicolumn{1}{c}{\textbf{Cand.}}  
         \\
         \multicolumn{1}{c}{(1)}  &  
         \multicolumn{1}{c}{(2)}   &  
         \multicolumn{1}{c}{(3)}   & 
         \multicolumn{1}{c}{(4)}  &
         \multicolumn{1}{c}{(5)}   &  
         \multicolumn{1}{c}{(6)}   &  
         \multicolumn{1}{c}{(7)} 
         \\ \midrule
           \multicolumn{1}{c}{\cite{banados2016pan}}  &
         \multicolumn{1}{c}{28}  &
         \multicolumn{1}{c}{3}  & 
         \multicolumn{1}{c}{39}  &
         \multicolumn{1}{c}{$42\%$} &
         \multicolumn{1}{c}{$90\%$} &
         \multicolumn{1}{c}{28526} 
         \\ \midrule
           \multicolumn{1}{c}{\cite{wang2019exploring}}  &
         \multicolumn{1}{c}{3 } &
         \multicolumn{1}{c}{28 } &
         \multicolumn{1}{c}{3 } &
         \multicolumn{1}{c}{$40\%$ } &
         \multicolumn{1}{c}{$11\%$} &
         \multicolumn{1}{c}{1892 }
         \\ \midrule
           \multicolumn{1}{c}{\cite{yang2023desi}}  &
         \multicolumn{1}{c}{51}  &
         \multicolumn{1}{c}{18}  &
         \multicolumn{1}{c}{7}  &
         \multicolumn{1}{c}{$88\%$ } &
         \multicolumn{1}{c}{$74\%$} &
         \multicolumn{1}{c}{39924} 
         \\ \midrule
           \multicolumn{1}{c}{\cite{wolf2024srg}}  &
         \multicolumn{1}{c}{18}  &
         \multicolumn{1}{c}{13}  &
         \multicolumn{1}{c}{53}  &
         \multicolumn{1}{c}{$25\%$}  &
         \multicolumn{1}{c}{$58\%$} &
         \multicolumn{1}{c}{2344} 
         \\ \midrule
           \multicolumn{1}{c}{CL+SED fitting}  &
         \multicolumn{1}{c}{52}  &
         \multicolumn{1}{c}{17}  &
         \multicolumn{1}{c}{10}  &
         \multicolumn{1}{c}{$84\%$}  &
         \multicolumn{1}{c}{$75\%$} &
         \multicolumn{1}{c}{1438 }
           \\ \bottomrule
        \end{tabular}
        \end{adjustbox}
        \begin{tablenotes}
        \small
        \item \textbf{Notes:} Col (1): Candidate selection; Col (2): True positives; Col (3): False negatives; Col (4): False positives; Col (5): Precision; Col (6): Completeness; Col (7): Number of candidates selected.
        \end{tablenotes}
   \end{table}

\vspace{3mm}

\section{List of confirmed UCDs}

In Table \ref{ultracool_dwarfs} we compile the sample of quasar candidates observed during our spectroscopic campaign that did not exhibit a Ly-$\alpha$ break and present a likely brown dwarf spectrum. A detailed analysis of the spectrum of these sources is beyond the scope of this paper.

  \begin{table*}[ht]
      \caption[]{Spectroscopically confirmed to not be high-z quasars.}
         \label{ultracool_dwarfs}
         \centering
         \begin{tabular}{@{}ccccccc@{}}
         \toprule
         \multicolumn{1}{c}{\textbf{Name}}  &  
         \multicolumn{1}{c}{\textbf{RA}}     & \multicolumn{1}{c}{\textbf{DEC}}   &
         \multicolumn{1}{c}{\textbf{Date}} &
         \multicolumn{1}{c}{\textbf{Telescope}} &
         \multicolumn{1}{c}{\textbf{Type}} &
         \multicolumn{1}{c}{$z_{\mathrm{ml}}$} 
        \\
        \multicolumn{1}{c}{(1)}  &  
         \multicolumn{1}{c}{(2)}    & 
         \multicolumn{1}{c}{(3)}   &
         \multicolumn{1}{c}{(4)} &
         \multicolumn{1}{c}{(5)} &
         \multicolumn{1}{c}{(6)} &
         \multicolumn{1}{c}{(7)} 
         \\  \midrule 
        \multicolumn{1}{c}{LS J012806.35-070949.28}  &
         \multicolumn{1}{c}{22.02645} & 
         \multicolumn{1}{c}{-7.16369} &
         \multicolumn{1}{c}{2024 Sep. 8} &
         \multicolumn{1}{c}{Hale/DBSP} &
         \multicolumn{1}{c}{EXP} &
         \multicolumn{1}{c}{6.89}
           \\ 
        \multicolumn{1}{c}{LS J013153.69-091341.41}  &
         \multicolumn{1}{c}{22.97369} &
         \multicolumn{1}{c}{-9.22817} &
         \multicolumn{1}{c}{2024 Oct. 4} &
         \multicolumn{1}{c}{Hale/DBSP} &
         \multicolumn{1}{c}{PSF} &
         \multicolumn{1}{c}{6.60}
           \\ 
        \multicolumn{1}{c}{LS J023459.52-184732.42}  &
         \multicolumn{1}{c}{38.74801} &
         \multicolumn{1}{c}{-18.79234} &
         \multicolumn{1}{c}{2024 Dec. 2} &
         \multicolumn{1}{c}{LBT/MODS} &
         \multicolumn{1}{c}{PSF} &
         \multicolumn{1}{c}{7.03}
         \\ 
        \multicolumn{1}{c}{LS J040409.74-010111.57}  &
         \multicolumn{1}{c}{61.04060} &
         \multicolumn{1}{c}{-1.01988} &
         \multicolumn{1}{c}{2024 Dec. 2} &
         \multicolumn{1}{c}{Hale/DBSP} &
         \multicolumn{1}{c}{PSF} &
         \multicolumn{1}{c}{7.07}
         \\ 
        \multicolumn{1}{c}{LS J040940.43-044153.29}  &
         \multicolumn{1}{c}{62.41846} &
         \multicolumn{1}{c}{-4.69797} &
         \multicolumn{1}{c}{2024 Oct. 4} &
         \multicolumn{1}{c}{Hale/DBSP} &
         \multicolumn{1}{c}{PSF} &
         \multicolumn{1}{c}{6.64}
           \\ 
         \multicolumn{1}{c}{LS J082810.33-024514.58}  &
         \multicolumn{1}{c}{127.04305} &
         \multicolumn{1}{c}{-2.75405} &
         \multicolumn{1}{c}{2024 Dec. 2} &
         \multicolumn{1}{c}{LBT/MODS} &
         \multicolumn{1}{c}{REX} &
         \multicolumn{1}{c}{6.29}
         \\ 
         \multicolumn{1}{c}{LS J091505.37-162230.52}  &
         \multicolumn{1}{c}{138.77237} &
         \multicolumn{1}{c}{-16.37507} &
         \multicolumn{1}{c}{2025 May. 1} &
         \multicolumn{1}{c}{Clay/LDSS3} &
         \multicolumn{1}{c}{PSF} &
         \multicolumn{1}{c}{7.08}
         \\ 
         \multicolumn{1}{c}{LS J093306.52-004703.13}  &
         \multicolumn{1}{c}{143.27715} &
         \multicolumn{1}{c}{-0.78420} &
         \multicolumn{1}{c}{2024 Dec. 5} &
         \multicolumn{1}{c}{LBT/MODS} &
         \multicolumn{1}{c}{PSF} &
         \multicolumn{1}{c}{5.96}
         \\ 
         \multicolumn{1}{c}{LS J101348.72-244710.12}  &
         \multicolumn{1}{c}{153.45299} &
         \multicolumn{1}{c}{-24.78617} &
         \multicolumn{1}{c}{2025 May. 1} &
         \multicolumn{1}{c}{Clay/LDSS3} &
         \multicolumn{1}{c}{PSF} &
         \multicolumn{1}{c}{7.05}
         \\ 
         \multicolumn{1}{c}{LS J113244.38+153820.65}  &
         \multicolumn{1}{c}{173.18492} &
         \multicolumn{1}{c}{15.63907} &
         \multicolumn{1}{c}{2025 April. 24} &
         \multicolumn{1}{c}{LBT/MODS} &
         \multicolumn{1}{c}{PSF} &
         \multicolumn{1}{c}{6.34}
         \\
         \multicolumn{1}{c}{LS J115515.06+024328.27}  &
         \multicolumn{1}{c}{178.81276} &
         \multicolumn{1}{c}{2.72452} &
         \multicolumn{1}{c}{2025 March. 5} &
         \multicolumn{1}{c}{Keck/LRIS} &
         \multicolumn{1}{c}{PSF} &
         \multicolumn{1}{c}{6.98}
         \\ 
         \multicolumn{1}{c}{LS J130209.38+064549.53}  &
         \multicolumn{1}{c}{195.53908} &
         \multicolumn{1}{c}{6.76376} &
         \multicolumn{1}{c}{2025 April. 24} &
         \multicolumn{1}{c}{LBT/MODS} &
         \multicolumn{1}{c}{PSF} &
         \multicolumn{1}{c}{7.02}
         \\
         \multicolumn{1}{c}{LS J132838.18-340410.74}  &
         \multicolumn{1}{c}{202.15908} &
         \multicolumn{1}{c}{-34.06965} &
         \multicolumn{1}{c}{2025 May. 1} &
         \multicolumn{1}{c}{Clay/LDSS3} &
         \multicolumn{1}{c}{REX} &
         \multicolumn{1}{c}{6.73}
         \\ 
         \multicolumn{1}{c}{LS J133245.75-391414.68}  &
         \multicolumn{1}{c}{203.19063} &
         \multicolumn{1}{c}{-39.23741} &
         \multicolumn{1}{c}{2025 May. 1} &
         \multicolumn{1}{c}{Clay/LDSS3} &
         \multicolumn{1}{c}{PSF} &
         \multicolumn{1}{c}{7.13}
         \\ 
         \multicolumn{1}{c}{LS J140247.81+102132.87}  &
         \multicolumn{1}{c}{210.69923} & 
         \multicolumn{1}{c}{10.35913} &
         \multicolumn{1}{c}{2025 April. 20} &
         \multicolumn{1}{c}{LBT/MODS} &
         \multicolumn{1}{c}{PSF} &
         \multicolumn{1}{c}{6.56}
         \\
         \multicolumn{1}{c}{LS J150722.36-232544.21}  &
         \multicolumn{1}{c}{226.84315} &
         \multicolumn{1}{c}{-23.42895} &
         \multicolumn{1}{c}{2025 May. 1} &
         \multicolumn{1}{c}{Clay/LDSS3} &
         \multicolumn{1}{c}{PSF} &
         \multicolumn{1}{c}{6.01}
         \\ 
         \multicolumn{1}{c}{LS J152443.10-262001.87}  &
         \multicolumn{1}{c}{231.17957} &
         \multicolumn{1}{c}{-26.33393} &
         \multicolumn{1}{c}{2025 May. 1} &
         \multicolumn{1}{c}{Clay/LDSS3} &
         \multicolumn{1}{c}{PSF} &
         \multicolumn{1}{c}{6.85}
         \\ 
         \multicolumn{1}{c}{LS J160255.50-082640.24}  &
         \multicolumn{1}{c}{240.73126} &
         \multicolumn{1}{c}{-8.44451} &
         \multicolumn{1}{c}{2025 May. 1} &
         \multicolumn{1}{c}{Clay/LDSS3} &
         \multicolumn{1}{c}{PSF} &
         \multicolumn{1}{c}{5.94}
         \\ 
         \multicolumn{1}{c}{LS J190808.07-691041.42}  &
         \multicolumn{1}{c}{287.03364} &
         \multicolumn{1}{c}{-69.17804} &
         \multicolumn{1}{c}{2025 May. 1} &
         \multicolumn{1}{c}{Clay/LDSS3} &
         \multicolumn{1}{c}{PSF} &
         \multicolumn{1}{c}{5.94}
         \\ 
         \multicolumn{1}{c}{LS J193812.38-570418.00}  &
         \multicolumn{1}{c}{294.55158} &
         \multicolumn{1}{c}{-57.07166} &
         \multicolumn{1}{c}{2025 May. 1} &
         \multicolumn{1}{c}{Clay/LDSS3} &
         \multicolumn{1}{c}{PSF} &
         \multicolumn{1}{c}{6.97}
         \\ 
         \multicolumn{1}{c}{LS J200926.15-600141.31}  &
         \multicolumn{1}{c}{302.35896} &
         \multicolumn{1}{c}{-60.02814} &
         \multicolumn{1}{c}{2025 May. 1} &
         \multicolumn{1}{c}{Clay/LDSS3} &
         \multicolumn{1}{c}{PSF} &
         \multicolumn{1}{c}{6.00}
         \\ 
         \multicolumn{1}{c}{LS J230435.88-800915.08}  &
         \multicolumn{1}{c}{346.14950} &
         \multicolumn{1}{c}{-80.15419} &
         \multicolumn{1}{c}{2025 May. 1} &
         \multicolumn{1}{c}{Clay/LDSS3} &
         \multicolumn{1}{c}{PSF} &
         \multicolumn{1}{c}{5.93}
           \\ \bottomrule
        \end{tabular}
        \begin{tablenotes}
        \small
        \item \textbf{Notes:} Col (1): Brown dwarf name with convention: "LS" for DESI Legacy Survey DR10; Col (2) and Col (3): coordinates (J2000) in degrees, Col (4): Date of the observations; Col (5): Telescope and instrument used for the optical spectroscopic follow-up; Col (6): Morphological model from LS DR10 catalog: PSF=stellar, REX=round exponential galaxy, or EXP=exponential; Col (7): maximum-likelihood photometric redshift estimated by \textit{eazy-py}.
        \end{tablenotes}
   \end{table*}

\section{SED fitting results and images for all the new quasars}

Here we report the optical and near-IR images and the SED fitting results of the 16 new quasars (see Figs \ref{stamps_0}, \ref{stamps_1}, \ref{stamps_2}, \ref{stamps_3}, \ref{stamps_4} and \ref{stamps_5}).

\begin{figure*}[ht!]
   \centering
   \textbf{LS J2301-1530}\par\medskip
   \vspace{-2mm}
   \includegraphics[width = 0.57\linewidth]{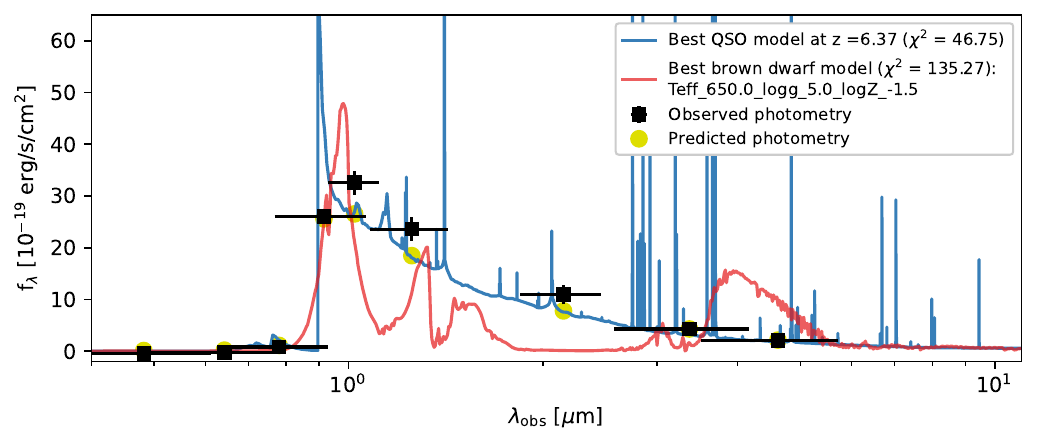}
   \includegraphics[width = 0.203\linewidth, trim = 0 {-3mm} 0 0]{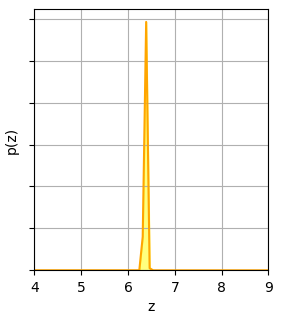} \\
   \includegraphics[width = 0.9\linewidth]{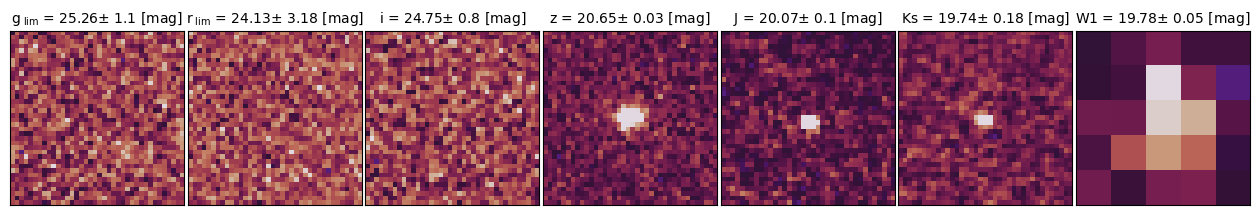} 
   \caption{Continued from Fig. \ref{SEDfitting_PSO1.55}. SED fitting results and postage stamps of the quasar LS J230129.72-153020.4. }
    \label{stamps_0}%
\end{figure*}

\onecolumn

\begin{figure*}
   \centering
   \textbf{LS J0208-6647}\par\medskip
   \vspace{-2mm}
   \includegraphics[width = 0.57\linewidth]{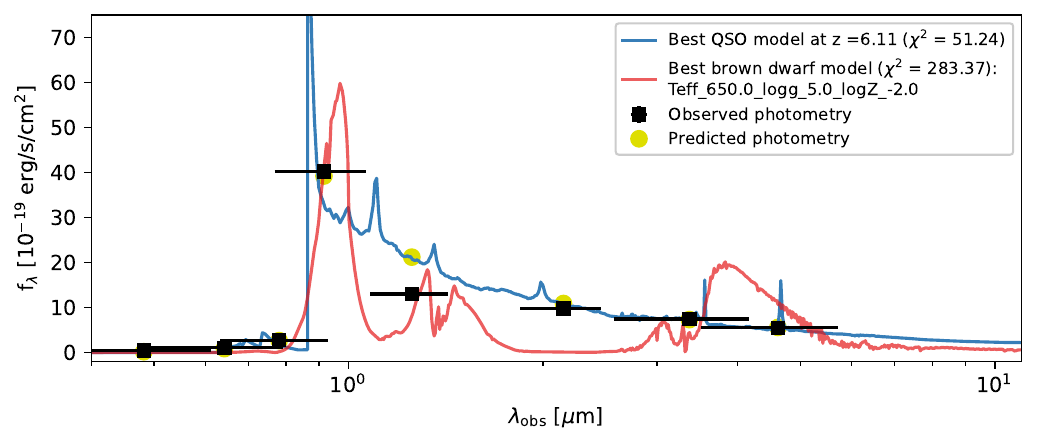}
   \includegraphics[width = 0.203\linewidth, trim = 0 {-3mm} 0 0]{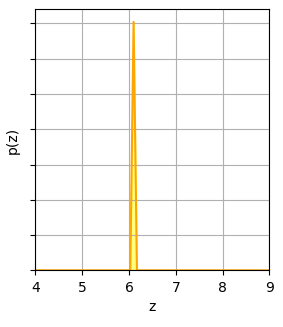} \\
   \includegraphics[width = 0.9\linewidth]{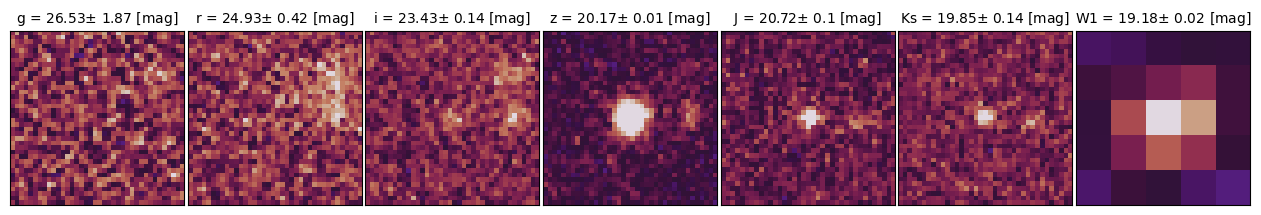} \\
   
   \vspace{2mm}
   \textbf{LS J2223-3815}\par\medskip
   \vspace{-2mm}
   \includegraphics[width = 0.57\linewidth]{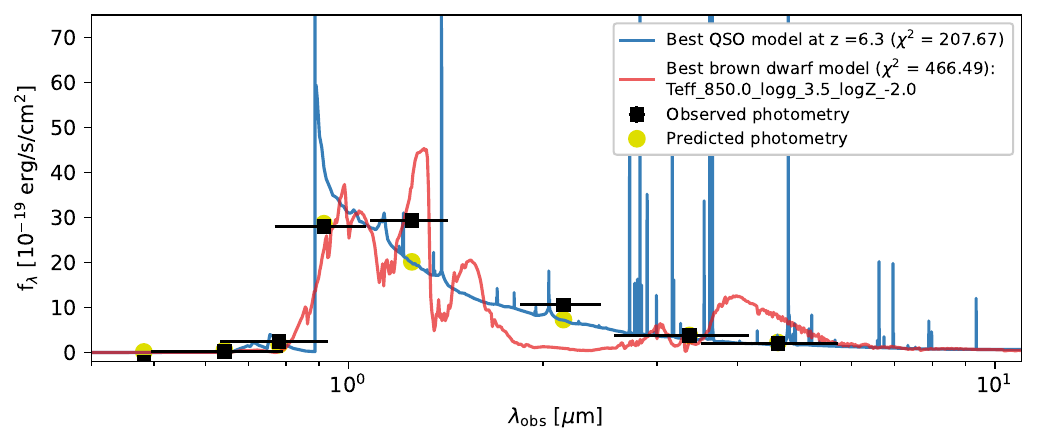}
   \includegraphics[width = 0.203\linewidth, trim = 0 {-3mm} 0 0]{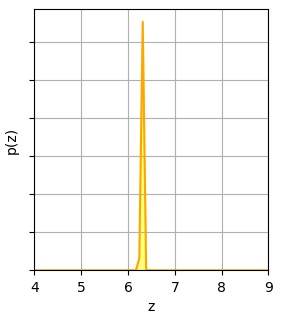} \\
   \includegraphics[width = 0.9\linewidth]{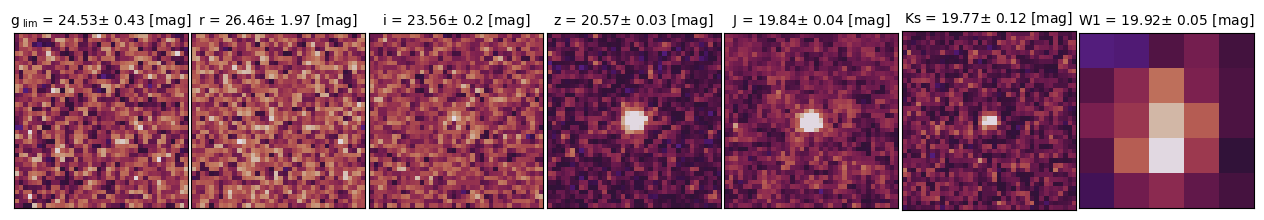} \\

   \vspace{2mm}
   \textbf{LS J0104-6857}\par\medskip
   \vspace{-2mm}
   \includegraphics[width = 0.57\linewidth]{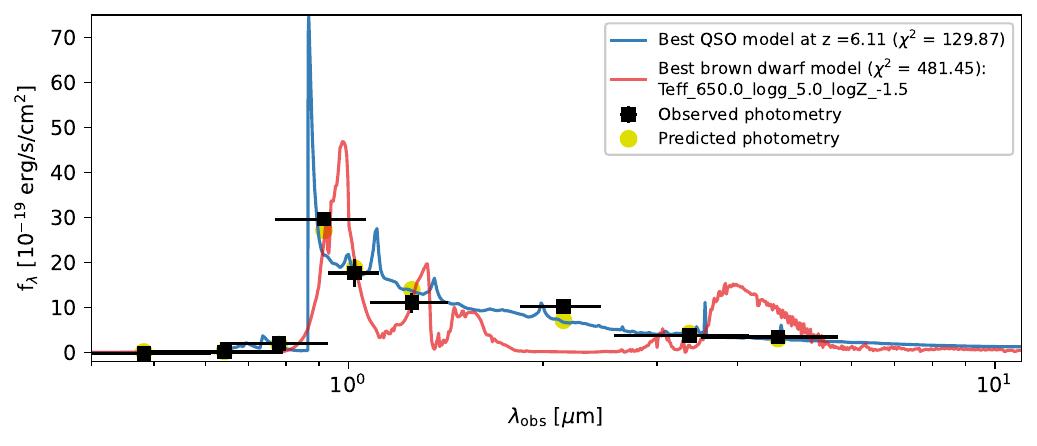}
   \includegraphics[width = 0.203\linewidth, trim = 0 {-3mm} 0 0]{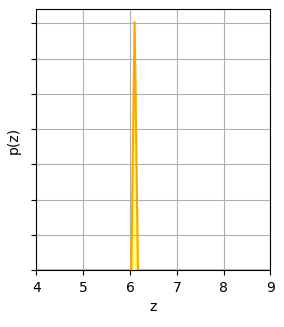} \\
   \includegraphics[width = 0.9\linewidth]{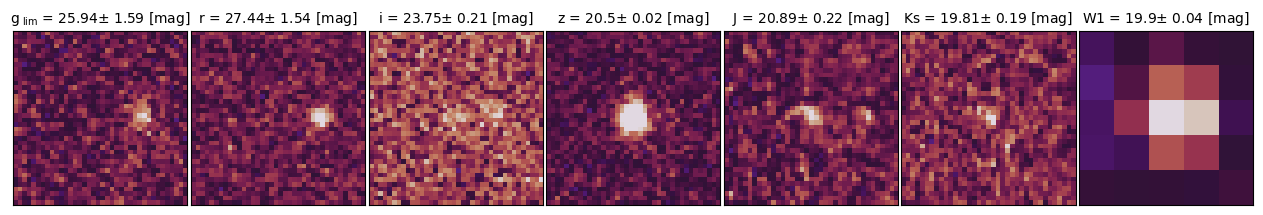} \\
   \caption{Continued from Fig. \ref{SEDfitting_PSO1.55}. SED fitting results and postage stamps of the quasars: LS J020801.31-664713.7 (top panel), LS J222343.78-381526.8 (central panel) and LS J010449.12-685756.8 (bottom panel). }
    \label{stamps_1}%
\end{figure*}

\begin{figure*}
   \centering

   \textbf{LS J1035-0515}\par\medskip
   \vspace{-2mm}
   \includegraphics[width = 0.57\linewidth]{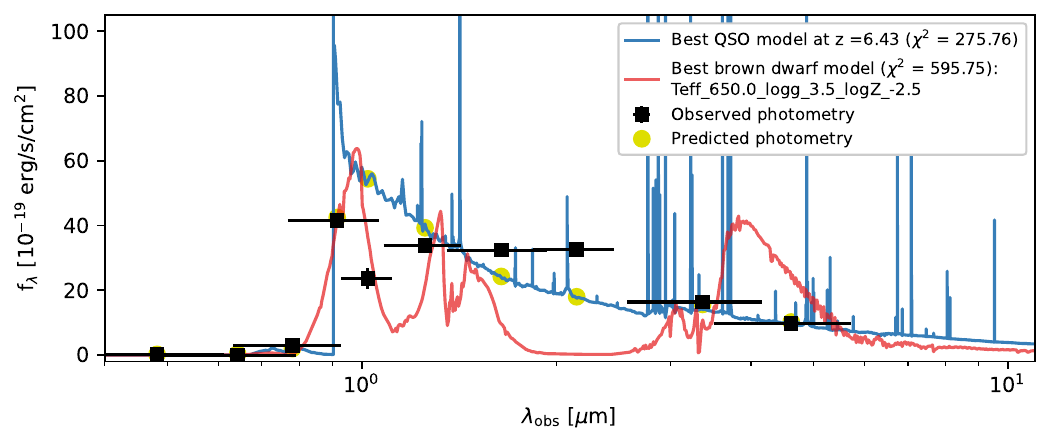}
   \includegraphics[width = 0.203\linewidth, trim = 0 {-3mm} 0 0]{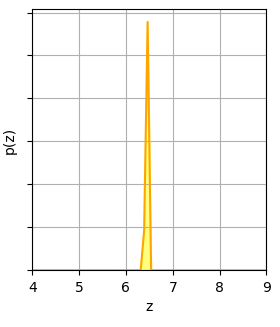} \\
   \includegraphics[width = 0.9\linewidth]{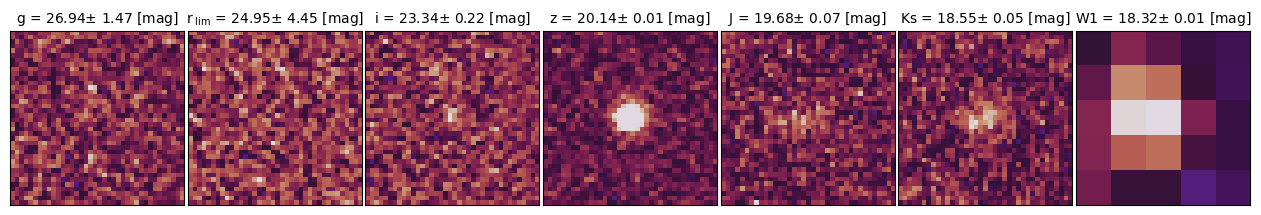} \\
   
   \vspace{2mm}
   \textbf{LS J1130+1420}\par\medskip
   \vspace{-2mm}
   \includegraphics[width = 0.57\linewidth]{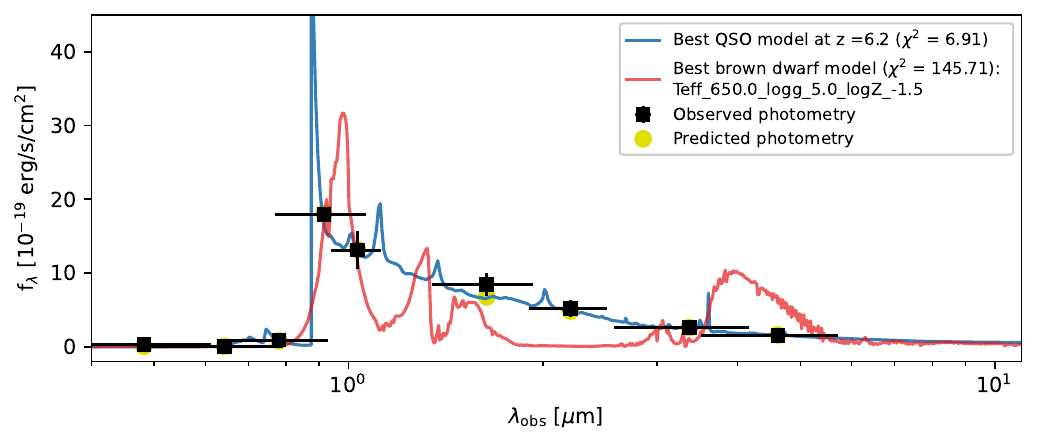}
   \includegraphics[width = 0.203\linewidth, trim = 0 {-3mm} 0 0]{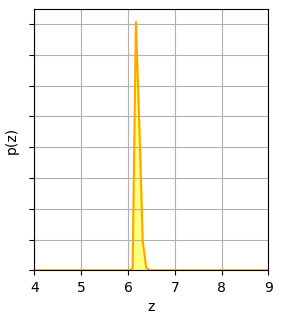} \\
   \includegraphics[width = 0.9\linewidth]{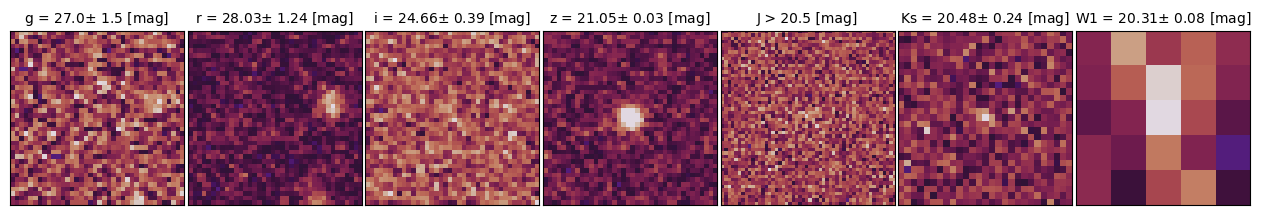} \\

   \vspace{2mm}
   \textbf{LS J1332+1102}\par\medskip
   \vspace{-2mm}
   \includegraphics[width = 0.57\linewidth]{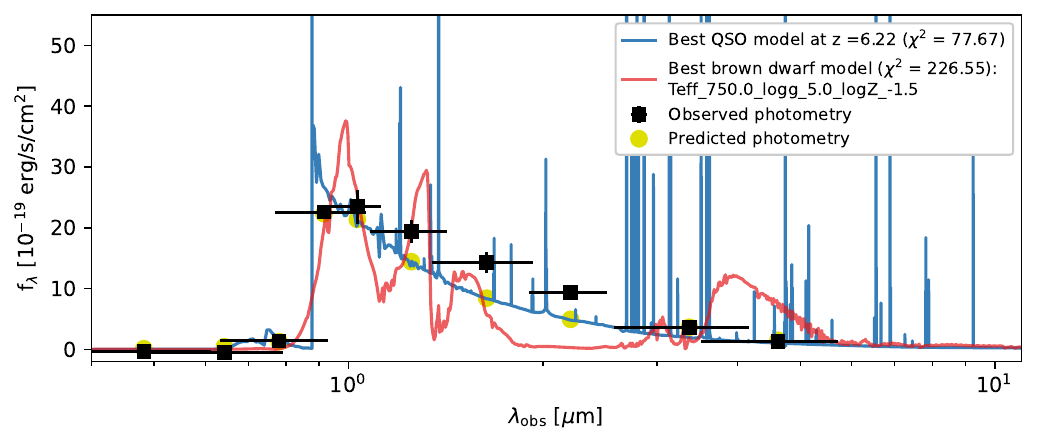}
   \includegraphics[width = 0.203\linewidth, trim = 0 {-3mm} 0 0]{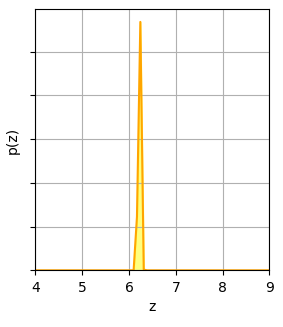} \\
   \includegraphics[width = 0.9\linewidth]{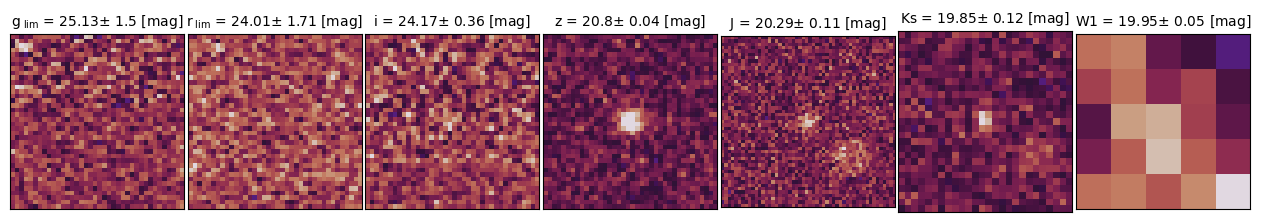}
   
   \caption{Continued from Fig. \ref{stamps_1}. SED fitting results and postage stamps of the quasars: LS J103511.29-051537.9 (top panel), LS J113000.56+142043.97 (central panel) and LS J133204.89+110208.94 (bottom panel).}
    \label{stamps_2}
\end{figure*}

\begin{figure*}
   \centering

   \textbf{LS J1435-1053}\par\medskip
   \vspace{-2mm}
   \includegraphics[width = 0.57\linewidth]{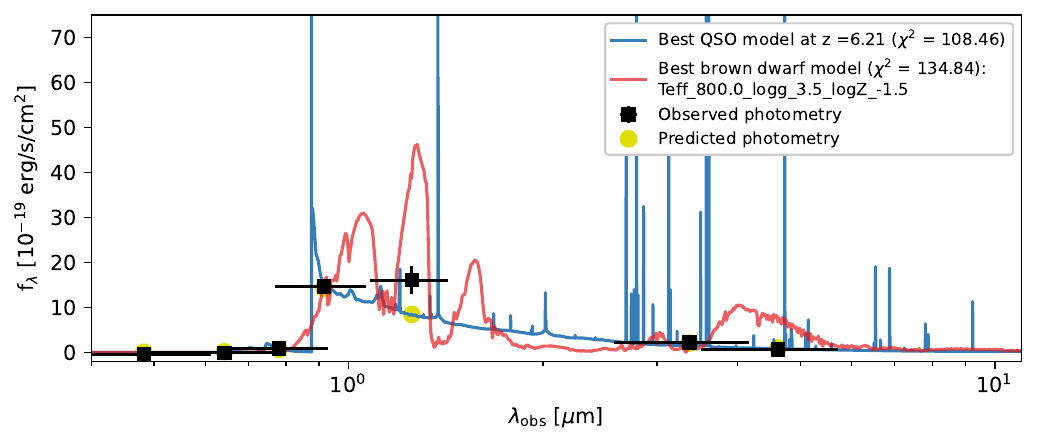}
   \includegraphics[width = 0.203\linewidth, trim = 0 {-3mm} 0 0]{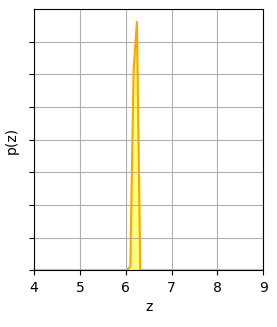} \\
   \includegraphics[width = 0.9\linewidth]{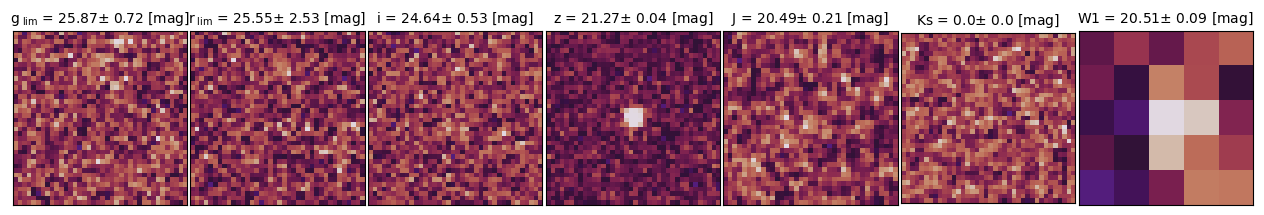} \\
   
   \vspace{2mm}
   \textbf{LS J1141+1006}\par\medskip
   \vspace{-2mm}
   \includegraphics[width = 0.57\linewidth]{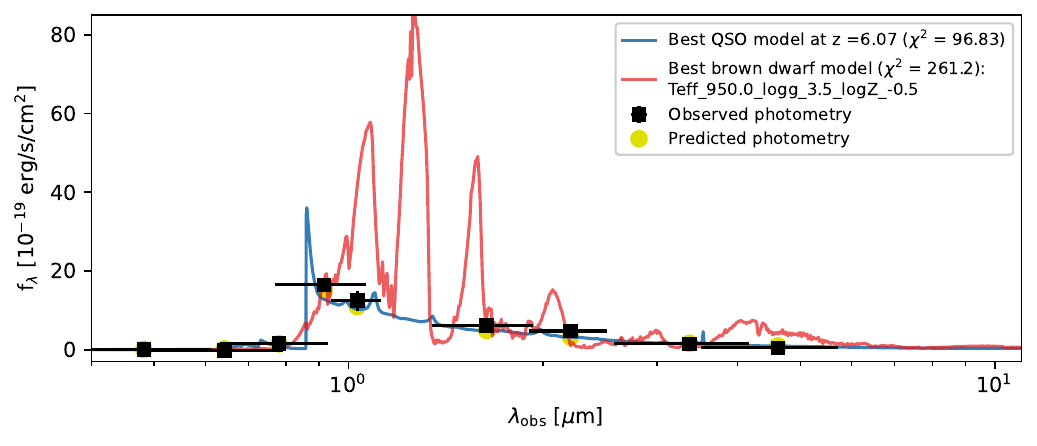}
   \includegraphics[width = 0.21\linewidth, trim = 0 {-3mm} 0 0]{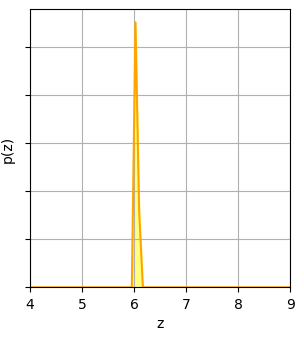} \\
   \includegraphics[width = 0.9\linewidth]{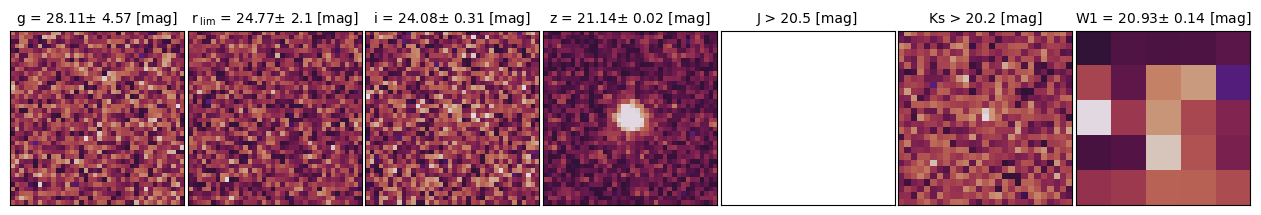} \\
   
   \vspace{2mm}
   \textbf{LS J0139-5209}\par\medskip
   \vspace{-2mm}
   \includegraphics[width = 0.57\linewidth]{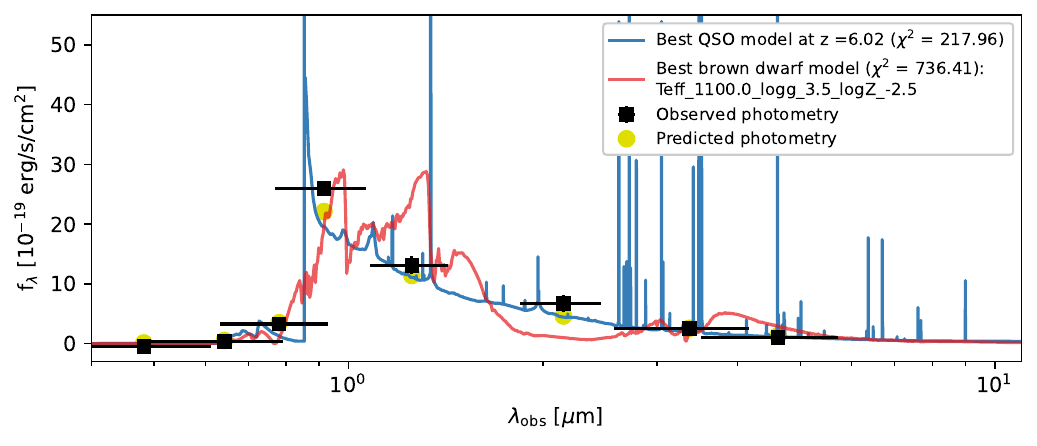}
   \includegraphics[width = 0.203\linewidth, trim = 0 {-3mm} 0 0]{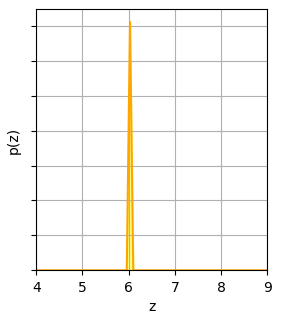} \\
   \includegraphics[width = 0.9\linewidth]{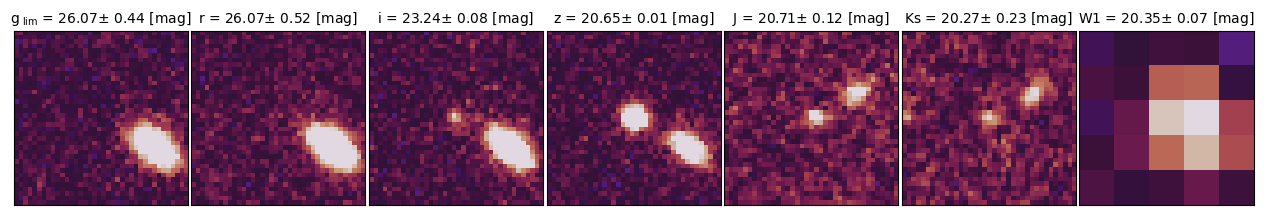}
   \caption{Continued from Fig. \ref{stamps_2}. SED fitting results and postage stamps of the quasars: LS J143510.65-105325.11 (top panel), LS J114156.14+100636.90 (central panel) and LS J013938.24-520945.73 (bottom panel).}
    \label{stamps_3}
\end{figure*}

\begin{figure*}
   \centering

   \textbf{LS J1330-4025}\par\medskip
   \vspace{-2mm}
   \includegraphics[width = 0.57\linewidth]{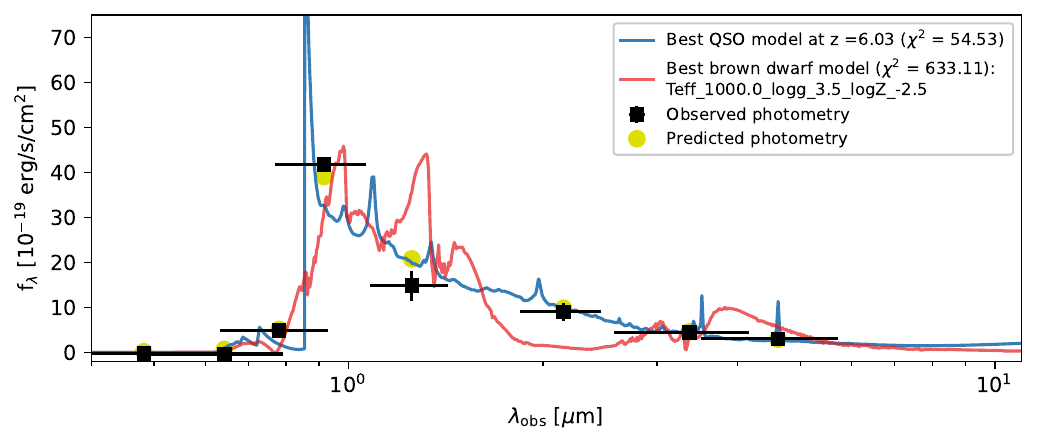}
   \includegraphics[width = 0.203\linewidth, trim = 0 {-3mm} 0 0]{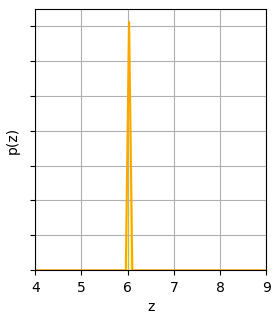} \\
   \includegraphics[width = 0.9\linewidth]{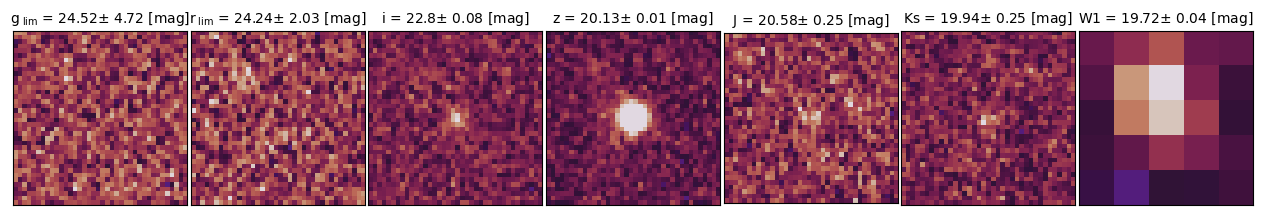} \\
   
   \vspace{2mm}
   \textbf{LS J2011-4436}\par\medskip
   \vspace{-2mm}
   \includegraphics[width = 0.57\linewidth]{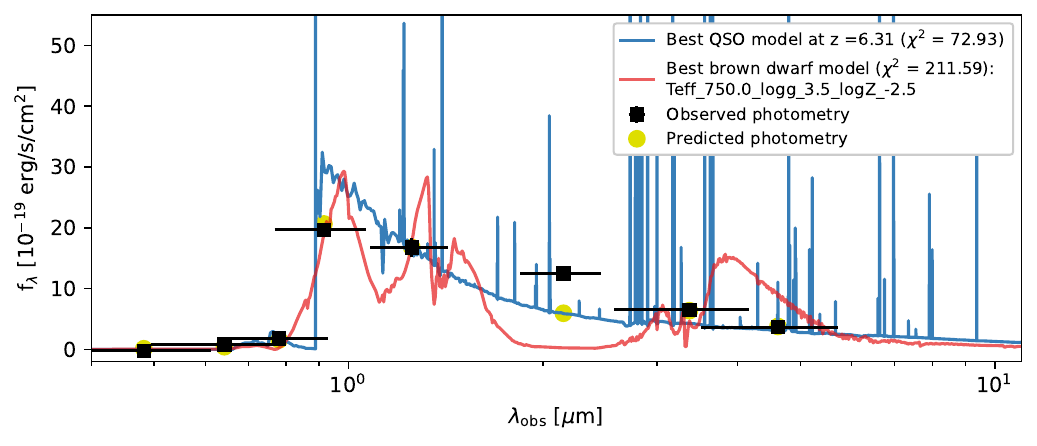}
   \includegraphics[width = 0.203\linewidth, trim = 0 {-3mm} 0 0]{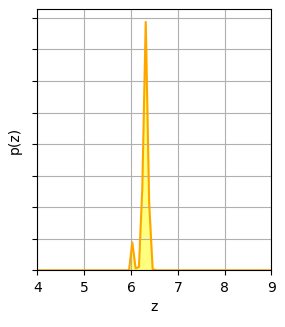} \\
   \includegraphics[width = 0.9\linewidth]{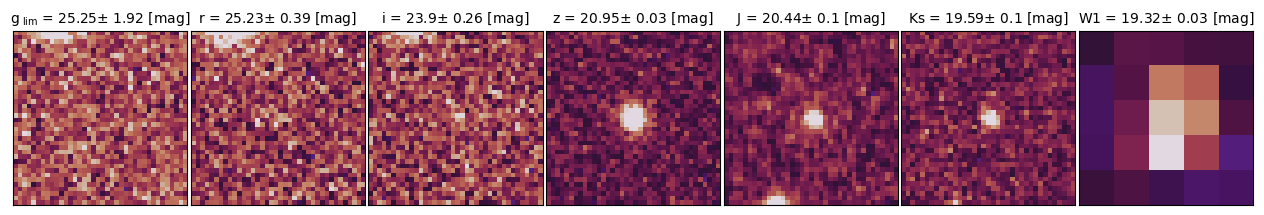} \\
   
   \vspace{2mm}
   \textbf{LS J2037-5152}\par\medskip
   \vspace{-2mm}
   \includegraphics[width = 0.57\linewidth]{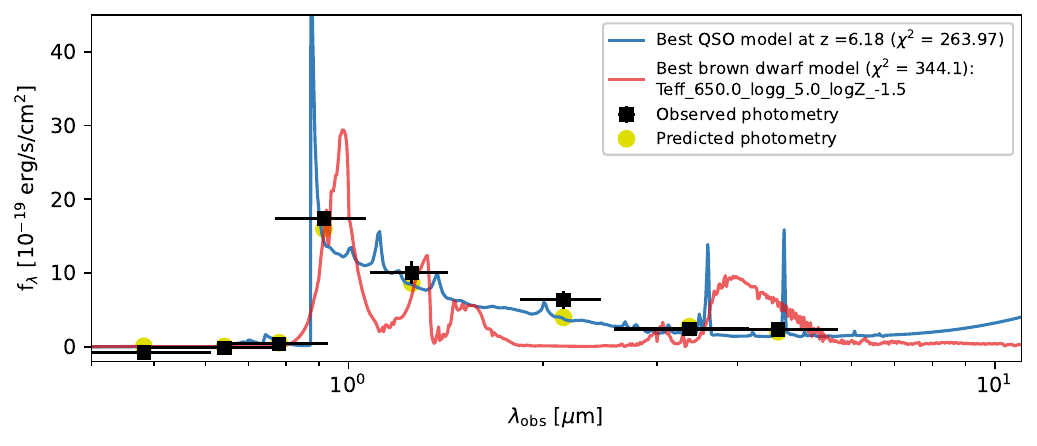}
   \includegraphics[width = 0.203\linewidth, trim = 0 {-3mm} 0 0]{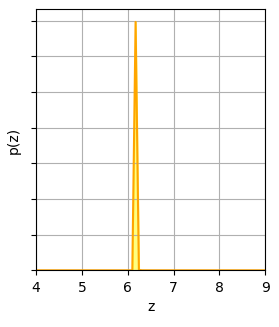} \\
   \includegraphics[width = 0.9\linewidth]{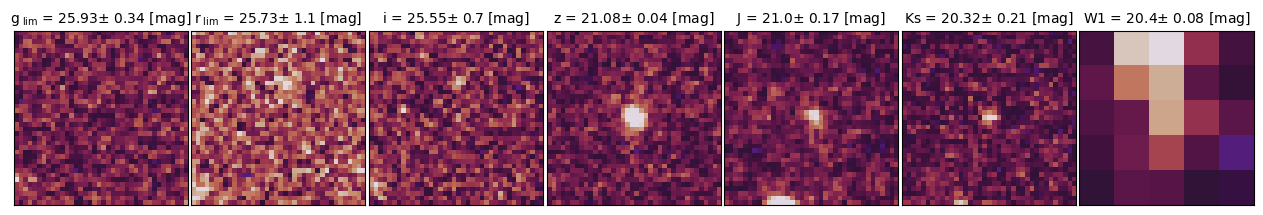} \\
   \caption{Continued from Fig. \ref{stamps_3}. SED fitting results and postage stamps of the quasars: LS J133014.01-402508.92 (top panel), LS J201119.04-443609.39 (central panel) and LS J203704.37-515240.27 (bottom panel).}
    \label{stamps_4}
\end{figure*}

\begin{figure*}
   \centering

   \textbf{LS J2155-5111}\par\medskip
   \vspace{-2mm}
   \includegraphics[width = 0.57\linewidth]{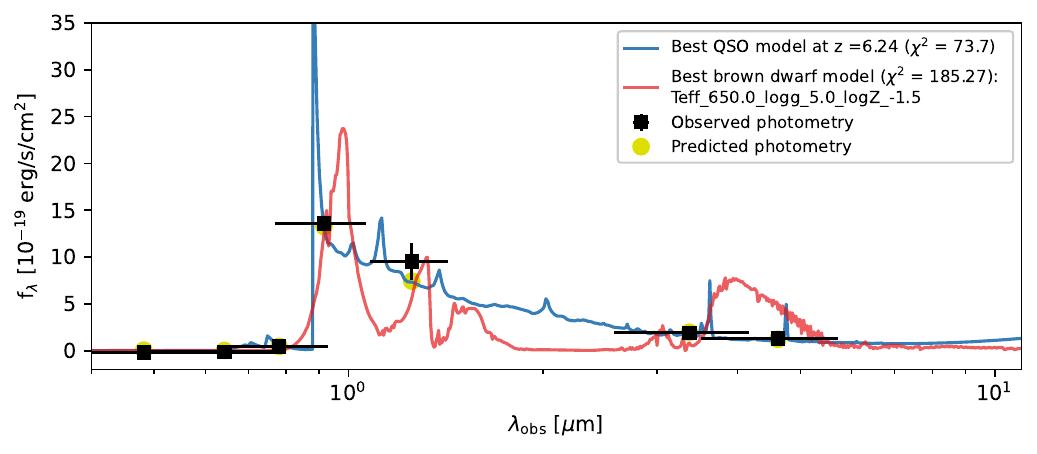}
   \includegraphics[width = 0.203\linewidth, trim = 0 {-3mm} 0 0]{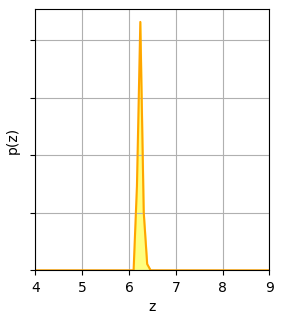} \\
   \includegraphics[width = 0.9\linewidth]{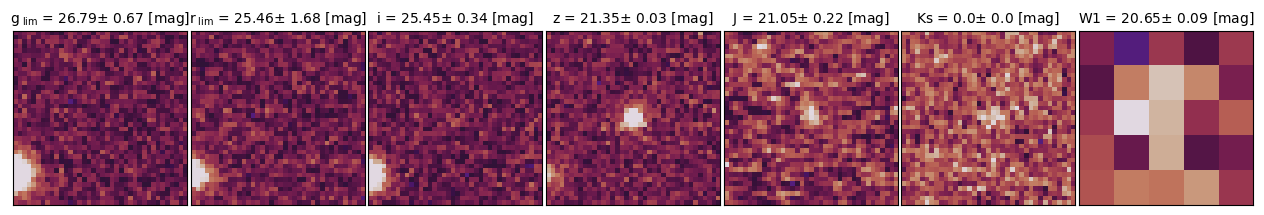} \\
   
   \vspace{2mm}
   \textbf{LS J1451-0445}\par\medskip
   \vspace{-2mm}
   \includegraphics[width = 0.57\linewidth]{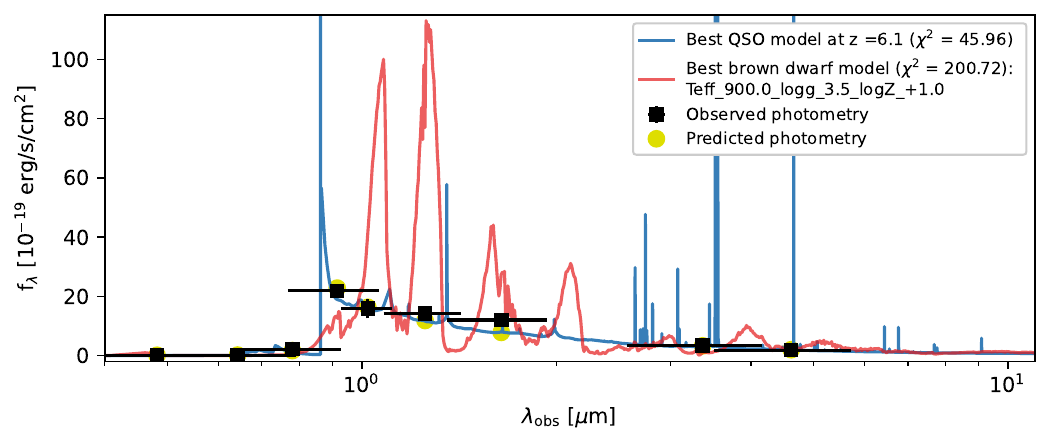}
   \includegraphics[width = 0.203\linewidth, trim = 0 {-3mm} 0 0]{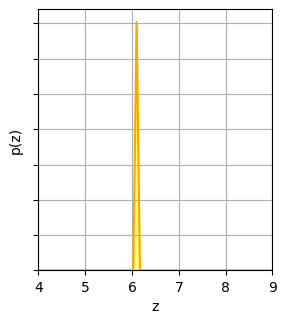} \\
   \includegraphics[width = 0.9\linewidth]{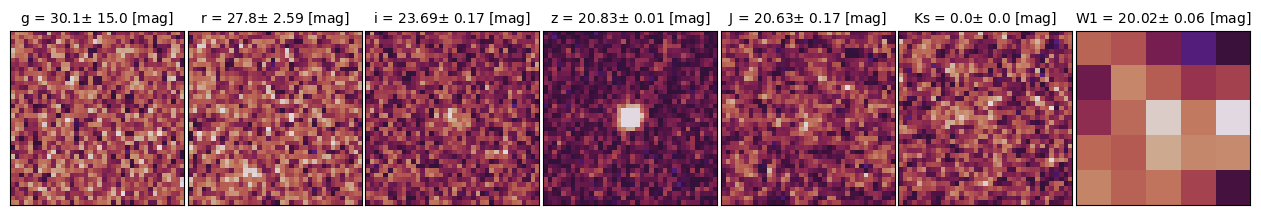}
   \caption{Continued from Fig. \ref{stamps_4}. SED fitting results and postage stamps of the quasars: LS J215501.13-511151.11 (top panel) and LS J145109.79-044542.12 (bottom panel).}
    \label{stamps_5}
\end{figure*}

\FloatBarrier 
\twocolumn
\section{Acknowledgements}

We thank the anonymous referee for their valuable suggestions and constructive report, which improved the quality and clarity of this paper. We gratefully acknowledge support from the National Agency for Research and Development (ANID) under the fellowship ANID Becas/Doctorado Nacional, \#21220337 (LNMR-R), Millennium Science Initiative Program - ICN12\_009 (LNM-R, FEB), CATA-BASAL - FB210003 (LNM-R, FEB, ET, CM), and FONDECYT Regular - \#1200495 (LNM-R, FEB) and - \#1250821 (ET);  and the Vicerrector\'ia de Investigaci\'on of Pontificia Universidad Cat\'olica de Chile under the fellowship Stay of Doctoral Co-tutelage Abroad, leading to double degree.
C.M. acknowledges support from Fondecyt Iniciacion grant 11240336.
REH acknowledges support by the German Aerospace Center (DLR) and the Federal Ministry for Economic Affairs and Energy (BMWi) through program 50OR2403 ‘RUBIES’.
The work of DS was carried out at the Jet Propulsion Laboratory, California Institute of Technology, under a contract with the National Aeronautics and Space Administration (80NM0018D0004).
RAM acknowledges support from the Swiss National Science Foundation (SNSF) through project grant 200020\_207349.
FL acknowledges support from the INAF 2022/2023 "Ricerca Fondamentale" Mini grant. 
This work is based on observations collected at the European Southern Observatory under ESO programmes  114.27MF.001 and 115.2816.001. This paper includes data from the LBT under the programmes MPIA-2024B-003 and MPIA-2025A-005. The LBT is an international collaboration among institutions in the United States, Italy, and Germany. The LBT Corporation partners are: The University of Arizona on behalf of the Arizona university system; Istituto Nazionale di Astrofisica, Italy; LBT Beteiligungsgesellschaft, Germany, representing the Max Planck Society, the Astrophysical Institute Potsdam, and Heidelberg University; The Ohio State University; The Research Corporation, on behalf of The University of Notre Dame, University of Minnesota and University of Virginia. This work includes data gathered with the 6.5 m Magellan Telescopes located at Las Campanas Observatory, Chile. This paper is based on observations collected with the Magellan/LDSS3 under the programme allocated by the Chilean Telescope Allocation Committee (CNTAC) no: CN2024B-55 and  CN2025A-26.
The Pan-STARRS1 Surveys (PS1) and the PS1 public science archive have been made possible through contributions by the Institute for Astronomy, the University of Hawaii, the Pan-STARRS Project Office, the Max-Planck Society and its participating institutes, the Max Planck Institute for Astronomy, Heidelberg and the Max Planck Institute for Extra-terrestrial Physics, Garching, The Johns Hopkins University, Durham University, the University of Edinburgh, the Queen’s University Belfast, the Harvard-Smithsonian Center for Astrophysics, the Las Cumbres Observatory Global Telescope Network Incorporated, the National Central University of Taiwan, the Space Telescope Science Institute, the National Aeronautics and Space Administration under Grant No. NNX08AR22G issued through the Planetary Science Division of the NASA Science Mission Directorate, the National Science Foundation Grant No. AST1238877, the University of Maryland, Eotvos Lorand University (ELTE), the Los Alamos National Laboratory, and the Gordon and Betty Moore Foundation. 
The Legacy Surveys consist of three individual and complementary projects: the Dark Energy Camera Legacy Survey (DECaLS; Proposal ID 2014B-0404; PIs: David Schlegel and Arjun Dey), the Beijing-Arizona Sky Survey (BASS; NOAO Prop. ID 2015A-0801; PIs: Zhou Xu and Xiaohui Fan), and the Mayall z-band Legacy Survey (MzLS; Prop. ID 2016A-0453; PI: Arjun Dey). DECaLS, BASS and MzLS together include data obtained, respectively, at the Blanco telescope, Cerro Tololo Inter-American Observatory, NSF’s NOIRLab; the Bok telescope, Steward Observatory, University of Arizona; and the Mayall telescope, Kitt Peak National Observatory, NOIRLab. Pipeline processing and analyses of the data were supported by NOIRLab and the Lawrence Berkeley National Laboratory (LBNL). The Legacy Surveys project is honored to be permitted to conduct astronomical research on Iolkam Du’ag (Kitt Peak), a mountain with particular significance to the Tohono O’odham Nation. NOIRLab is operated by the Association of Universities for Research in Astronomy (AURA) under a cooperative agreement with the National Science Foundation. LBNL is managed by the Regents of the University of California under contract to the U.S. Department of Energy. 
This project used data obtained with the Dark Energy Camera (DECam), which was constructed by the Dark Energy Survey (DES) collaboration. Funding for the DES Projects has been provided by the U.S. Department of Energy, the U.S. National Science Foundation, the Ministry of Science and Education of Spain, the Science and Technology Facilities Council of the United Kingdom, the Higher Education Funding Council for England, the National Center for Supercomputing Applications at the University of Illinois at Urbana-Champaign, the Kavli Institute of Cosmological Physics at the University of Chicago, Center for Cosmology and Astro-Particle Physics at the Ohio State University, the Mitchell Institute for Fundamental Physics and Astronomy at Texas A\&M University, Financiadora de Estudos e Projetos, Fundacao Carlos Chagas Filho de Amparo, Financiadora de Estudos e Projetos, Fundacao Carlos Chagas Filho de Amparo a Pesquisa do Estado do Rio de Janeiro, Conselho Nacional de Desenvolvimento Cientifico e Tecnologico and the Ministerio da Ciencia, Tecnologia e Inovacao, the Deutsche Forschungsgemeinschaft and the Collaborating Institutions in the Dark Energy Survey. The Collaborating Institutions are Argonne National Laboratory, the University of California at Santa Cruz, the University of Cambridge, Centro de Investigaciones Energeticas, Medioambientales y Tecnologicas- Madrid, the University of Chicago, University College London, the DES-Brazil Consortium, the University of Edinburgh, the Eidgenossische Technische Hochschule (ETH) Zurich, Fermi National Accelerator Laboratory, the University of Illinois at Urbana-Champaign, the Institut de Ciencies de l’Espai (IEEC/CSIC), the Institut de Fisica d’Altes Energies, Lawrence Berkeley National Laboratory, the Ludwig Maximilians Universitat Munchen and the associated Excellence Cluster Universe, the University of Michigan, NSF’s NOIRLab, the University of Nottingham, the Ohio State University, the University of Pennsylvania, the University of Portsmouth, SLAC National Accelerator Laboratory, Stanford University, the University of Sussex, and Texas A\&M University. BASS is a key project of the Telescope Access Program (TAP), which has been funded by the National Astronomical Observatories of China, the Chinese Academy of Sciences (the Strategic Priority Research Program “The Emergence of Cosmological Structures” Grant \# XDB09000000), and the Special Fund for Astronomy from the Ministry of Finance. The BASS is also supported by the External Cooperation Program of Chinese Academy of Sciences (Grant \# 114A11KYSB20160057), and Chinese National Natural Science Foundation (Grant \# 12120101003, \#11433005). The Legacy Survey team makes use of data products from the Near-Earth Object Wide-field Infrared Survey Explorer (NEOWISE), which is a project of the Jet Propulsion Laboratory/California Institute of Technology. NEOWISE is funded by the National Aeronautics and Space Administration. The Legacy Surveys imaging of the DESI footprint is supported by the Director, Office of Science, Office of High Energy Physics of the U.S. Department of Energy under Contract No. DE-AC02-05CH1123, by the National Energy Research Scientific Computing Center, a DOE Office of Science User Facility under the same contract; and by the U.S. National Science Foundation, Division of Astronomical Sciences under Contract No. AST-0950945 to NOAO. This project used public archival data from the Dark Energy Survey (DES). Funding for the DES Projects has been provided by the U.S. Department of Energy, the U.S. National Science Foundation, the Ministry of Science and Education of Spain, the Science and Technology Facilities Council of the United Kingdom, the Higher Education Funding Council for England, the National Center for Super-computing Applications at the University of Illinois at Urbana- Champaign, the Kavli Institute of Cosmological Physics at the University of Chicago, the Center for Cosmology and Astro-Particle Physics at the Ohio State University, the Mitchell Institute for Fundamental Physics and Astronomy at Texas A\&M University, Financiadora de Estudos e Projetos, Fundação Carlos Chagas Filho de Amparo \'a Pesquisa do Estado do Rio de Janeiro, Conselho Nacional de Desenvolvimento Cient\'ifico e Tecnol\'ogico and the Minist\'erio da Ciencia, Tecnologia e Inovac\~ao, the Deutsche Forschungsgemeinschaft, and the Collaborating Institutions in the Dark Energy Survey. The Collaborating Institutions are Argonne National Laboratory, the University of California at Santa Cruz, the University of Cambridge, Centro de Investigaciones Energ\'eticas, Medioambientales y Tecnol\'ogicas-Madrid, the University of Chicago, University College London, the DES-Brazil Consortium, the University of Edinburgh, the Eidgenössische Technische Hochschule (ETH) Zurich, Fermi National Accelerator Laboratory, the University of Illinois at Urbana-Champaign, the Institut de Ciències de l’Espai (IEEC/CSIC), the Institut de Física d’Altes Energies, Lawrence Berkeley National Laboratory, the Ludwig- Maximilians Universit\"at M\"unchen and the associated Excellence Cluster Universe, the University of Michigan, the National Optical Astronomy Observatory, the University of Nottingham, The Ohio State University, the OzDES Membership Consortium, the University of Pennsylvania, the University of Portsmouth, SLAC National Accelerator Laboratory, Stanford University, the University of Sussex, and Texas A\&M University. Based in part on observations at Cerro Tololo Inter-American Observatory, National Optical Astronomy Observatory, which is operated by the Association of Universities for Research in Astronomy (AURA) under a cooperative agreement with the National Science Foundation. This publication makes use of data products from the Wide-field Infrared Survey Explorer, which is a joint project of the University of California, Los Angeles, and the Jet Propulsion Laboratory/Caltech, funded by the National Aeronautics and Space Administration. Based on observation obtained as part of the VISTA Hemisphere Survey, ESO Program, 179.A-2010 (PI: McMahon).
The VISTA Data Flow System pipeline processing and science archive are described in \cite{irwin2004vista} and \cite{hambly2008wfcam}. This research made use of Astropy, a community-developed core Python package for Astronomy \citep{price2018astropy}.

\end{appendix}

\end{document}